\documentclass[journal=acscii, manuscript=article]{achemso}
\usepackage[version=3]{mhchem}
\usepackage{amsmath}
\usepackage{textcomp}
\usepackage{booktabs}
\usepackage{multirow}
\usepackage[labelfont=bf]{caption}
\usepackage{hyperref}
\usepackage{xr}
\usepackage{xurl}
\usepackage[super]{nth}
\makeatletter
\newcommand*{\addFileDependency}[1]{
  \typeout{(#1)}
  \@addtofilelist{#1}
  \IfFileExists{#1}{}{\typeout{No file #1.}}
}
\makeatother
\newcommand*{\myexternaldocument}[1]{
    \externaldocument{#1}
    \addFileDependency{#1.tex}
    \addFileDependency{#1.aux}
}
\myexternaldocument{si}

\title{Assessing Thermodynamic Selectivity of Solid-State Reactions for the Predictive Synthesis of Inorganic Materials}

\author{Matthew J. McDermott}
\affiliation{Materials Sciences Division, Lawrence Berkeley National Laboratory, Berkeley, CA}
\alsoaffiliation{Department of Materials Science and Engineering, University of California, Berkeley, CA}
\altaffiliation{These authors contributed equally to this work.}
\author{Brennan C. McBride}
\affiliation{Department of Chemistry, Colorado State University, Fort Collins, CO}
\altaffiliation{These authors contributed equally to this work.}
\author{Corlyn Regier}
\affiliation{Department of Chemistry, Colorado State University, Fort Collins, CO}
\author{Gia Thinh Tran}
\affiliation{Department of Chemistry, Colorado State University, Fort Collins, CO}
\author{Yu Chen}
\affiliation{Materials Sciences Division, Lawrence Berkeley National Laboratory, Berkeley, CA}
\alsoaffiliation{Department of Materials Science and Engineering, University of California, Berkeley, CA}
\author{Adam A. Corrao}
\affiliation{Department of Chemistry, Stony Brook University, Stony Brook, NY}
\author{Max C. Gallant}
\affiliation{Materials Sciences Division, Lawrence Berkeley National Laboratory, Berkeley, CA}
\alsoaffiliation{Department of Materials Science and Engineering, University of California, Berkeley, CA}
\author{Gabrielle E. Kamm}
\affiliation{Department of Chemistry, Stony Brook University, Stony Brook, NY}
\author{Christopher J. Bartel}
\affiliation{Department of Chemical Engineering and Materials Science, University of Minnesota, Minneapolis, MN}
\author{Karena W. Chapman}
\affiliation{Department of Chemistry, Stony Brook University, Stony Brook, NY}
\author{Peter G. Khalifah}
\affiliation{Department of Chemistry, Stony Brook University, Stony Brook, NY}
\alsoaffiliation{Chemistry Division, Brookhaven National Laboratory, Upton, NY}
\author{Gerbrand Ceder}
\affiliation{Materials Sciences Division, Lawrence Berkeley National Laboratory, Berkeley, CA}
\alsoaffiliation{Department of Materials Science and Engineering, University of California, Berkeley, CA}
\author{James R. Neilson} 
\affiliation{Department of Chemistry, Colorado State University, Fort Collins, CO}
\author{Kristin A. Persson}
\affiliation{Molecular Foundry, Lawrence Berkeley National Laboratory, Berkeley, CA}
\alsoaffiliation{Department of Materials Science and Engineering, University of California, Berkeley, CA}
\email{kapersson@lbl.gov}

\begin{document}

\pagebreak

\begin{abstract}
Synthesis is a major challenge in the discovery of new inorganic materials. Currently, there is limited theoretical guidance for identifying optimal solid-state synthesis procedures. We introduce two selectivity metrics, primary and secondary competition, to assess the favorability of target/impurity phase formation in solid-state reactions. We used these metrics to analyze 3,520 solid-state reactions in the literature, ranking existing approaches to popular target materials. Additionally, we implemented these metrics in a data-driven synthesis planning workflow and demonstrated its application in the synthesis of barium titanate (\ce{BaTiO3}). Using an 18-element chemical reaction network with first-principles thermodynamic data from the Materials Project, we identified 82,985 possible \ce{BaTiO3} synthesis reactions and selected nine for experimental testing. Characterization of reaction pathways via synchrotron powder X-ray diffraction reveals that our selectivity metrics correlate with observed target/impurity formation. We discovered two efficient reactions using unconventional precursors (\ce{BaS}/\ce{BaCl2} and \ce{Na2TiO3}) that produce \ce{BaTiO3} faster and with fewer impurities than conventional methods, highlighting the importance of considering complex chemistries with additional elements during precursor selection. Our framework provides a foundation for predictive inorganic synthesis, facilitating the optimization of existing recipes and the discovery of new materials, including those not easily attainable with conventional precursors.
\end{abstract}

\pagebreak

\section{Introduction}

The predictive synthesis of inorganic materials remains a grand challenge in modern chemistry and materials science.\cite{Kovnir2021} Unlike organic synthesis, which is often described via discrete reaction steps or mechanisms, inorganic materials synthesis reactions cannot be deconstructed into elementary steps,\cite{Disalvo1990, Kohlmann2019} hindering the analogous development of retrosynthetic analysis techniques \cite{Corey1989} and computer-aided synthesis planning.\cite{Coley2018} This lack of successful mechanistic models has made the synthesis of predicted new materials a critical bottleneck in high-throughput computational materials design efforts,\cite{Kim2017} with many proposed materials having yet to be successfully synthesized. \cite{Narayan2016, Rong2017, Gardner2019}

\begin{figure}[ht!]
\begin{center}
\includegraphics[width=0.9\textwidth]{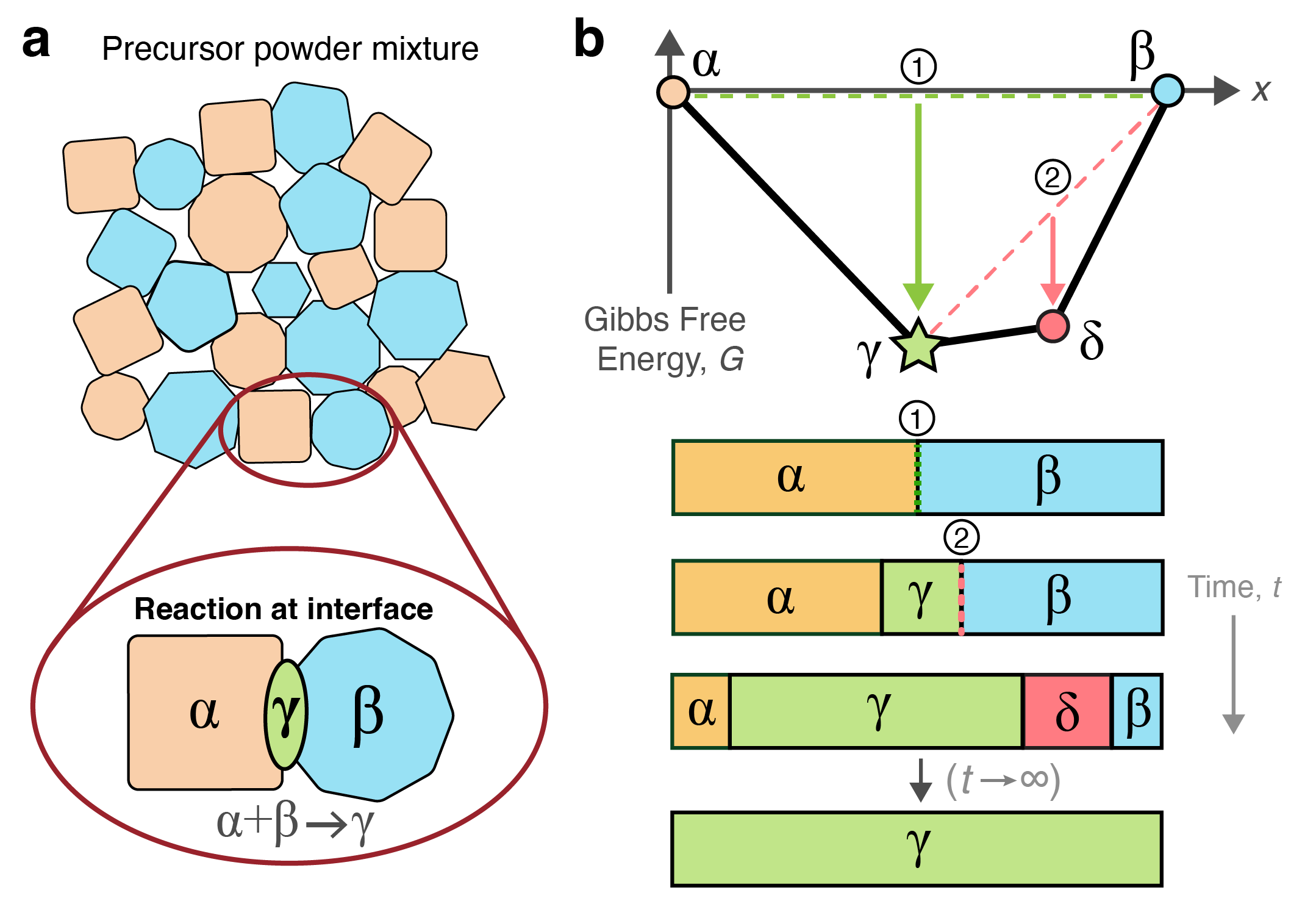}
\caption{\label{fig:interface_model} \textbf{Modeling chemical reactions at heterogeneous solid interfaces in a binary/quasibinary chemical system}. (a) Cartoon model of a powder reaction between the hypothetical precursors: $\alpha$ (orange) and $\beta$ (blue). The nucleation of a new phase, $\gamma$ (green), occurs at the $\alpha|\beta$ interface according to the reaction \ce{$\alpha$ + $\beta$ -> $\gamma$}. (b) A possible reaction pathway for the powder system in which a secondary reaction of the equilibrium phase ($\gamma$) yields an impurity phase, $\delta$ (red). The interface reaction hull (top) shows available interfacial reactions and their corresponding Gibbs free energies, $G$, and mixing ratios, $x$. The one-dimensional spatial model (bottom) shows reaction steps beginning from an equal mixture of $\alpha$ and $\beta$. The impurity phase, $\delta$, may be kinetically retained in a local equilibrium state; however, with infinite time, the system should approach the global equilibrium state composed entirely of $\gamma$.}
\end{center} 
\end{figure}

While there are numerous inorganic synthesis methods (e.g., hydrothermal, mechanochemical, sol-gel, intercalation, etc.),\cite{West2014} we limit the scope of this work to bulk solid-state synthesis via powder reactions. This choice has been motivated by the straightforward and scalable nature of working with bulk powders, which makes solid-state synthesis suitable for applications ranging from one-off laboratory synthesis to industrial mass manufacturing. In powder reactions, product formation proceeds via nucleation and growth at interfacial contact areas in the powder mixture (Figure \ref{fig:interface_model}a).\cite{Miura2021} The equilibrium phases of the reacting system can be predicted by constructing a convex hull in free energy and composition space, where the composition axis is a mixing ratio between the two precursor compositions (Figure \ref{fig:interface_model}b).\cite{Richards2015} Here, we calculate the convex hull exclusively using normalized compositions and energies (i.e., on a per-atom basis). This construction, which we refer to herein as the ``interface reaction hull'', is a subsection of the compositional phase diagram for binary systems and a ``quasibinary'' two-dimensional slice of the full phase diagram for chemical systems with three or more elements. The exact product species, and the sequence in which they appear, cannot be predicted with thermodynamics alone; to do so requires intimate knowledge of the kinetic rates of all physically feasible reactions. However, a commonly adopted theoretical simplification assumes that the reaction product(s) with the most negative \textit{pairwise} reaction energy will be the first to nucleate and grow as a powder mixture is heated.\cite{Bianchini2020, Miura2021} This hypothesis is based on two principles: 1) the random packing of solid crystallites results in very few locations where three or more particles are simultaneously in contact, and 2) the activation energy barrier to nucleation scales inversely with free energy as $\Delta G^\dagger \propto \frac{\gamma^3}{\Delta G_{rxn}^2}$. The surface energy, $\gamma$, is particularly important when comparing the feasibility of reactions with similar energies.\cite{Aykol2021} Despite this, many solid-state reactions are likely not nucleation-limited but rather \textit{transport-limited} due to the relative sluggishness of solid-state diffusion and the often large driving forces of these reactions (10-100 kJ/mol).\cite{dHeurle1988}

The complex interplay between thermodynamics and kinetics makes solid-state synthesis prone to the unpredictable and undesirable formation of impurity phases.\cite{Bianchini2020} A classic example of a nonselective synthesis is that of the prototypical multiferroic bismuth ferrite, \ce{BiFeO3}, via the standard reaction from binary oxides: \ce{Bi2O3 + Fe2O3 -> 2\ BiFeO3}. This reaction typically yields impurity phases \ce{Bi2Fe4O9} and \ce{Bi25FeO39}, which are challenging to isolate and remove.\cite{Kumar2000, Bernardo2011} Unfortunately, the presence of an impurity phase is difficult to predict \textit{a priori}, and is typically attributed to ``kinetic'' factors or changes in phase equilibria related to precursor purity, morphology, volatility, or processing conditions. To optimize the performance of solid-state reactions and maximize conversion to the desired target, the experimentalist frequently relies on intuition and heuristic rules to choose the 1) precursor compositions (typically off-the-shelf binary phases such as carbonates, oxides, etc.), 2) grinding/milling protocol, 3) synthesis annealing temperature, 4) synthesis atmosphere (e.g., vacuum, flowing \ce{O2}), 5) synthesis time, and 6) cooling protocol. Heuristics include well-known rules such as Tamman's Rule for estimating reaction onset temperature (i.e., two-thirds the melting temperature of the precursor with the lowest melting point),\cite{Merkle2005} as well as ``chemical intuition'' or human-biased experimental protocols (e.g., selecting synthesis times based on common increments, such as four hours, eight hours, etc.).\cite{Huo2022} Unfortunately, these heuristics may be insufficient to achieve successful synthesis on the first attempt(s), necessitating follow-up experiments that can be time-intensive and costly. In the worst cases, human-biased heuristics lead to lower success rates than randomly-generated experimental protocols.\cite{Jia2019}

The \textit{a priori} calculation of reaction selectivity in solid-state synthesis permits the ranking of synthesis approaches based on their thermodynamic likelihood of success, thereby circumventing the current time-consuming trial-and-error (Edisonian) approach. The assessment of reaction selectivity is particularly relevant in the proposal of optimal synthesis precursors;\cite{He2023a} in several cases, improved navigation of the phase diagram was shown to lead to a more practical synthesis \cite{Jiang2017, Miura2020, Miura2021, Todd2021, Chen2023}. However, no solid-state reaction selectivity metric has been formally established. In recent work,\cite{Aykol2021} Aykol et al.\ demonstrated a computational workflow for ranking solid-state synthesis reactions by two performance metrics: 1) a catalytic nucleation barrier factor incorporating structural similarity and epitaxial matching, and 2) the number of known competing phases. These metrics perform well in rationalizing successful syntheses in the literature but lack generality; for example, the nucleation metric is derived assuming all reactions are nucleation-limited, which, as discussed, is not true for many solid-state reactions. Additionally, while a metric based on the total \textit{number} of competing phases is significant as it hints at a measure of reaction selectivity, such a scheme does not account for the \textit{relative} stability of these competing phases. A count-based selectivity metric is also biased by how many phases are known to exist at the present time and the extent to which various structural configurations (e.g., disordered or defective phases) have been enumerated within the data.

In this work, we address the longstanding issue of assessing the selectivity of solid-state reactions by deriving two complementary thermodynamic metrics measuring the degree of phase competition from the interface reaction model. We incorporate these competition metrics into a computational synthesis planning workflow for identifying and ranking synthesis reactions, which builds upon the high-throughput reaction enumeration tools we previously developed for constructing solid-state chemical reaction networks\cite{McDermott2021} from large materials databases such as the Materials Project.\cite{Jain2013} Our selectivity metrics, computational workflow, literature analysis, and experimental findings yield a framework for generating more optimal and efficient solid-state synthesis routes, providing a foundation for the predictive synthesis of inorganic materials. The suggestion of nonstandard precursors, particularly those involving additional elements beyond those in the target composition, expands the synthetic capabilities of the solid-state approach.

\section{Results and Discussion}

\subsection{Derivation of Selectivity Metrics}

\subsubsection{The Interface Reaction Hull}

To construct the interface reaction hull (Figure \ref{fig:interface_model}b), one begins with thermodynamic data for the reacting system (i.e., a set of relevant phases and their compositions and energies). In this work, we acquire formation enthalpies, $\Delta H_f$, from the Materials Project database \cite{Jain2013} and extend them to Gibbs free energies of formation, $\Delta G_f (T)$, through the use of a prior machine learning model \cite{Bartel2018} and supplemental experimental thermochemistry data \cite{Chase1998} (see Methods). For systems with two elements, the interface reaction hull is equivalent to the binary compositional phase diagram, where each vertex represents a single phase. However, for systems with three or more elements, the non-precursor vertices include both single phases and mixtures of phases. These mixtures are stoichiometric combinations of phases representing the products of balanced reactions of the precursors. The balanced reactions can be determined via 1) computing slices of the full compositional phase diagram along the tie-line connecting the precursors \cite{Richards2015} or 2) combinatorial reaction enumeration.\cite{McDermott2021} More specifically, the maximum number of products for a particular vertex is one less than the number of elements in the system. The interface reaction hull is thus generalized such that all non-precursor vertices correspond to reactions with coordinates given by the atomic mixing ratio of precursors, $x$, and the Gibbs free energy of the reaction, $\Delta G_\text{rxn}$. This model can be further generalized to environmental conditions other than fixed temperature and pressure by constructing the hull with the appropriate thermodynamic potential. For example, in open systems, one would use the grand potential energy, $\Phi$. Note that in these systems, the hull vertices may include additional open elemental reactants/products (e.g., \ce{O2}) that do not factor into the determination of $x$.

As a model for solid-state reactions, the interface reaction hull construction also rationalizes the formation of impurity phases. To demonstrate this, we revisit the binary system in Figure \ref{fig:interface_model}. In this system, the target phase ($\gamma$) is predicted to form first because it is the phase with the highest driving force of formation (most negative $\Delta G_\textrm{rxn}$), irrespective of the average composition of the total system \cite{Richards2015}. When the $\gamma$ phase forms, however, it also introduces two additional interfaces: $\alpha \vert \gamma$ and $\gamma \vert \beta$. These secondary interfaces can be modeled via the construction of new interface reaction hulls or, more simply, by splitting the original hull into two subsections (i.e., to the left and the right of the target). Figure \ref{fig:interface_model}b suggests that the $\gamma \vert \beta$ interface should produce an impurity phase via the exergonic reaction, $\gamma + \beta \longrightarrow \delta$. Hence, the full conversion of reactants to the target phase is impeded while $\delta$ persists. In a ``one-dimensional'' solid-state reaction (Figure \ref{fig:interface_model}b, bottom), local thermodynamic equilibrium may be achieved when the system reaches a state in which all stable product phases on the interface reaction hull have formed and the growth of the product layer(s) slows down until it ceases entirely. This situation has been observed in previous experimental studies on diffusion couples.\cite{Schmalzried1981, vanLoo1990} This observed mixture of products may be kinetically ``stable'' (i.e., with locally stable interfaces), but it is not the global equilibrium state of the system. Rather, the global equilibrium state is the combination of phases that minimizes the free energy given the composition of the entire powder mixture; in Figure \ref{fig:interface_model}b, this corresponds to entirely $\gamma$. In powder reactions, access to the global equilibrium state is often provided by re-grinding steps in which new interfaces are exposed and mixed to facilitate complete conversion to the equilibrium products. However, in reacting systems with significant phase competition and slow transport, high temperatures and long heating times may be necessary but impractical; pure target synthesis may not be achievable if the desired products are unstable at high temperatures. These situations can be avoided entirely by proposing alternative precursors that are more selective (i.e., those with interface reaction hulls containing few to no competing phases).

\subsubsection{Measuring Phase Competition}

\begin{figure}[ht!]
\begin{center}
\includegraphics[width=0.4\textwidth]{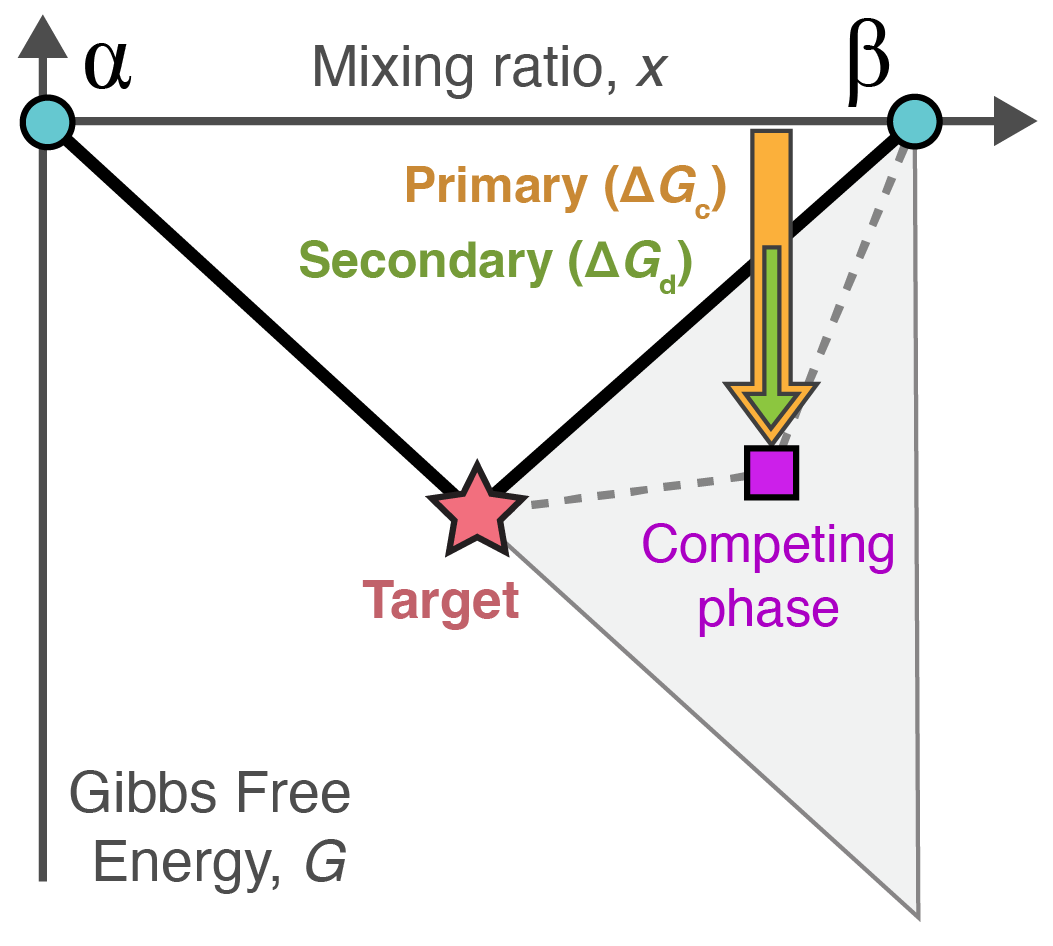}
\caption{\label{fig:competition} \textbf{Origin of primary and secondary competition in solid-state reactions}. In this simple interface reaction hull for a binary (two-element) system, two interface reactions can form a competing phase (magenta square). The primary reaction (yellow arrow) occurs at the interface between the two precursors $\alpha$ and $\beta$, whereas the secondary reaction (green arrow) occurs between the target phase (pink star) and the remaining $\beta$ precursor, leading to a smaller driving force (arrow length). The coordinates of the competing phase, which must lie within the illustrated bounds (gray triangle) if the target phase is to be stable, determine the relative values of the primary and secondary reaction energies.}
\end{center} 
\end{figure}

To predict the thermodynamic selectivity of a solid-state reaction, we propose two complementary metrics for assessing phase competition using the interface reaction hull: primary ($C_1$) and secondary ($C_2$) competition. Although both metrics measure the relative energetic favorability of competing reactions, they model different mechanisms for impurity formation. The primary competition measures the favorability of competing reactions of the original precursors, while the secondary competition measures the favorability of \textit{subsequent} competing reactions between the precursors and target phase(s). The origin of the two competition mechanisms is illustrated in Figure \ref{fig:competition}. 

Primary competition, $C_1$, is measured via calculation of the relative thermodynamic advantage of the most exergonic competing reaction from the original precursors, as assessed through an energy difference:
\begin{equation}
 C_1 = \Delta G_\textrm{rxn} - \min_i(\Delta G_{c_i}).
 \label{eq:c1}
\end{equation}
Here, $\Delta G_\textrm{rxn}$ is the energy of the target synthesis reaction, and $\Delta G_{c_i}$ are the energies of possible competing reactions from the precursors. Lower $C_1$ values are favorable and result in more selective target formation. When $C_1$ is positive, the target reaction is less energetically favorable than the competing reaction with the greatest driving force (most negative energy), suggesting that a competing phase is likely to form. On the other hand, when $C_1$ is negative, the target reaction is predicted to have the greatest driving force of any reaction on the hull. By considering only the single most competitive reaction, this functional form avoids the aforementioned bias related to using the total \textit{number} of competing reactions. When no exergonic competing reactions are predicted for an interface, the competing reaction energy term is assigned a value of zero, representing the scenario in which the precursors do not react (e.g., \ce{$\alpha$ -> $\alpha$}). This results in the limiting condition: $C_1 \geq \Delta G_\textrm{rxn}$.

Secondary competition, $C_2$, assesses the favorability of impurity phase formation via secondary reactions between the target and precursor(s). This metric is important and distinct from primary competition because it measures the relative stability of the products of the target synthesis reaction with respect to decomposition into the competing phase(s). Their relative stability can be measured by computing the ``inverse distance to the hull'' (Figure \ref{fig:competition}), which for systems with one competing phase, is equivalent to the secondary reaction energy, $\Delta G_d$, at the precursor-target interface.

\begin{figure}[ht!]
\begin{center}
\includegraphics[width=1.0\textwidth]{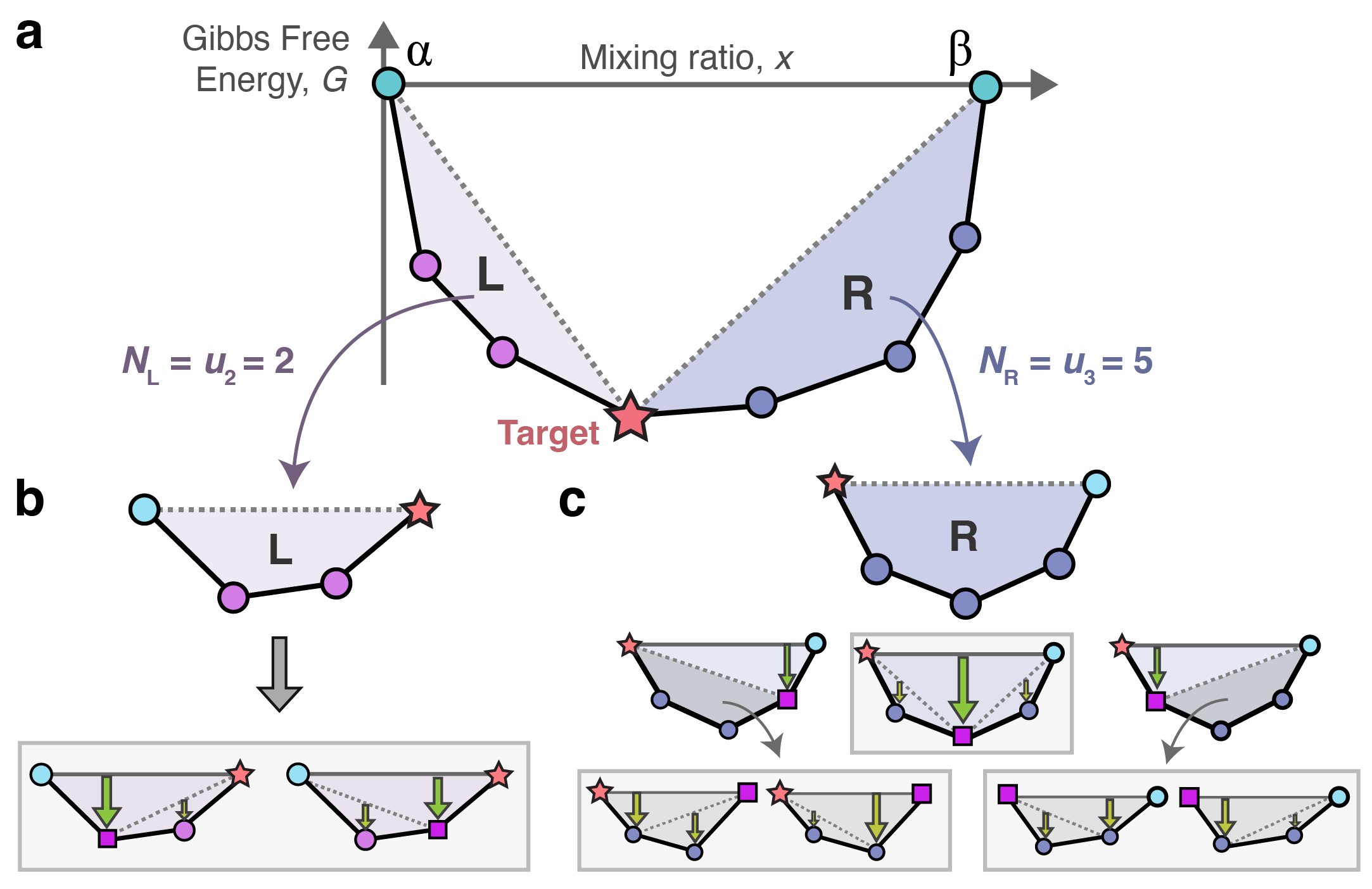}
\caption{\label{fig:secondary} \textbf{Secondary reaction sequences in an interface reaction hull}. (a) The hull is divided into two subsections to the left (L) and right (R) of the target, representing the two additional precursor-target interfaces. (b-c) Secondary reaction sequences on either side of the target, with gray boxes highlighting the final reaction sequences. The recursive, binary nature results in the number of unique sequences, $N(n)$, following the Catalan numbers $u_n$.}
\end{center} 
\end{figure}

In an interface reaction hull with several competing phases, a sequence of multiple secondary reactions may occur (Figure \ref{fig:secondary}). When the target phase is formed, it introduces two new precursor-target interfaces that divide the hull into subsections to the left and right of the target. A secondary reaction may occur in either subsection, exposing another two interfaces. If, at either interface, there is a remaining driving force to form additional competing phases, then this process may continue in a recursive fashion until all possible secondary reactions have occurred. There are multiple ways to draw a feasible secondary reaction sequence (Figures \ref{fig:secondary}b,c). Consider a particular secondary reaction sequence indexed $j$. This sequence has a total energy given by the sum of the energies of its $n$ steps,
\begin{equation}
\Delta G_{2,j}= \sum_{k=1}^n \Delta G_{d_{j,k}} = \Delta G_{d_{j,1}} + \Delta G_{d_{j,2}}  + ... + \Delta G_{d_{j,n}},
\label{eqn:secondary_seq}
\end{equation}
where the number of reaction steps ($n$) in the sequence also equals the number of non-reactant (interior) vertices in the hull subsection. 

If every secondary reaction step is required to be the one with the minimum energy (i.e., largest driving force), then only one unique reaction sequence exists in the left and right hull subsections. However, one must consider the situation where the minimum-energy principle does not hold due to kinetic limitations; this applies in particular to hulls where all secondary reactions have similar magnitude driving forces or a particular phase is kinetically limited from forming, perhaps due to an overall small driving force. Therefore, we must consider the alternative secondary reaction sequences shown in Figures \ref{fig:secondary}b,c. These alternative sequences are not necessarily less favorable; while each alternative sequence may feature a first reaction step with a smaller driving force (i.e., small $\Delta G_{d_{j,1}}$), the latter steps may have larger magnitude energies, resulting in comparable total energy ($\Delta G_{2,j}$) for the particular sequence.

To encompass all combinatorial possibilities in our estimation of secondary competition, we choose to compute the mean total energy of all feasible secondary reaction sequences:
\begin{equation}
\overline{\Delta G_{2}}= \frac{1}{N}\sum_{j=1}^N \Delta G_{2,j}.
\label{eqn:secondary_mean}
\end{equation}
Determining the total number of unique secondary reaction sequences, $N$, is mathematically equivalent to calculating the total number of full binary trees with $n$ interior nodes, which yields the Catalan number sequence, $u_n = 1, 1, 2, 5, 14, 42, 132, 429, ... (n=0,1,2, ...)$.\cite{Stanley1999} Using this connection to the Catalan numbers, we developed a non-recursive algorithm for calculating $\overline{\Delta G_{2}}$ that is significantly faster than the equivalent recursive solution (see Methods).

Finally, we formulate the $C_2$ metric such that it accounts for all possible reaction sequences in either hull subsection to the left (L) and right (R) of the target phase:
\begin{equation}
C_2 =-\left(\overline{\Delta G_{2,L}} +  \overline{\Delta G_{2,R}} \right).
\label{eq:c2}
\end{equation}
The negative factor is included so that a lower $C_2$ value corresponds to a more favorable selectivity. Because our definition of a secondary reaction assumes that $\Delta G_d \leq 0$, the secondary competition metric obeys the limiting behavior: $C_2 \geq 0$. 
 
We note that both $C_1$ and $C_2$ implicitly assume that the target phase is thermodynamically stable (``on the hull'') under the conditions for which the equilibrium phase diagram is derived. However, the competition metrics are still calculable for a metastable phase by manually decreasing its energy until it becomes stable.

\subsection{Application to Experimental Literature}

\begin{figure}[ht!]
\begin{center}
\includegraphics[width=1.0\textwidth]{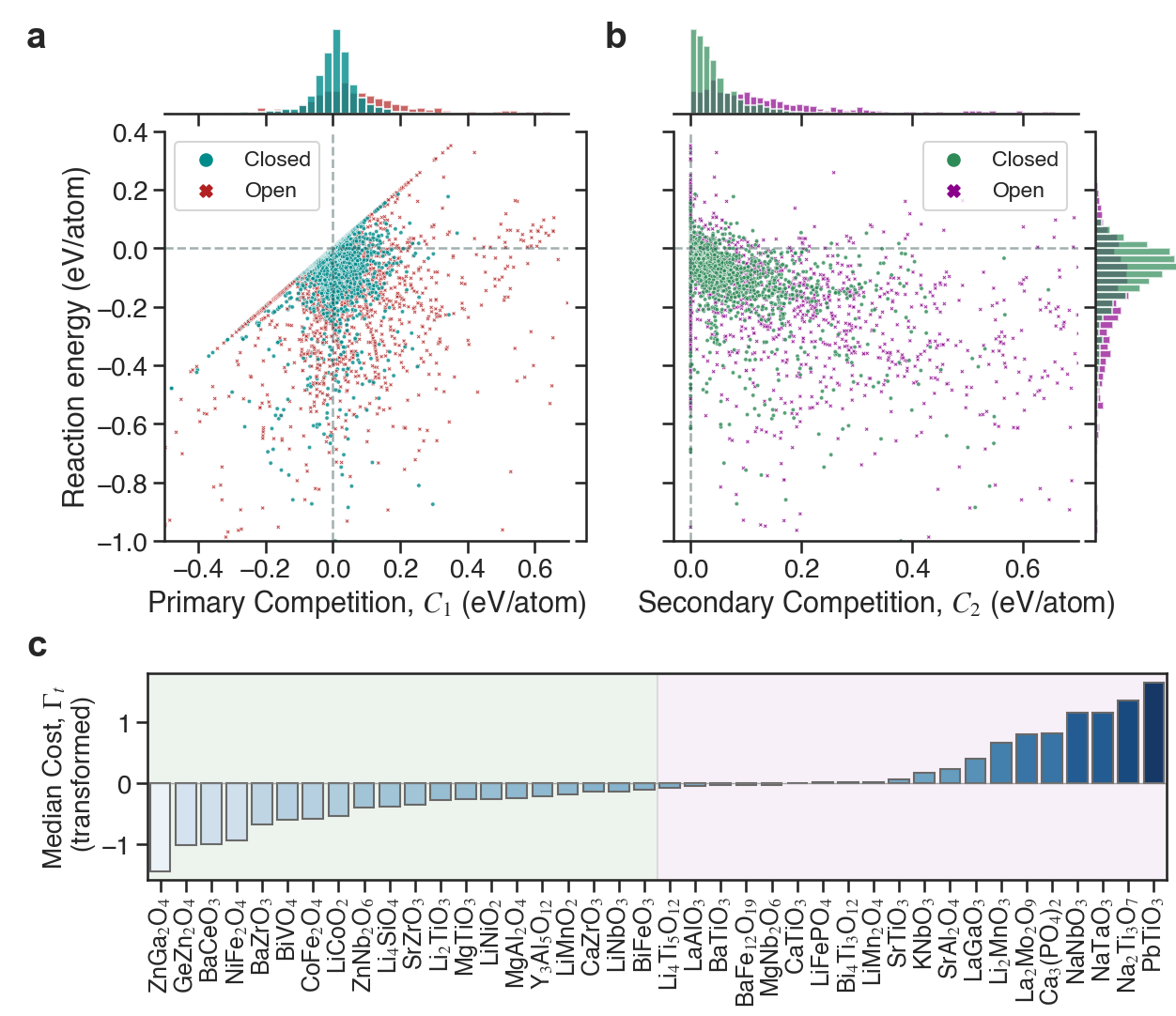}
\caption{\label{fig:literature} \textbf{Thermodynamic analysis of synthesis recipes in the experimental solid-state literature}. Synthesis maps of 3,520 literature reactions from the Kononova et al.\ dataset, plotted on a shared axis of reaction energy, $\Delta E_{\textrm{rxn}}$, and independent axes of (a) primary competition, $C_1$, and (b) secondary competition, $C_2$. These selectivity metrics are constrained by their lower bounds: $C_1=\Delta E_{\textrm{rxn}}$ (diagonal parity line) and $C_2=0$. (c) Median transformed cost ($\Gamma_t$) rankings of synthesis recipes for each of the 40 most popular targets in the dataset. The shading marks the targets with average recipes below (green) or above (pink) the median $\Gamma_t$ of all experiments in the dataset (-0.089). Selected recipes are discussed in Tables \ref{tab:lit_rxns_easy} and \ref{tab:lit_rxns_hard}. The full data set is provided in the Supporting Information.}
\end{center} 
\end{figure}

Using the competition metrics $C_1$ and $C_2$, we now assess the selectivities of solid-state reactions previously reported in the experimental literature and use these to rank synthesis recipes by their predicted thermodynamic optimality. Reaction energies and competition metrics were calculated for 3,520 unique experiments reported in the text-mined solid-state reaction literature dataset by Kononova et al. \cite{Kononova2019}. Each unique experiment corresponds to a particular balanced reaction, maximum synthesis temperature, and atmospheric environment (e.g., air, flowing \ce{O2}, etc.). We modeled all reactions containing up to two solid precursors and an optional gaseous reactant (see Methods). Reactions that were reported with no particular atmospheric environment are denoted as ``closed'' and modeled with Gibbs free energies ($\Delta G_\textrm{rxn}$), while those with a defined environment are denoted as ``open'' and modeled using grand potential energies ($\Delta \Phi_\text{rxn}$). The results of these calculations are shown in Figures \ref{fig:literature}a,b.

The synthesis ``maps'' (Figures \ref{fig:literature}a,b) allow one to identify favorable reactions by comparing the relative weights of the three reaction metrics: reaction energy ($\Delta E_{\textrm{rxn}}$), primary competition ($C_1$), and secondary competition ($C_2$). Thermodynamically optimal reactions are ones that minimize all three metrics, resulting in placement in the lower left region of each plot. According to our calculations, many reactions reported in the literature are predicted to be energetically favorable and selective.  Approximately 17.3\% of reactions have no exergonic competing reactions on the interface reaction hull (i.e., they are on the bounding line $C_1 = \Delta E_\textrm{rxn}$), and 23.0\% of reactions have negligible secondary competition ($C_2 \leq 0.001$ eV/atom). Assuming the literature reactions are experimentally feasible, one would expect all calculated reaction energies to be negative. Our thermodynamic modeling captures this within reasonable error: 82.8\% of reactions have a negative reaction energy, and 97.0\% of reactions have $\Delta E_\textrm{rxn} \leq 0.1$ eV/atom.  Of the reactions with positive energies, most (87.1\%) contain one or more common gases: \ce{O2} (85.1\%), \ce{CO2} (48.8\%), or \ce{H2O} (8.4\%). A major source of error in our reaction energy calculations is likely the disagreement between the assumed and actual gas partial pressures; it is often challenging to model the actual environmental conditions of synthesis, as it requires that they be both 1) accurately reported and 2) correctly extracted from the text. Another known source of error is systematic challenges in estimating GGA-calculated formation energies of carbonate compounds.\cite{Aykol2021} However, this has been addressed and partially mitigated with a fitted energy correction (see Methods).

To quantitatively assess thermodynamic optimality, we follow a similar approach to prior works\cite{McDermott2021, Todd2021} and define a cost function, $\Gamma$, that combines the driving force and reaction selectivity evaluated at a particular set of conditions (i.e., temperature and atmosphere). In this work, we opt to use a simple linear weighted summation of the reaction's energy and competition scores,
\begin{equation}
\Gamma = x_0\Delta E_{\textrm{rxn}}+ x_1C_1 +x_2C_2,
\label{eq:cost_function}
\end{equation}
where $\Delta E_\text{rxn}$ is the reaction energy (either $\Delta G_\textrm{rxn}$ or $\Delta \Phi_\textrm{rxn}$), and $x_0$, $x_1$, and $x_2$ are user-defined weights for each parameter. Due to the different scaling of each parameter, we find that a non-equal weighting of $x_0=0.10$, $x_1=0.45$, and $x_2=0.45$ produces reasonably diverse results that do not significantly favor one parameter over another. We note that this selection is arbitrary and subject to further optimization.

Unfortunately, closed and open reactions cannot be rigorously compared due to their different energy scales (i.e., Gibbs free energies vs.\ grand potential energies). This is further evidenced by differences in the reaction metric distributions (Figures \ref{fig:literature}a,b). Since no ideal solution exists for ranking and comparing reactions under different environmental conditions, we transform the cost function to account for the energy scale difference. To do this, we apply a power transformation to the cost distributions for the closed and open reactions, resulting in monotonically transformed costs, $\Gamma_t$, whose distributions are closer to standard normal distributions (Figure \ref{supp_fig:power_transform}). This new variable facilitates a fairer comparison between closed and open reactions, allowing for a more realistic ranking of synthesis recipes.

Figure \ref{fig:literature}c shows the median values of $\Gamma_t$ calculated from the synthesis recipes of the 40 most popular targets in the literature dataset (i.e., those with the most reactions extracted from the text). Given the extent of coverage of our thermodynamic data, we limit our analysis to only the targets for which we have at least five recipes successfully calculated (see Methods). In the following sections, we select several targets with costs below and above the median $\Gamma_t$ value, analyzing the factors leading to their optimal and suboptimal synthesis recipes, respectively.

\subsubsection{Optimal Literature Recipes}

Of the 40 most popular targets in the literature dataset (Figure \ref{fig:literature}c), twenty have synthesis recipes with an average cost value ($\Gamma_t$) below the dataset's median (-0.089), indicating generally favorable thermodynamic optimality. A selection of these targets and their highest/lowest-ranked recipes are provided in Table \ref{tab:lit_rxns_easy}, along with other selected reactions of interest. DOIs and raw (untransformed) costs for each reaction are provided in the Supporting Information.

For many targets in Table \ref{tab:lit_rxns_easy}, the conventional reaction involving off-the-shelf binary precursors is predicted to be thermodynamically optimal. For example, in the synthesis of the spinel \ce{ZnGa2O4}, the reaction between the binary oxides, \ce{ZnO + Ga2O3 -> ZnGa2O4}, is already perfectly selective ($C_1=\Delta E_\textrm{rxn}$ and $C_2=0$) over all temperatures in the dataset. Furthermore, these precursors appear in \textit{all} 17 calculated literature recipes for this target. The favorability of the conventional route seems to apply to several other targets presented here, including \ce{BiVO4}, \ce{CoFe2O4}, \ce{Y3Al5O12}, and \ce{BiFeO3}. Since the conventional precursors for these targets are more favorable than the explored alternatives, we can instead analyze which synthesis \textit{conditions} are most optimal. For \ce{CoFe2O4}, higher temperatures in an open oxygen environment appear to be the most favorable. On the other hand, lower temperatures in an open oxygen environment favor the production of \ce{BiVO4} and \ce{BiFeO3}. Finally, for \ce{Y3Al5O12}, there appears to be a tradeoff between selectivity (which is optimal at lower temperatures) and driving force (which is optimal at high temperatures); it appears that intermediate temperatures (600 \textcelsius{}) in a closed environment result in the most suitable compromise. We note, however, that there are many reasons to use specific processing conditions outside of the pursuit of target phase purity (e.g., improved density, annihilation of defects, optimization of crystallite sizes, etc.). These alternative reasons may explain some of the variability of conditions reported in the literature.

Interestingly, several targets in Table \ref{tab:lit_rxns_easy} feature synthesis recipes that appear extremely unfavorable. These often involve elemental precursors, such as \ce{Bi + 0.5\ V2O5 + 0.75\ O2 -> BiVO4}, or \ce{Bi + 0.5\ Fe2O3 + 0.75\ O2 -> BiFeO3}. Referencing the original articles from which these reactions were sourced \cite{Cooper2016, Selbach2009a} suggests that these precursors were not actually used, and the reaction's inclusion in the dataset is likely the result of an error in the literature extraction process. This explains some other impractical and highly suboptimal routes in the dataset, such as reactions involving alkali metal precursors (e.g., \ce{4\ Li + SiO2 + O2 -> Li4SiO4}).

For other targets such as \ce{LiCoO2}, \ce{Li4SiO4}, and \ce{LiNiO2}, there appears to be great variability in precursor selection in the literature. For \ce{LiCoO2}, the optimal synthesis recipe from our calculations is the reaction of \ce{LiOH} and \ce{Co3O4} in a flowing oxygen environment at low to intermediate temperatures (i.e., 700 \textcelsius{}). The use of \ce{Co3O4} leads to significantly greater performance than recipes using \ce{CoCO3}. Additionally, using \ce{LiOH} may offer a thermodynamic advantage over \ce{Li2CO3}, although with little effect on the reaction selectivity. A similar conclusion applies to the synthesis of \ce{Li4SiO4}, although the use of lithium carbonate appears to be more favorable at high temperatures in an open-oxygen environment. \ce{LiOH} (particularly the monohydrate) also appears to offer some advantage in the synthesis of \ce{LiNiO2}, especially in oxygen at low temperatures (480-600 \textcelsius{}).

\begin{table}[ht!]
\scriptsize
\caption{Thermodynamic analysis of optimal experimental synthesis recipes for selected popular targets in the literature. The ranking of reactions is determined by the transformed cost, $\Gamma_t$. The highest- and lowest-ranked reactions are shown along with selected reactions of interest. Raw costs and DOIs are available in the Supporting Information.}
\label{tab:lit_rxns_easy}
\begin{tabular}{ccp{5.7cm}cccccc}
\toprule
 &  & Reaction & Temp. & Energy & Open & C$_1$ & C$_2$ & $\Gamma_t$ \\
Target & Rank &  & (ºC) & (eV/at) &  & (eV/at) & (eV/at) & (a.u.) \\
\midrule
\multirow{3}{*}{\textbf{ZnGa$_2$O$_4$}} & \textbf{1} & ZnO + Ga$_{2}$O$_{3}$ $\longrightarrow$ ZnGa$_2$O$_4$ & 1400 & -0.218 & O & -0.218 & 0.000 & -1.622 \\
\textbf{} & \textbf{12} & ZnO + Ga$_{2}$O$_{3}$ $\longrightarrow$ ZnGa$_2$O$_4$ & 1200 & -0.088 &  & -0.088 & 0.000 & -0.969 \\
\textbf{} & \textbf{17} & ZnO + Ga$_{2}$O$_{3}$ $\longrightarrow$ ZnGa$_2$O$_4$ & 500 & -0.067 &  & -0.067 & 0.000 & -0.818 \\
\cline{1-9}
\multirow{3}{*}{\textbf{BiVO$_4$}} & \textbf{1} & 0.5 Bi$_{2}$O$_{3}$ + 0.5 V$_{2}$O$_{5}$ $\longrightarrow$ BiVO$_4$ & 600 & -0.529 & O & -0.210 & 0.171 & -1.156 \\
\textbf{} & \textbf{5} & 0.5 Bi$_{2}$O$_{3}$ + 0.5 V$_{2}$O$_{5}$ $\longrightarrow$ BiVO$_4$ & 500 & -0.169 &  & -0.061 & 0.034 & -0.713 \\
\textbf{} & \textbf{12} & Bi + 0.5 V$_{2}$O$_{5}$ + 0.75 O$_{2}$ $\longrightarrow$ BiVO$_4$ & 600 & -1.860 & O & 0.801 & 4.489 & 2.412 \\
\cline{1-9}
\multirow{4}{*}{\textbf{CoFe$_2$O$_4$}} & \textbf{1} & CoO + Fe$_{2}$O$_{3}$ $\longrightarrow$ CoFe$_2$O$_4$ & 1250 & -0.088 & O & -0.062 & 0.000 & -0.877 \\
\textbf{} & \textbf{3} & 0.3333 Co$_{3}$O$_{4}$ + Fe$_{2}$O$_{3}$ $\longrightarrow$ CoFe$_2$O$_4$ + 0.1667 O$_{2}$ & 950 & -0.058 & O & -0.018 & 0.011 & -0.662 \\
\textbf{} & \textbf{11} & Co + 2 Fe + 2 O$_{2}$ $\longrightarrow$ CoFe$_2$O$_4$ & 1400 & -1.738 & O & 0.223 & 2.957 & 2.111 \\
\textbf{} & \textbf{14} & CoCO$_{3}$ + Fe$_{2}$O$_{3}$ $\longrightarrow$ CoFe$_2$O$_4$ + CO$_{2}$ & 1400 & -0.215 &  & 0.156 & 0.370 & 2.781 \\
\cline{1-9}
\multirow{4}{*}{\textbf{LiCoO$_{2}$}} & \textbf{1} & LiOH + 0.3333 Co$_{3}$O$_{4}$ + 0.08333 O$_{2}$ $\longrightarrow$ LiCoO$_{2}$ + 0.5 H$_{2}$O & 700 & -0.120 & O & -0.012 & -0.000 & -0.727 \\
\textbf{} & \textbf{4} & 0.5 Li$_{2}$CO$_{3}$ + 0.3333 Co$_{3}$O$_{4}$ + 0.08333 O$_{2}$ $\longrightarrow$ LiCoO$_{2}$ + 0.5 CO$_{2}$ & 950 & -0.061 & O & -0.014 & -0.000 & -0.692 \\
\textbf{} & \textbf{23} & 0.5 Li$_{2}$CO$_{3}$ + CoCO$_{3}$ + 0.25 O$_{2}$ $\longrightarrow$ LiCoO$_{2}$ + 1.5 CO$_{2}$ & 300 & -0.121 & O & 0.154 & 0.357 & 0.532 \\
\textbf{} & \textbf{30} & LiOH$\cdot$H$_2$O + Co + 0.75 O$_{2}$ $\longrightarrow$ LiCoO$_{2}$ + 1.5 H$_{2}$O & 900 & -0.382 & O & 1.001 & 1.454 & 1.992 \\
\cline{1-9}
\multirow{4}{*}{\textbf{Li$_{4}$SiO$_{4}$}} & \textbf{1} & 2 Li$_{2}$CO$_{3}$ + SiO$_{2}$ $\longrightarrow$ Li$_{4}$SiO$_{4}$ + 2 CO$_{2}$ & 1445 & -0.040 & O & -0.017 & -0.000 & -0.683 \\
\textbf{} & \textbf{7} & 4 LiOH + SiO$_{2}$ $\longrightarrow$ Li$_{4}$SiO$_{4}$ + 2 H$_{2}$O & 700 & -0.049 &  & -0.004 & -0.000 & -0.412 \\
\textbf{} & \textbf{10} & 2 Li$_{2}$CO$_{3}$ + SiO$_{2}$ $\longrightarrow$ Li$_{4}$SiO$_{4}$ + 2 CO$_{2}$ & 1150 & -0.010 &  & -0.005 & -0.000 & -0.366 \\
\textbf{} & \textbf{18} & 4 Li + SiO$_{2}$ + O$_{2}$ $\longrightarrow$ Li$_{4}$SiO$_{4}$ & 800 & -2.133 & O & 0.163 & 2.830 & 2.038 \\
\cline{1-9}
\multirow{5}{*}{\textbf{LiNiO$_{2}$}} & \textbf{1} & LiOH$\cdot$H$_2$O + NiO + 0.25 O$_{2}$ $\longrightarrow$ LiNiO$_{2}$ + 1.5 H$_{2}$O & 480 & 0.008 & O & 0.010 & 0.002 & -0.559 \\
\textbf{} & \textbf{4} & 0.5 Li$_{2}$O + NiO + 0.25 O$_{2}$ $\longrightarrow$ LiNiO$_{2}$ & 800 & 0.032 & O & 0.058 & 0.021 & -0.344 \\
\textbf{} & \textbf{6} & LiOH + Ni(OH)$_2$ + 0.25 O$_{2}$ $\longrightarrow$ LiNiO$_{2}$ + 1.5 H$_{2}$O & 770 & 0.074 & O & 0.074 & -0.000 & -0.331 \\
\textbf{} & \textbf{24} & 0.5 Li$_{2}$O$_{2}$ + NiO $\longrightarrow$ LiNiO$_{2}$ & 800 & -0.035 &  & 0.067 & 0.147 & 0.973 \\
\textbf{} & \textbf{27} & Li + Ni(OH)$_2$ + 0.5 O$_{2}$ $\longrightarrow$ LiNiO$_{2}$ + H$_{2}$O & 700 & -0.536 & O & 1.770 & 3.044 & 2.396 \\
\cline{1-9}
\multirow{3}{*}{\textbf{Y$_{3}$Al$_{5}$O$_{12}$}} & \textbf{1} & 1.5 Y$_{2}$O$_{3}$ + 2.5 Al$_{2}$O$_{3}$ $\longrightarrow$ Y$_{3}$Al$_{5}$O$_{12}$ & 600 & -0.071 &  & -0.034 & 0.000 & -0.623 \\
\textbf{} & \textbf{15} & 1.5 Y$_{2}$O$_{3}$ + 2.5 Al$_{2}$O$_{3}$ $\longrightarrow$ Y$_{3}$Al$_{5}$O$_{12}$ & 1727 & -0.240 & O & -0.051 & 0.221 & -0.260 \\
\textbf{} & \textbf{36} & 1.5 Y$_{2}$O$_{3}$ + 5 Al(OH)$_3$ $\longrightarrow$ Y$_{3}$Al$_{5}$O$_{12}$ + 7.5 H$_{2}$O & 1300 & -0.120 &  & -0.007 & 0.118 & 0.202 \\
\cline{1-9}
\multirow{3}{*}{\textbf{BiFeO$_3$}} & \textbf{1} & 0.5 Bi$_{2}$O$_{3}$ + 0.5 Fe$_{2}$O$_{3}$ $\longrightarrow$ BiFeO$_3$ & 600 & -0.073 & O & 0.034 & 0.053 & -0.386 \\
\textbf{} & \textbf{4} & 0.5 Bi$_{2}$O$_{3}$ + 0.5 Fe$_{2}$O$_{3}$ $\longrightarrow$ BiFeO$_3$ & 100 & -0.011 &  & 0.000 & 0.004 & -0.312 \\
\textbf{} & \textbf{38} & Bi + 0.5 Fe$_{2}$O$_{3}$ + 0.75 O$_{2}$ $\longrightarrow$ BiFeO$_3$ & 900 & -1.285 & O & 1.087 & 3.558 & 2.364 \\
\cline{1-9}
\bottomrule
\end{tabular}
\end{table}

\subsubsection{Suboptimal Literature Recipes}

Reactions can still be successful despite high $C_1$ and $C_2$ values. However, our thermodynamic assessment suggests that these reactions are suboptimal and likely require some combination of 1) long heating times to promote thermodynamic equilibrium, 2) follow-up regrinding steps, or 3) fine-tuning of temperature and reaction atmosphere. In Table \ref{tab:lit_rxns_hard}, we highlight several popular target materials that are associated with higher-than-average costs for their synthesis recipes.

\begin{table}[ht!]
\scriptsize
\caption{Thermodynamic analysis of suboptimal experimental synthesis recipes for selected popular targets in the literature. The ranking of reactions is determined by the transformed cost, $\Gamma_t$. The highest- and lowest-ranked reactions are shown along with selected reactions of interest. Raw costs and DOIs are available in the Supporting Information.}
\label{tab:lit_rxns_hard}
\begin{tabular}{ccp{5.7cm}cccccc}
\toprule
 &  & Reaction & Temp. & Energy & Open & C$_1$ & C$_2$ & $\Gamma_t$ \\
Target & Rank &  & (ºC) & (eV/at) &  & (eV/at) & (eV/at) & (a.u.) \\
\midrule
\multirow{3}{*}{\textbf{PbTiO$_3$}} & \textbf{1} & PbCO$_{3}$ + TiO$_{2}$ $\longrightarrow$ PbTiO$_3$ + CO$_{2}$ & 1200 & -0.402 & O & 0.118 & 0.750 & 0.960 \\
\textbf{} & \textbf{2} & PbO + TiO$_{2}$ $\longrightarrow$ PbTiO$_3$ & 1100 & -0.360 & O & 0.186 & 0.694 & 0.987 \\
\textbf{} & \textbf{13} & PbO + TiO$_{2}$ $\longrightarrow$ PbTiO$_3$ & 800 & -0.163 &  & 0.136 & 0.288 & 2.172 \\
\cline{1-9}
\multirow{3}{*}{\textbf{Na$_{2}$Ti$_{3}$O$_{7}$}} & \textbf{1} & 2 NaOH + 3 TiO$_{2}$ $\longrightarrow$ Na$_{2}$Ti$_{3}$O$_{7}$ + H$_{2}$O & 750 & -0.058 & O & 0.079 & 0.046 & -0.273 \\
\textbf{} & \textbf{3} & Na$_{2}$CO$_{3}$ + 3 TiO$_{2}$ $\longrightarrow$ Na$_{2}$Ti$_{3}$O$_{7}$ + CO$_{2}$ & 1100 & 0.009 &  & 0.108 & 0.104 & 1.025 \\
\textbf{} & \textbf{8} & Na$_{2}$CO$_{3}$ + 3 TiO$_{2}$ $\longrightarrow$ Na$_{2}$Ti$_{3}$O$_{7}$ + CO$_{2}$ & 80 & 0.053 &  & 0.202 & 0.200 & 2.344 \\
\cline{1-9}
\multirow{3}{*}{\textbf{NaTaO$_{3}$}} & \textbf{1} & 0.5 Na$_{2}$CO$_{3}$ + 0.5 Ta$_{2}$O$_{5}$ $\longrightarrow$ NaTaO$_{3}$ + 0.5 CO$_{2}$ & 1150 & -0.088 &  & 0.035 & 0.085 & 0.300 \\
\textbf{} & \textbf{4} & 0.5 Na$_{2}$CO$_{3}$ + 0.5 Ta$_{2}$O$_{5}$ $\longrightarrow$ NaTaO$_{3}$ + 0.5 CO$_{2}$ & 1200 & -0.233 & O & 0.285 & 0.665 & 1.098 \\
\textbf{} & \textbf{9} & 0.5 Na$_{2}$CO$_{3}$ + 0.5 Ta$_{2}$O$_{5}$ $\longrightarrow$ NaTaO$_{3}$ + 0.5 CO$_{2}$ & 900 & -0.194 & O & 0.441 & 0.866 & 1.435 \\
\cline{1-9}
\multirow{5}{*}{\textbf{Ca$_{3}$(PO$_{4}$)$_{2}$}} & \textbf{1} & CaCO$_{3}$ + Ca$_{2}$P$_{2}$O$_{7}$ $\longrightarrow$ Ca$_{3}$(PO$_{4}$)$_{2}$ + CO$_{2}$ & 800 & -0.068 &  & -0.013 & -0.000 & -0.494 \\
\textbf{} & \textbf{3} & 3 CaCO$_{3}$ + 2 NH$_4$H$_2$PO$_4$ $\longrightarrow$ Ca$_{3}$(PO$_{4}$)$_{2}$ + 3 CO$_{2}$ + 3 H$_{2}$O + 2 H$_{3}$N & 1150 & -0.199 & H & 0.078 & 0.269 & 0.187 \\
\textbf{} & \textbf{4} & 3 CaCO$_{3}$ + 2 (NH$_4$)$_2$HPO$_4$ $\longrightarrow$ Ca$_{3}$(PO$_{4}$)$_{2}$ + 3 CO$_{2}$ + 3 H$_{2}$O + 4 H$_{3}$N & 1200 & -0.156 & N & 0.105 & 0.374 & 0.463 \\
\textbf{} & \textbf{10} & 3 CaO + P$_{2}$O$_{5}$ $\longrightarrow$ Ca$_{3}$(PO$_{4}$)$_{2}$ & 550 & -0.641 &  & -0.007 & 0.463 & 1.673 \\
\textbf{} & \textbf{12} & 3 CaCO$_{3}$ + 2 (NH$_4$)$_2$HPO$_4$ $\longrightarrow$ Ca$_{3}$(PO$_{4}$)$_{2}$ + 3 CO$_{2}$ + 3 H$_{2}$O + 4 H$_{3}$N & 1200 & -0.143 &  & 0.094 & 0.336 & 2.232 \\
\cline{1-9}
\multirow{6}{*}{\textbf{Li$_{2}$MnO$_{3}$}} & \textbf{1} & Li$_{2}$CO$_{3}$ + MnO$_{2}$ $\longrightarrow$ Li$_{2}$MnO$_{3}$ + CO$_{2}$ & 975 & -0.096 & O & -0.015 & 0.015 & -0.670 \\
\textbf{} & \textbf{4} & 2 LiOH + MnO$_{2}$ $\longrightarrow$ Li$_{2}$MnO$_{3}$ + H$_{2}$O & 650 & -0.092 &  & -0.024 & 0.010 & -0.533 \\
\textbf{} & \textbf{5} & 2 LiOH$\cdot$H$_2$O + MnO$_{2}$ $\longrightarrow$ Li$_{2}$MnO$_{3}$ + 3 H$_{2}$O & 70 & -0.059 &  & 0.001 & 0.016 & -0.295 \\
\textbf{} & \textbf{14} & Li$_{2}$CO$_{3}$ + MnCO$_{3}$ + 0.5 O$_{2}$ $\longrightarrow$ Li$_{2}$MnO$_{3}$ + 2 CO$_{2}$ & 500 & -0.196 & O & 0.262 & 0.545 & 0.942 \\
\textbf{} & \textbf{15} & 2 LiOH$\cdot$H$_2$O + MnCO$_{3}$ + 0.5 O$_{2}$ $\longrightarrow$ Li$_{2}$MnO$_{3}$ + CO$_{2}$ + 3 H$_{2}$O & 450 & -0.187 & O & 0.253 & 0.554 & 0.943 \\
\textbf{} & \textbf{25} & 2 Li + MnO$_{2}$ + 0.5 O$_{2}$ $\longrightarrow$ Li$_{2}$MnO$_{3}$ & 700 & -1.971 & O & 0.400 & 2.399 & 1.989 \\
\cline{1-9}
\multirow{5}{*}{\textbf{LiMn$_{2}$O$_{4}$}} & \textbf{1} & 0.5 Li$_{2}$CO$_{3}$ + 2 MnO$_{2}$ $\longrightarrow$ LiMn$_{2}$O$_{4}$ + 0.5 CO$_{2}$ + 0.25 O$_{2}$ & 700 & -0.032 & O & 0.041 & 0.051 & -0.347 \\
\textbf{} & \textbf{2} & LiOH + 2 MnO$_{2}$ $\longrightarrow$ LiMn$_{2}$O$_{4}$ + 0.5 H$_{2}$O + 0.25 O$_{2}$ & 1000 & -0.052 & O & 0.031 & 0.067 & -0.343 \\
\textbf{} & \textbf{12} & 0.5 Li$_{2}$CO$_{3}$ + Mn$_{2}$O$_{3}$ + 0.25 O$_{2}$ $\longrightarrow$ LiMn$_{2}$O$_{4}$ + 0.5 CO$_{2}$ & 950 & -0.073 & O & 0.030 & 0.110 & -0.240 \\
\textbf{} & \textbf{47} & 0.5 Li$_{2}$CO$_{3}$ + 2 MnCO$_{3}$ + 0.75 O$_{2}$ $\longrightarrow$ LiMn$_{2}$O$_{4}$ + 2.5 CO$_{2}$ & 400 & -0.319 & O & 0.148 & 0.488 & 0.664 \\
\textbf{} & \textbf{63} & 0.5 Li$_{2}$O$_{2}$ + Mn$_{2}$O$_{3}$ $\longrightarrow$ LiMn$_{2}$O$_{4}$ & 800 & -0.138 &  & 0.121 & 0.283 & 2.072 \\
\cline{1-9}
\multirow{5}{*}{\textbf{LiFePO$_{4}$}} & \textbf{1} & 0.3333 Li$_{3}$PO$_{4}$ + 0.3333 Fe$_{3}$(PO$_{4}$)$_{2}$ $\longrightarrow$ LiFePO$_{4}$ & 600 & -0.020 &  & -0.008 & 0.002 & -0.386 \\
\textbf{} & \textbf{2} & LiPO$_{3}$ + 0.5 Fe$_{2}$O$_{3}$ $\longrightarrow$ LiFePO$_{4}$ + 0.25 O$_{2}$ & 1200 & -0.151 & O & 0.108 & 0.120 & -0.060 \\
\textbf{} & \textbf{3} & 0.3333 Li$_{3}$PO$_{4}$ + 0.3333 Fe$_3$(PO$_4$)$_2$$\cdot$8H$_2$O $\longrightarrow$ LiFePO$_{4}$ + 2.667 H$_{2}$O & 800 & -0.065 &  & 0.000 & 0.068 & 0.012 \\
\textbf{} & \textbf{4} & LiOH$\cdot$H$_2$O + Fe(PO$_4$)$\cdot$2H$_2$O $\longrightarrow$ LiFePO$_{4}$ + 3.5 H$_{2}$O + 0.25 O$_{2}$ & 650 & -0.111 & O & 0.062 & 0.247 & 0.149 \\
\textbf{} & \textbf{5} & 0.5 Li$_{2}$CO$_{3}$ + FePO$_{4}$ $\longrightarrow$ LiFePO$_{4}$ + 0.5 CO$_{2}$ + 0.25 O$_{2}$ & 700 & 0.006 & O & 0.189 & 0.122 & 0.207 \\
\cline{1-9}
\multirow{4}{*}{\textbf{BaTiO$_{3}$}} & \textbf{1} & BaCO$_{3}$ + TiO$_{2}$ $\longrightarrow$ BaTiO$_{3}$ + CO$_{2}$ & 1050 & -0.056 & O & 0.075 & 0.065 & -0.230 \\
\textbf{} & \textbf{16} & BaCO$_{3}$ + TiO$_{2}$ $\longrightarrow$ BaTiO$_{3}$ + CO$_{2}$ & 1100 & -0.028 &  & 0.025 & 0.026 & -0.045 \\
\textbf{} & \textbf{37} & BaO + TiO$_{2}$ $\longrightarrow$ BaTiO$_{3}$ & 800 & -0.239 &  & 0.061 & 0.168 & 0.778 \\
\textbf{} & \textbf{40} & BaO + TiO$_{2}$ $\longrightarrow$ BaTiO$_{3}$ & 1300 & -0.225 &  & 0.080 & 0.167 & 0.916 \\
\cline{1-9}
\bottomrule
\end{tabular}
\end{table}

Our findings suggest that the targets in Table \ref{tab:lit_rxns_hard} require a more judicious synthesis due to inherently greater competition in their phase space. Many targets are ranked poorly because the conventional reaction is predicted to be suboptimal. This appears to be true for lead titanate (\ce{PbTiO3}), which, on average, has the most suboptimal recipe of any target investigated (Figure \ref{fig:literature}c). The high $C_1$ and $C_2$ values for \ce{PbTiO3} synthesis seem to be almost entirely related to the instability of the \ce{PbO} precursor, both with respect to decomposition and to the formation of a solid-solution phase. In our calculations, the latter manifests as the predicted stabilization of a theoretical \ce{Pb15TiO17} phase. Our modeling of \ce{PbO} instability supports experimental observations suggesting that the volatility of \ce{PbO} at higher temperatures results in a challenging \ce{PbTiO3} synthesis.\cite{Forrester2004} The use of \ce{PbCO3} as an alternative appears to perform similarly, albeit with a slightly more favorable driving force and lower $C_1$.

For \ce{Ca3(PO4)2}, \ce{Li2MnO3}, and \ce{LiFePO4}, the highest-ranked synthesis recipes appear to be nearly thermodynamically optimal already (i.e., low $C_1$ and $C_2$). The optimal recipe for \ce{Ca3(PO4)2} is the reaction of stoichiometric amounts of \ce{CaCO3} and \ce{Ca2P2O7} at moderate temperatures (800 \textcelsius{}) in a closed environment. For \ce{Li2MnO3}, the reaction of \ce{Li2CO3} and \ce{MnO2} in open-oxygen environments at moderately high temperatures (900-975 \textcelsius{}) is favorable, and the use of \ce{LiOH} in a closed environment at lower temperatures (650 \textcelsius{}) appears to be similarly advantageous.\cite{Paik2002} Finally, for the synthesis of \ce{LiFePO4}, the use of \ce{Li3PO4} and \ce{Fe3(PO4)2} in a closed environment at lower temperatures (600 \textcelsius{})\cite{Herstedt2003} appears to be highly selective and greatly outperforms the other recipes analyzed in the literature dataset. 

For the remaining targets \ce{Na2Ti3O7}, \ce{NaTaO3}, \ce{LiMn2O4}, and \ce{BaTiO3}, even the highest-ranked literature synthesis recipes are theoretically suboptimal, which suggests that these materials are suitable candidates for future synthesis optimization efforts. Recipes for \ce{Na2Ti3O7} generally feature poor driving forces and low selectivities; however, using \ce{NaOH} as an alternative precursor mitigates this some. All \ce{NaTaO3} recipes in our dataset used the same precursors, but those with open-oxygen environments resulted in substantially higher phase competition. \ce{LiMn2O4} synthesis appears to follow similar trends as those discussed previously for \ce{LiMn2O3}; the lithium carbonate route in open air appears to be the most optimal of the recipes explored, despite a somewhat low driving force and moderate competition.\cite{Kitamura2009} The use of a \ce{MnCO3} precursor is not recommended due to greatly decreased selectivity. Finally, for \ce{BaTiO3}, the conventional \ce{BaCO3} route at high temperatures (1050 \textcelsius{}) and open oxygen conditions appears to be the most favorable; however, with this route, the driving force is still somewhat low and features a moderate amount of phase competition. The \ce{BaTiO3} system will be examined extensively in the following sections as an experimental case study for assessing reaction selectivity.

The literature reaction data reveal an inherent tradeoff between driving force and selectivity. Reactions with \textit{only} elemental precursors (e.g., Li metal, \ce{O2} gas) typically have significantly greater driving forces but also much higher competition scores (i.e., lower selectivity), with median values of $\Delta G_\textrm{rxn}=-0.272$, $C_1=0.016$, and $C_2=0.112$ eV/atom for the closed reactions. For reactions with at least one binary (two-element) precursor, these values shift to $\Delta G_\textrm{rxn}=-0.037$, $C_1=0.006$, and $C_2=0.020$ eV/atom. In other words, sourcing an element from a precursor with pre-formed bonds yields an interface reaction hull representing a more selective slice of the phase diagram, but at the expense of sacrificing some of the available reaction energy. Fortunately, this tradeoff is not universal and can be circumvented by considering more complex precursor chemistries containing additional elements other than those in the target composition. For example, in oxide synthesis, the use of hydroxides, carbonates, and salts (as in metathesis reactions) often permits the formation of thermodynamically favorable byproducts that increase the reaction's driving force and selectivity. The use of additional elements beyond those found in the target composition has been dubbed ``hyperdimensional chemistry''\cite{Neilson2023} due to its connection to phase diagram geometry; adding a new compositional axis greatly increases the number of ways the phase space can be sliced, allowing one to thermodynamically ``shortcut'' otherwise unavoidable competing impurity phases. However, not all elemental precursors are unselective, and the necessity of these alternative routes depends on the degree of phase competition in the chemical system of interest. For example, it is typically favorable to synthesize a binary target comprised of elements commonly existing in only a single oxidation state; this is exemplified by the reaction \ce{Cd + Te -> CdTe}, which has both perfect selectivity ($C_1=-0.576$, $C_2=0.000$ eV/atom) and a very large driving force ($\Delta G_\textrm{rxn}= -0.576$ eV/atom) at 800 \textcelsius{}.

\subsection{Case Study: Synthesis of Barium Titanate (\ce{BaTiO3})}

To investigate whether our reaction selectivity metrics accurately describe impurity formation, we designed and performed experiments testing the influence of precursor selection on the reaction pathway observed during solid-state synthesis. To suggest these experiments, we developed a computational synthesis planning workflow that integrates our proposed selectivity metrics with previous methods for computing solid-state reaction networks.\cite{McDermott2021} The workflow has five user inputs: 1) target composition, 2) additional elements, 3) temperature, 4) thermodynamic stability cutoff (i.e., maximum $E_\textrm{hull}$), and 5) chemical potential of an open element, if any. The workflow returns a ranked list of possible synthesis reactions based on the cost function implemented in Equation \ref{eq:cost_function}.

We selected the ferroelectric barium titanate (\ce{BaTiO3}) as a case study system for testing our predictions. In addition to being a technologically important and well-studied material in the literature, \ce{BaTiO3} is an ideal target for investigating the thermodynamic selectivity of solid-state reactions due to the high number of competing phases in the Ba--Ti--O chemical system (Figure \ref{supp_fig:batio3_irh}). Conventional routes from binary oxides are well-known to proceed through intermediates \cite{Bierach1988}, and at least 12 unique ternary compositions have been experimentally observed, including the compositional neighbors \ce{BaTi2O5} and \ce{Ba2TiO4}, which frequently appear during synthesis (the latter in particular). Other commonly observed minor impurities include \ce{Ba4Ti13O30}, \ce{BaTi4O9}, and \ce{Ba6Ti17O40}. Additional compositions in this space, such as \ce{Ba2Ti9O20}, \ce{BaTi5O11}, \ce{Ba4Ti12O27}, and \ce{Ba2Ti6O13}, have been previously synthesized but are less commonly observed as intermediates or impurity phases.\cite{Bierach1988}  Another motivation for selecting \ce{BaTiO3} is the kinetic accessibility of its chemical system; even with such a high number of competing Ba--Ti--O phases, the solid-state synthesis of barium titanate is generally considered facile. The ease with which Ba--Ti--O phases can be synthesized suggests that kinetic factors do not significantly divert reaction outcomes from those that are thermodynamically favorable, providing greater justification for our attempts at assessing the likelihood of intermediate/impurity formation through a purely thermodynamic lens.

\begin{figure}[ht!]
\begin{center}
\includegraphics[width=1.0\textwidth]{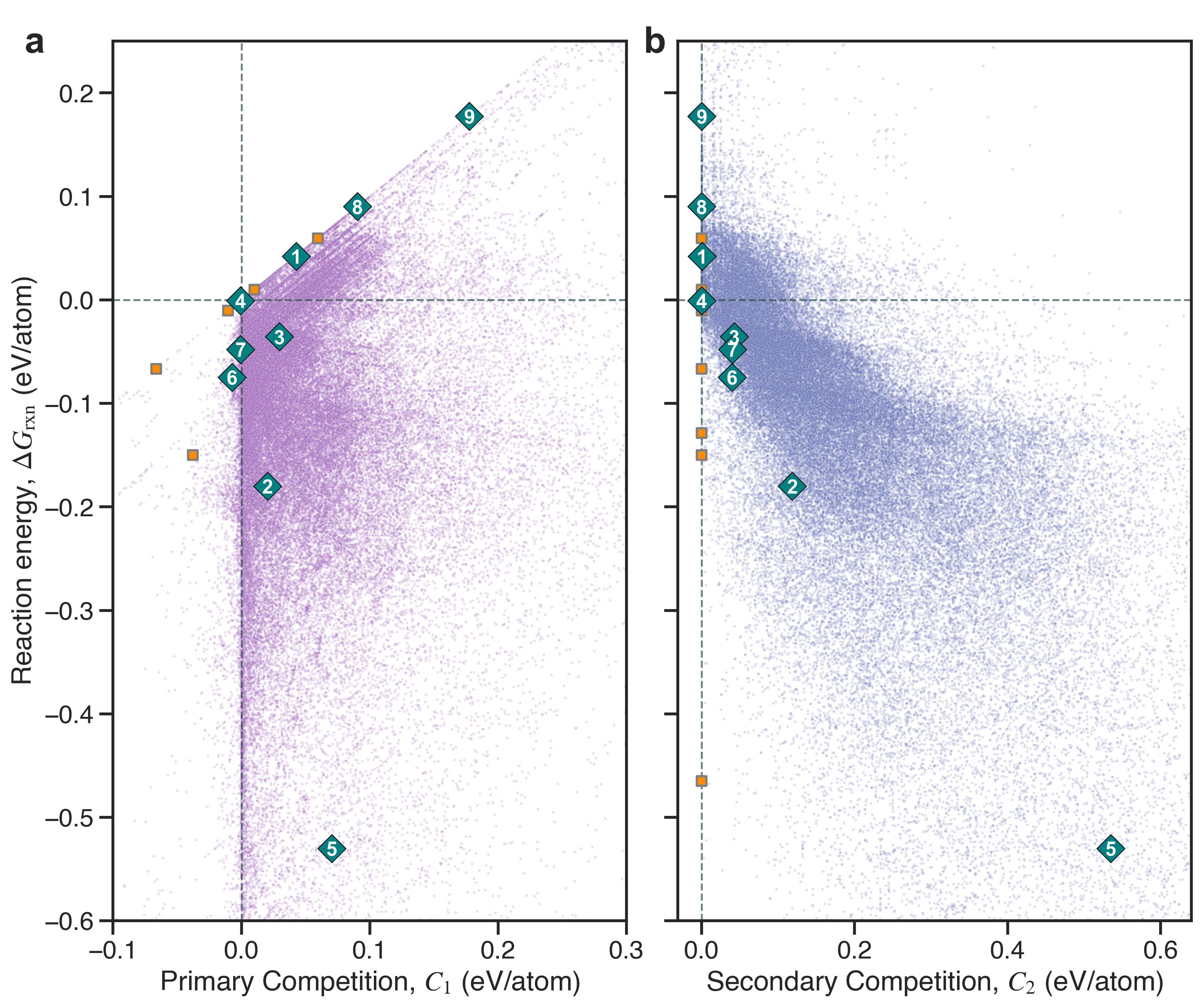}
\caption{\label{fig:batio3_calcs} \textbf{Synthesis maps of 82,985 calculated reactions producing \ce{BaTiO3}}. Target reactions and their competition scores are extracted from an 18-element network of 2,536,160 reactions modeled at a temperature of $T=600$ \textcelsius{}. Reactions are plotted on a shared axis of reaction energy, $\Delta G_\text{rxn}$, and on independent axes of (a) primary competition, $C_1$, and (b) secondary competition, $C_2$. The sharp boundaries are lower-bound limits of $C_1=\Delta G_{\textrm{rxn}}$ and $C_2=0$.  Blue diamonds indicate selected reactions experimentally tested in this work. Orange squares represent reactions on the three-dimensional Pareto front of $\Delta G_\text{rxn}$, $C_1$, and $C_2$.} 
\end{center} 
\end{figure}

Using our synthesis planning workflow, we computed and ranked 82,985 binary (two-precursor) synthesis reactions producing \ce{BaTiO3}, selected from a massive 18-element reaction network composed of all 2,536,160 enumerated binary reactions among 2,417 phases with energies $\leq 0.050$ meV/atom above the hull. To capture alternative chemistries (e.g., metathesis reactions, gas-forming reactions, ion exchange reactions, etc.), we selected a chemical system consisting of the target elements (Ba, Ti, O), alkali metals (Li, Na, K) and alkaline earth metals (Mg, Ca, Sr), halogens (F, Cl, Br), chalcogens (S), pnictogens (N, P), and other common elements (B, C, H). The $\Delta G_\textrm{rxn}$, $C_1$, and $C_2$ values for all calculated synthesis reactions are shown in the synthesis maps illustrated in Figure \ref{fig:batio3_calcs}. We calculated these values at an intermediate temperature of $T=600$ \textcelsius{}, which is near the median of our experimentally accessible temperature range (see Methods). We also considered the modeling of open-\ce{O2} reactions; similar results are shown in Figure \ref{supp_fig:batio3_open} for 62,133 open reactions in a more constrained subsystem.

Determining an optimal synthesis reaction can be formulated as an optimization problem of simultaneous minimization of the three reaction metrics: $\Delta G_\text{rxn}$, $C_1$, and $C_2$. A common approach for multi-objective optimization in synthesis planning is identifying the Pareto front.\cite{Aykol2021} Here, we calculate a three-dimensional Pareto front for the \ce{BaTiO3} synthesis reactions (Table \ref{tab:bto_closed_pareto}). The Pareto front reactions for the open-\ce{O2} system are provided in the Supporting Information.

\begin{table}[ht!]
\scriptsize
\caption{Pareto front reactions to \ce{BaTiO3} and their associated Gibbs free energies, $\Delta G_{\text{rxn}}$ ($T=600$ \textcelsius{}), primary competition scores, $C_1$, secondary competition scores, $C_2$, and costs, $\Gamma$. All units are in eV/atom.}
\label{tab:bto_closed_pareto}
\begin{tabular}{cp{8.2cm}ccccc}
\toprule
 & Reaction & $\Delta G_{\text{rxn}}$ & C$_1$ & C$_2$ & $\Gamma$ & Theoretical \\
Rank &  &  &  &  &  &  \\
\midrule
\textbf{1} & 2 TiP + 4.5 BaCO$_{3}$ $\longrightarrow$ BaTiO$_{3}$ + Ba$_{3}$(PO$_{4}$)$_{2}$ + 0.5 BaTi$_{2}$O$_{5}$ + 4.5 C & -0.465 & -0.465 & -0.000 & -0.256 &  \\
\textbf{2} & 3 Ba(NO$_{3}$)$_{2}$ + 3.333 LiTi$_{2}$N$_{3}$ $\longrightarrow$ BaTiO$_{3}$ + 2 BaTi$_{2}$O$_{5}$ + 1.667 Li$_{2}$TiO$_{3}$ + 8 N$_{2}$ & -0.868 & -0.302 & -0.000 & -0.223 & \ce{LiTi2N3} \\
\textbf{4} & MgTiN$_{2}$ + 0.6 Ba(NO$_{3}$)$_{2}$ $\longrightarrow$ MgO + 0.2 BaTiO$_{3}$ + 0.4 BaTi$_{2}$O$_{5}$ + 1.6 N$_{2}$ & -0.901 & -0.223 & 0.008 & -0.187 &  \\
\textbf{6} & 8 O$_{2}$ + Ba$_{4}$TiP$_{4}$ $\longrightarrow$ BaTiO$_{3}$ + Ba$_{3}$P$_{4}$O$_{13}$ & -2.330 & 0.062 & 0.136 & -0.144 &  \\
\textbf{7} & 4 BaO$_{2}$ + 2 Ti $\longrightarrow$ Ba$_{3}$TiO$_{5}$ + BaTiO$_{3}$ & -1.489 & -0.021 & 0.036 & -0.142 &  \\
\textbf{18} & 0.25 Mg$_{4}$TiN$_{4}$ + 0.3 Ba(NO$_{3}$)$_{2}$ $\longrightarrow$ MgO + 0.2 BaTiO$_{3}$ + 0.05 Ba$_{2}$TiO$_{4}$ + 0.8 N$_{2}$ & -0.983 & -0.057 & 0.078 & -0.089 & \ce{Mg4TiN4} \\
\textbf{28} & TiO + BaO$_{2}$ $\longrightarrow$ BaTiO$_{3}$ & -1.044 & -0.034 & 0.099 & -0.075 &  \\
\textbf{33} & 0.75 Ti$_{7}$P$_{4}$ + 0.3333 Ba$_{5}$P$_{3}$O$_{12}$F $\longrightarrow$ BaTiO$_{3}$ + 0.25 Ba$_{2}$TiO$_{4}$ + 4 TiP + 0.1667 BaF$_{2}$ & -0.129 & -0.129 & -0.000 & -0.071 &  \\
\textbf{54} & MgTi(SO$_{4}$)$_{3}$ + 4 BaMg$_{2}$ $\longrightarrow$ BaTiO$_{3}$ + 9 MgO + 3 BaS & -1.515 & -0.005 & 0.225 & -0.052 & \ce{MgTi(SO4)3} \\
\textbf{63} & 4 Ti(NO$_{3}$)$_{4}$ + 5 Ba$_{4}$P$_{2}$O $\longrightarrow$ Ba$_{2}$TiO$_{4}$ + 3 BaTiO$_{3}$ + 5 Ba$_{3}$(PO$_{4}$)$_{2}$ + 8 N$_{2}$ & -1.570 & -0.018 & 0.261 & -0.047 &  \\
\textbf{85} & 2.25 TiC + 0.5 Ba$_{3}$(PO$_{4}$)$_{2}$ $\longrightarrow$ BaTiO$_{3}$ + 0.25 Ba$_{2}$TiO$_{4}$ + TiP + 2.25 C & -0.067 & -0.067 & -0.000 & -0.037 &  \\
\textbf{90} & 1.25 Ba$_{5}$(TiN$_{3}$)$_{2}$ + 1.125 Ti(NO$_{3}$)$_{4}$ $\longrightarrow$ BaTiO$_{3}$ + 2.625 Ba$_{2}$TiO$_{4}$ + 6 N$_{2}$ & -1.287 & -0.029 & 0.236 & -0.035 &  \\
\textbf{108} & 0.75 Ti$_{7}$P$_{4}$ + 0.5 Ba$_{3}$(PO$_{4}$)$_{2}$ $\longrightarrow$ BaTiO$_{3}$ + 0.25 Ba$_{2}$TiO$_{4}$ + 4 TiP & -0.150 & -0.038 & -0.000 & -0.032 &  \\
\textbf{199} & 1.45 Ti(ClO$_{4}$)$_{4}$ + 0.8 Ba$_{6}$Mg$_{23}$ $\longrightarrow$ BaTiO$_{3}$ + 0.45 Ba$_{2}$TiO$_{4}$ + 18.4 MgO + 2.9 BaCl$_{2}$ & -2.540 & -0.008 & 0.527 & -0.020 &  \\
\textbf{259} & 3 BaO$_{2}$ + Ti$_{2}$N$_{2}$O $\longrightarrow$ BaTiO$_{3}$ + Ba$_{2}$TiO$_{4}$ + N$_{2}$ & -0.913 & -0.099 & 0.268 & -0.015 & \ce{Ti2N2O} \\
\textbf{498} & TiNCl + 1.5 BaO$_{2}$ $\longrightarrow$ BaTiO$_{3}$ + 0.5 BaCl$_{2}$ + 0.5 N$_{2}$ & -0.967 & -0.150 & 0.350 & -0.007 &  \\
\textbf{547} & Li$_{2}$TiO$_{3}$ + 0.3333 Ba$_{3}$(PO$_{4}$)$_{2}$ $\longrightarrow$ BaTiO$_{3}$ + 0.6667 Li$_{3}$PO$_{4}$ & -0.011 & -0.011 & -0.000 & -0.006 &  \\
\textbf{552} & 1.5 BaO$_{2}$ + TiNF $\longrightarrow$ BaTiO$_{3}$ + 0.5 BaF$_{2}$ + 0.5 N$_{2}$ & -0.960 & -0.153 & 0.354 & -0.006 & \ce{TiNF} \\
\textbf{651} & TiBrN + 1.5 BaO$_{2}$ $\longrightarrow$ BaTiO$_{3}$ + 0.5 BaBr$_{2}$ + 0.5 N$_{2}$ & -0.963 & -0.152 & 0.357 & -0.004 &  \\
\textbf{1850} & Ba$_{3}$(PO$_{4}$)$_{2}$ + Ca$_{4}$Ti$_{3}$O$_{10}$ $\longrightarrow$ Ca$_{4}$P$_{2}$O$_{9}$ + 3 BaTiO$_{3}$ & 0.010 & 0.010 & -0.000 & 0.005 &  \\
\textbf{2825} & 0.05882 Ba$_{2}$Mg$_{17}$ + 0.1118 Ti(NO$_{3}$)$_{4}$ $\longrightarrow$ MgO + 0.1059 BaTiO$_{3}$ + 0.005882 Ba$_{2}$TiO$_{4}$ + 0.2235 N$_{2}$ & -1.896 & -0.004 & 0.447 & 0.010 &  \\
\textbf{6598} & 3 Ti(SO$_{4}$)$_{2}$ + 8 BaMg$_{2}$ $\longrightarrow$ BaTiO$_{3}$ + 16 MgO + 6 BaS + BaTi$_{2}$O$_{5}$ & -1.628 & -0.006 & 0.413 & 0.020 &  \\
\textbf{7657} & 2 BaO$_{2}$ + 0.25 NaTi$_{5}$(NCl)$_{5}$ $\longrightarrow$ BaTiO$_{3}$ + 0.25 Ba$_{2}$TiO$_{4}$ + 0.25 NaCl + 0.5 BaCl$_{2}$ + 0.625 N$_{2}$ & -0.932 & -0.079 & 0.336 & 0.023 & \ce{NaTi5(NCl)5} \\
\textbf{8896} & 0.5 Ba$_{5}$(TiN$_{3}$)$_{2}$ + 3 KNOF$_{2}$ $\longrightarrow$ BaTiO$_{3}$ + 3 KF + 1.5 BaF$_{2}$ + 3 N$_{2}$ & -1.170 & -0.040 & 0.356 & 0.025 &  \\
\textbf{13191} & TiO$_{2}$ + 0.5 Ba$_{2}$Ca(BO$_{2}$)$_{6}$ $\longrightarrow$ BaTiO$_{3}$ + 0.5 Ca(B$_{3}$O$_{5}$)$_{2}$ & 0.060 & 0.060 & -0.000 & 0.033 &  \\
\textbf{15339} & 0.5 Ba$_{5}$(TiN$_{3}$)$_{2}$ + 3 NOF $\longrightarrow$ BaTiO$_{3}$ + 1.5 BaF$_{2}$ + 3 N$_{2}$ & -1.667 & -0.058 & 0.509 & 0.036 &  \\
\textbf{39153} & 0.5 Ba$_{5}$(TiN$_{3}$)$_{2}$ + 1.5 SO$_{2}$ $\longrightarrow$ BaTiO$_{3}$ + 1.5 BaS + 1.5 N$_{2}$ & -1.102 & -0.042 & 0.450 & 0.074 &  \\
\textbf{45245} & 0.5 Ti$_{2}$S + 1.5 BaO$_{2}$ $\longrightarrow$ BaTiO$_{3}$ + 0.5 BaS & -1.366 & -0.068 & 0.563 & 0.086 &  \\
\textbf{57844} & 0.8 Ba$_{5}$(TiN$_{3}$)$_{2}$ + 0.45 Ti(ClO$_{4}$)$_{4}$ $\longrightarrow$ BaTiO$_{3}$ + 1.05 Ba$_{2}$TiO$_{4}$ + 0.9 BaCl$_{2}$ + 2.4 N$_{2}$ & -1.729 & -0.077 & 0.728 & 0.120 &  \\
\textbf{64696} & 5 MgTiH$_{4}$ + 3 Ba(NO$_{3}$)$_{2}$ $\longrightarrow$ BaTiO$_{3}$ + 5 MgO + 10 H$_{2}$ + 2 BaTi$_{2}$O$_{5}$ + 3 N$_{2}$ & -1.142 & -0.078 & 0.659 & 0.147 &  \\
\textbf{65291} & 0.8 Ba$_{5}$(TiN$_{3}$)$_{2}$ + 1.8 Br$_{2}$O$_{3}$ $\longrightarrow$ BaTiO$_{3}$ + 0.6 Ba$_{2}$TiO$_{4}$ + 1.8 BaBr$_{2}$ + 2.4 N$_{2}$ & -1.810 & -0.055 & 0.790 & 0.150 &  \\
\textbf{65313} & 0.5 Ba$_{5}$(TiN$_{3}$)$_{2}$ + 1.5 BrO$_{2}$F $\longrightarrow$ BaTiO$_{3}$ + 1.5 BaBrF + 1.5 N$_{2}$ & -1.878 & -0.053 & 0.804 & 0.150 &  \\
\textbf{66253} & 3.5 Ba$_{5}$(TiN$_{3}$)$_{2}$ + 9 ClO$_{3}$ $\longrightarrow$ BaTiO$_{3}$ + 6 Ba$_{2}$TiO$_{4}$ + 4.5 BaCl$_{2}$ + 10.5 N$_{2}$ & -1.841 & -0.021 & 0.775 & 0.155 &  \\
\bottomrule
\end{tabular}
\end{table}

Many Pareto-optimal reactions feature unconventional reactants and byproducts. \ce{BaO2} and \ce{Ba5(TiN3)2} are the most commonly appearing precursors (eight times each), followed by \ce{Ba(NO3)2} and \ce{Ba3(PO4)2} (four times each). Nearly all reactions (31 of 33) involve precursors containing additional elements other than Ba, Ti, and O. In particular, nitrogen is used in over half of the reactions (19 of 33), each featuring \ce{N2} gas formation. While unconventional, the high prevalence of nitride precursors in the Pareto front is not theoretically unreasonable; nitrides generally have less negative formation energies than oxides, making oxide formation with \ce{N2} gas evolution both energetically and entropically favorable.

However, many of the reactions appearing on the Pareto front are impractical from an experimental standpoint. For example, the aforementioned nitride precursors are likely challenging to synthesize and handle. Other Pareto front reactions involve theoretical phases (e.g., \ce{LiTi2N3}), uncommon or toxic precursors (e.g., \ce{Ba5(TiN3)2}, \ce{SO2}), difficult-to-remove byproducts (e.g., \ce{Ba2TiO4}), or refractory precursors (e.g., \ce{TiC}). Some of these suggested reactions can be removed easily with user-applied filters that account for specific experimental restrictions. For example, one can supply a list of available precursor compositions (i.e., ``off-the-shelf'' phases) or composition types to be avoided (e.g., sulfides, acids, etc.). In our provided code (see Methods), we support functionality for the former by including a list of hundreds of common precursors compiled from the catalogs of chemical suppliers. Filtering by these commonly available precursors (e.g., \ce{BaCO3}, \ce{Ba3(PO4)2}, \ce{TiO2}) reduces the full set of 82,985 \ce{BaTiO3} synthesis reactions to 478, making the generated recipes more easily parseable and readily testable. The filtered \ce{BaTiO3} reactions, including the corresponding open-\ce{O2} reactions, are provided in the Supporting Information. While filtering by conventional precursors is practically convenient, we consider the unorthodox nature of the \textit{unfiltered} reactions an advantage of our approach, as this permits synthesis recommendations that expand beyond traditional chemical intuition. Still, synthesis recipes must be screened for reactivity, volatility, safety, and material costs. These challenges can be mitigated through the use of additional data or models; for example, reactivity can be approximated through surrogate data, such as defect formation energies or physical properties (melting points, hardness, etc.).\cite{Hong2022b}

We selected nine reactions from Figure \ref{fig:batio3_calcs} to test experimentally. The calculated thermodynamic metrics for these reactions are provided in Table \ref{tab:bto_closed_selected}. We intentionally selected reactions spanning various precursor chemistries, free energies, and competition scores. To prioritize the study of impurity-forming reactions and avoid the aforementioned practicality challenges, we did not explicitly include any reactions on the Pareto front. The conventional synthesis route, \ce{BaCO3 + TiO2 -> BaTiO3 + CO2}, was chosen as a baseline reference (Expt.\ 1). A positive energy is calculated for this reaction ($\Delta G_\textrm{rxn}$=+0.042 eV/atom at $T=600$ \textcelsius{}); however, this is likely due to residual uncorrected error in the calculated energy of \ce{BaCO3}, which is not included in the NIST-JANAF dataset. We did not test the analogous reaction with \ce{BaO} precursor due to its hygroscopic nature, which makes it difficult to handle. However, we did test the alternative reaction from barium peroxide, \ce{BaO2} (Expt.\ 2). We included two reactions that form \ce{BaTiO3} directly from at least one other ternary phase (Expts.\ 3, 4). A reaction with a Ti metal precursor was selected due to its extremely high $C_2$ score (Expt.\ 5). Two metathesis reactions (Expts.\ 6, 7) were selected due to their predicted high performance, including the unconventional use of a sulfide precursor (\ce{BaS}). The final two reactions (Expts.\ 8, 9) were selected for being endergonic ($\Delta G_\text{rxn}>0$) to validate the accuracy of our free energy predictions. We note that several experiments feature precursors that are not easily purchasable from a chemical supplier (e.g., \ce{Ba2TiO4}, \ce{Na2TiO3}); these phases were synthesized following recipes reported in the literature (see Methods).

\begin{table}[ht!]
\footnotesize
\caption{Selected experimental \ce{BaTiO3} synthesis reactions and their associated Gibbs free energies, $\Delta G_{\text{rxn}}$ ($T=600$ \textcelsius{}), primary competition scores, $C_1$, secondary competition scores, $C_2$, and costs, $\Gamma$.}
\label{tab:bto_closed_selected}
\begin{tabular}{cccccc}
\toprule
 & Reaction & $\Delta G_{\text{rxn}}$ & C$_1$ & C$_2$ & $\Gamma$ \\
Expt. &  & (eV/at) & (eV/at) & (eV/at) &  (eV/at) \\
\midrule
\textbf{1} & BaCO$_{3}$ + TiO$_{2}$ $\longrightarrow$ BaTiO$_{3}$ + CO$_{2}$ & 0.042 & 0.043 & 0.000 & 0.024 \\
\textbf{2} & BaO$_{2}$ + TiO$_{2}$ $\longrightarrow$ BaTiO$_{3}$ + 0.5 O$_{2}$ & -0.180 & 0.021 & 0.119 & 0.045 \\
\textbf{3} & Ba$_{2}$TiO$_{4}$ + TiO$_{2}$ $\longrightarrow$ 2 BaTiO$_{3}$ & -0.036 & 0.030 & 0.043 & 0.029 \\
\textbf{4} & Ba$_{2}$TiO$_{4}$ + BaTi$_{2}$O$_{5}$ $\longrightarrow$ 3 BaTiO$_{3}$ & -0.001 & -0.001 & 0.000 & -0.001 \\
\textbf{5} & Ba(OH)$_2\cdot$H$_2$O + 3.666 Ti $\longrightarrow$ BaTiO$_{3}$ + 1.333 Ti$_{2}$H$_{3}$ & -0.530 & 0.070 & 0.534 & 0.219 \\
\textbf{6} & BaCl$_{2}$ + Na$_{2}$TiO$_{3}$ $\longrightarrow$ BaTiO$_{3}$ + 2 NaCl & -0.075 & -0.007 & 0.040 & 0.007 \\
\textbf{7} & BaS + Na$_{2}$TiO$_{3}$ $\longrightarrow$ BaTiO$_{3}$ + Na$_{2}$S & -0.048 & -0.001 & 0.041 & 0.013 \\
\textbf{8} & 2 BaS + 3 TiO$_{2}$ $\longrightarrow$ 2 BaTiO$_{3}$ + TiS$_{2}$ & 0.090 & 0.090 & 0.000 & 0.050 \\
\textbf{9} & BaSO$_{4}$ + 2 TiO$_{2}$ $\longrightarrow$ BaTiO$_{3}$ + TiOSO$_{4}$ & 0.178 & 0.178 & 0.000 & 0.098 \\
\bottomrule
\end{tabular}
\end{table}

\begin{figure}[ht!]
\begin{center}
\includegraphics[width=1.0\textwidth]{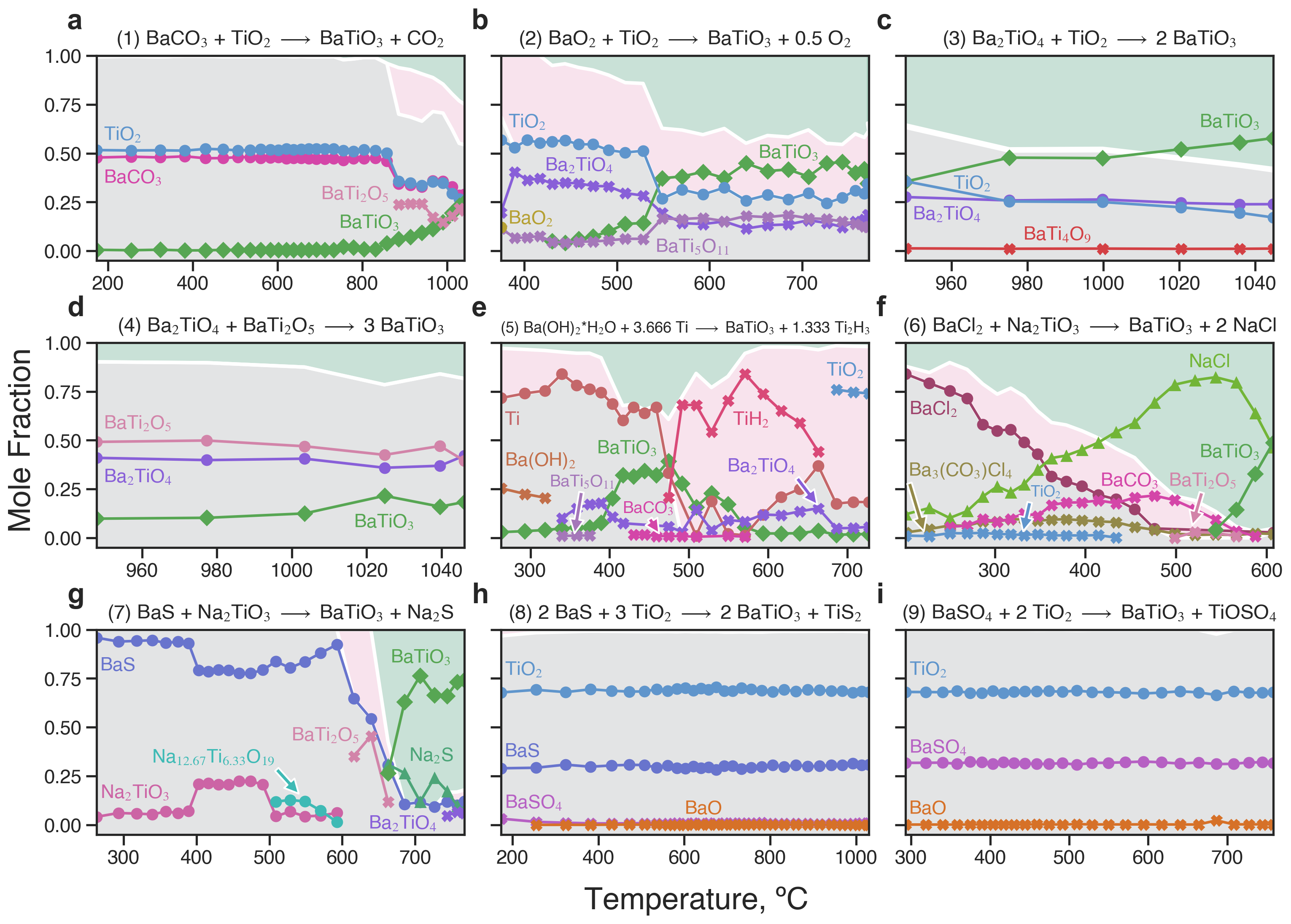}
\caption{\textbf{Reaction pathways for selected \ce{BaTiO3} synthesis experiments.} (a-i) Mole fractions of observed phases for Experiments 1-9, as determined through Rietveld refinements of \textit{ex post facto} SPXRD data. Phase types are distinguished by shape: precursors (circles), targets (diamonds), byproducts (triangles), and impurities (exes). Background shading denotes the total mole fractions of precursor (gray), impurity (pink), and target/byproduct (green). The median total reaction time was $\sim$67 min; exact times for each experiment are provided in Table \ref{tab:exp_times}.}
\label{fig:batio3_expts} 
\end{center} 
\end{figure}

The nine synthesis experiments were completed using a gradient furnace (see Methods),\cite{ONolan2020} allowing the observation of reaction products over a wide range of temperatures ($\sim$200-1000 \textcelsius{}) with \textit{ex post facto} SPXRD. The experimental results are summarized in Figure \ref{fig:batio3_expts}. Mole fractions of each phase were determined via Rietveld refinement of SPXRD patterns acquired at various positions (temperatures) along the length of the sample after heating and subsequent cooling to ambient temperature. The \textit{ex post facto} phase fraction plots can be interpreted similarly to those constructed with \textit{in situ} data, as each effectively illustrates the reaction pathway during heating. More rigorously, however, each point in Figure \ref{fig:batio3_expts} is pseudo-independent and best interpreted as the result of an isothermal (\textit{ex situ}) reaction at its associated temperature. Shorter reaction times were selected to ensure capture of the onset of short-lived intermediate phases critical to assessing reaction pathway selectivities (Table \ref{tab:exp_times}). Selected Rietveld analysis results for each experiment are provided in the Supporting Information (Figures \ref{supp_fig:refinement1}-\ref{supp_fig:refinement9}). 

The observed reaction pathways demonstrate significant variation in target and impurity formation. Visualizing the interface reaction hulls of the selected experiments helps to rationalize their predicted and observed performance (Figure \ref{supp_fig:all_irh}). The most complex of these pathways is that of Expt.\ 5, which features the formation of \ce{BaTiO3} at an intermediate temperature range (400-500 \textcelsius{}) before impurities \ce{Ba2TiO4}, \ce{TiH2}, and \ce{TiO2} begin to dominate. Indeed, Expt.\ 5 exhibits both the largest driving force (-0.530 eV/atom) and $C_2$ value (0.534 eV/atom) of any reaction, supporting the observation of a complex reaction pathway containing many impurity phases. 

The conventional synthesis reaction between \ce{BaCO3} and \ce{TiO2} (Expt.\ 1) was largely incomplete after 60 min at 1000 \textcelsius{} and exhibited significant formation of \ce{BaTi2O5}. The interface reaction hull for this system (Figure \ref{supp_fig:all_irh}a) suggests that the formation of \ce{BaTi2O5} is the most favorable reaction outcome, supporting our observations. In Expt.\ 2, the reaction of \ce{TiO2} with \ce{BaO2} appears to significantly decrease the reaction onset temperature but also features substantial impurity formation. In three of the experiments (Expts.\ 4, 6, and 7), the \ce{BaTiO3} synthesis reaction is the most favorable reaction on the hull, resulting in $C_1<0$. Notably, the use of ternary precursor(s) in Expts.\ 3 and 4 (\ce{Ba2TiO4}, \ce{BaTi2O5}) results in low (or zero) $C_2$, but also very little driving force. As a result, we observe near-perfect selectivity (i.e., very few visible impurities) at the expense of slowing down the reactions substantially. When the reaction energy is above zero, impurities associated with exergonic competing reactions may still form. For example, the hull for Expt.\ 8 indicates a significant degree of competition, including several reactions with only slightly positive energies ($\Delta G_\textrm{rxn}< 0.01$ eV/atom). Experiments 8 and 9 are indeed largely unreacted as predicted but feature minor impurities (\ce{BaSO4} and/or \ce{BaO}).

The metathesis reactions (Expts.\ 6, 7) show the overall greatest performance, yielding primarily \ce{BaTiO3} and the predicted byproducts at moderately low temperatures (600-700 \textcelsius{}). While metathesis reactions producing alkali halides are well-known for their optimal performance \cite{Martinolich2017a}, it is a notable and surprising result that the sulfide-based reaction (\ce{BaS + Na2TiO3 -> BaTiO3 + Na2S}) achieves such pure and direct synthesis of \ce{BaTiO3}, as predicted. Its success further highlights the importance of considering more complex chemistries involving additional elements besides those in the target phase (i.e., hyperdimensional chemistries) \cite{Neilson2023}. Some impurity formation, however, is evident in both metathesis reactions, particularly at lower temperatures. \ce{BaTi2O5} forms in both experiments and small amounts of \ce{Ba2TiO4} form in Expt.\ 7. However, this observation is supported by the calculated $C_1$ and $C_2$ scores, which indicate that neither reaction should be perfectly selective. Unexpectedly, the dominant impurities in Expt.\ 6 are carbonate compounds: \ce{Ba3(CO3)Cl4}, and \ce{BaCO3}. We presume this results from minor contamination of the precursors via reaction with \ce{CO2} in the air; some \ce{Na2CO3} observed in the precursor (see Methods) may have also contributed to the formation of the barium carbonate impurities via energetically favorable Ba/Na ion exchange reactions. For completeness, we have accounted for these unexpected impurities in our experimental analysis even though they were not explicitly considered in the selectivity calculations. However, we did exclude from consideration any Si-containing impurities such as \ce{Ba2TiSi2O8}, which formed in small amounts due to reaction with the quartz capillaries; these Si impurities were minor ($<$2 mol \%) and did not significantly affect our analysis.

\begin{figure}[ht!]
\begin{center}
\includegraphics[width=1.0\textwidth]{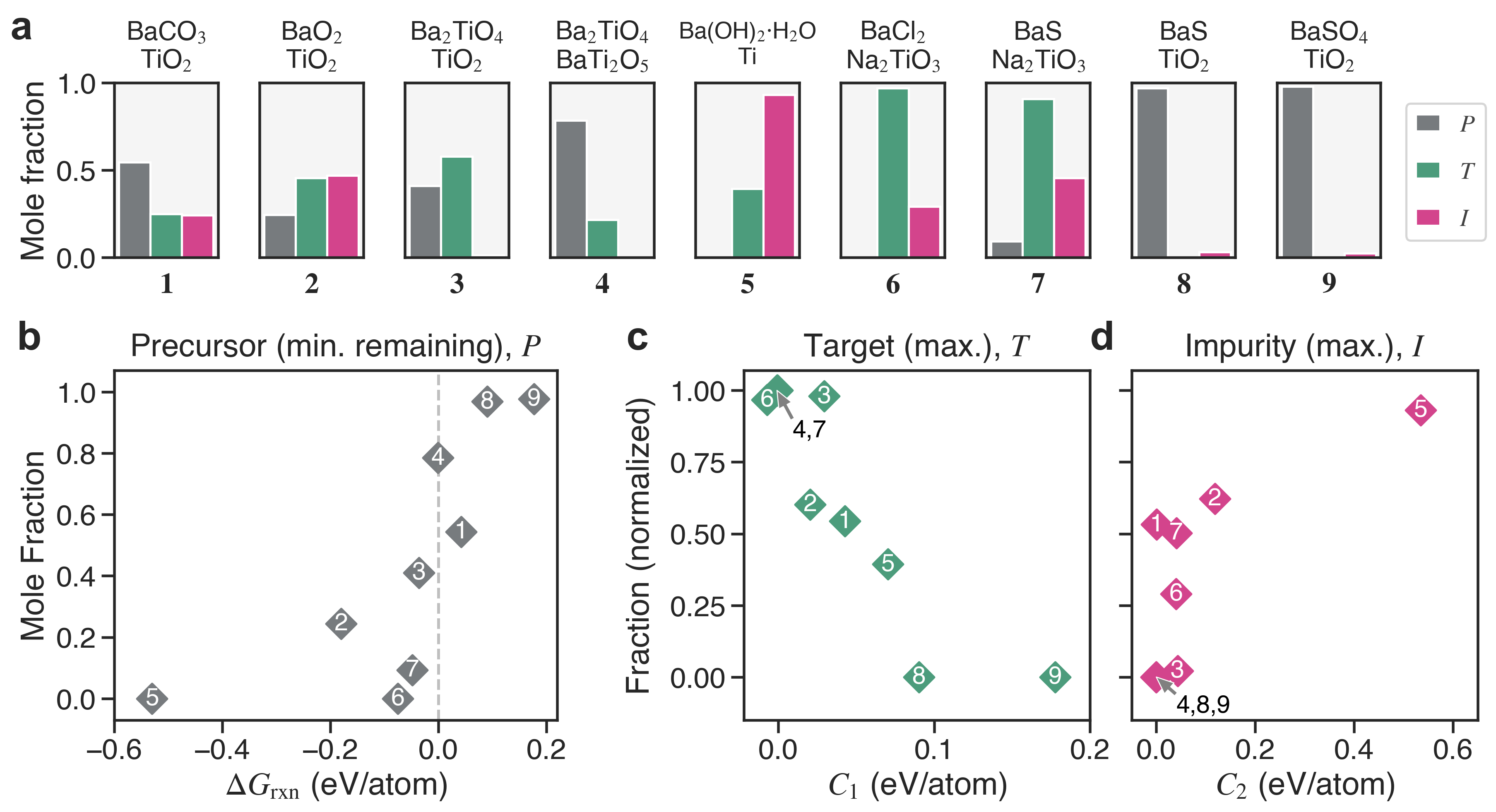}
\caption{\textbf{Summary of experimental results and correlations with calculated reaction metrics.} (a) Precursor-Target-Impurity (PTI) plots, where the height of each bar indicates the relevant mole fraction captured at the most representative position (i.e., temperature) along the length of the sample after heating (and cooling). The different color bars correspond to the minimum amount of precursor remaining ($P$, gray), the maximum amount of target/byproducts formed ($T$, green), and the maximum amount of impurity formed ($I$, pink). (b) Positive correlation between $P$ and the Gibbs free energy of the reaction, $\Delta G_\textrm{rxn}$. (c) Negative correlation between $T$ and $C_1$. (d) Positive correlation between $I$ and $C_2$. The $T$ and $I$ mole fractions have been normalized by the maximum amount of precursor consumed in the experiment, $1-P$, for enhanced visualization of the trend. The small impurity amounts detected in Expts.\ 8 and 9 are treated as yielding $I=0$ to more accurately reflect the lack of observed reaction.}
\label{fig:metric_performance} 
\end{center} 
\end{figure}

To quantitatively assess the performance of our predictions in determining the outcomes of experimental reactions, we propose three reaction outcome metrics summarizing the behavior of a reaction pathway: the minimum precursor remaining ($P$), the maximum target/byproduct formed ($T$), and the maximum impurity formed ($I$). Each metric is a mole fraction value taken from any data point (i.e., temperature) within the reaction pathway. This permits the capture of key features of the pathway independent of the kinetics of that reaction and is necessary given the range of chemistries explored. Our outcome metrics are visualized on the Precursor-Target-Impurity (PTI) plots in Figure \ref{fig:metric_performance}a. The first quantity, $P$ (minimum precursor remaining), gives insight into the reactivity and kinetics of the reaction: high values indicate reactions that did not complete at any temperature within the reaction timeframe. The second quantity, $T$ (maximum target/byproduct formed), provides a measure of the success of the reaction in producing \ce{BaTiO3} and predicted byproduct(s). Finally, the third quantity, $I$ (maximum impurity formed), measures the selectivity of the reaction pathway, indicating the maximum fraction of intermediate/impurity phases synthesized at any temperature. Note that the temperature-independent PTI metrics are not required to sum to one for a particular experiment. This is intentional and advantageous because it permits the capture of poor selectivity in even nominally well-performing reactions; i.e., both $T$ and $I$ can be high ($\sim$1) within the same experiment.

Figures \ref{fig:metric_performance}b-d show selected correlations between reaction outcomes ($P$, $T$, and $I$) and the calculated reaction metrics ($\Delta G_\textrm{rxn}$, $C_1$, and $C_2$). The full set of all (3x3) pairwise correlation plots is available in Figure \ref{supp_fig:metric_performance}. We observe that the calculated reaction energy ($\Delta G_\textrm{rxn}$) correlates most strongly with the minimum amount of precursor remaining ($P$) at the conclusion of the experiment (Figure \ref{fig:metric_performance}b). With infinite reaction times, we would theoretically expect this distribution to resemble a step function: $P=0$ for reactions with $\Delta G_\textrm{rxn}<0$, and $P=1$ for those with $\Delta G_\textrm{rxn}>0$. In our work, the distribution is less defined, given the shorter reaction times and different chemistries explored. The coordinates of Expts. 1 and 2 appear to deviate the most from a step-like distribution. Difficulty in modeling the energetics of carbonate reactions was previously discussed and likely explains the deviation of Expt.\ 1. In contrast, the deviation of Expt.\ 2 is likely kinetic in nature, as the reaction appears to stall (Figure \ref{fig:batio3_expts}b); this may suggest that the maximum temperature (750-800 \textcelsius{}) is too low to achieve sufficient reaction completion.

The primary ($C_1$) and secondary ($C_2$) competition metrics display negative and positive correlations with the maximum amounts of target ($T$) and impurity ($I$) phases formed, respectively (Figures \ref{fig:metric_performance}c,d). We note that these trends are not strictly obeyed in a monotonic fashion. Still, the correlations are significant and can be theoretically rationalized by their derivation from the interface reaction hull. It is reasonable that $C_1$ correlates most strongly with $T$, as this competition metric effectively measures the relative favorability of the target reaction over competing reactions. Similarly, it is sensible that $C_2$ should correlate most strongly with $I$ given its derivation as a measure of the \textit{relative} stability of impurity phases with respect to the target phase. To be precise, the experimental correlation between $C_2$ and $I$ suggests that one should consider not only the sum of the inverse hull distance energies for all competing phases but the \textit{total} energy of the entire secondary reaction sequence containing them (Equation \ref{eqn:secondary_seq}). By definition, this quantity includes, and thus will always be larger than or equal to, the sum of all competing inverse hull distance values for a particular interface reaction hull. Regarding the functional form of $C_2$, the question arises as to whether a more simple summation of the \textit{maximum} energy secondary reactions in the left and right hull subsections is a sufficient measure for secondary competition. While the maximum energy secondary reactions tend to account for much of the value of $C_2$, on average, the full $C_2$ metric is 0.077 eV/atom greater than considering the maximum energy secondary reactions alone (Figure \ref{supp_fig:correlations_secondary}), suggesting that $C_2$ is a more conservative metric. For completeness, we also tested an alternative formulation of the secondary competition metric using the enclosed ``area'' to the hull. While this metric correlates with $C_2$, its calculation is more numerically unstable, and its units are less interpretable. Hence, we generally recommend the approach of modeling full secondary reaction sequences, which is straightforward to implement using our secondary competition algorithm (see Methods).

In a previous study,\cite{Todd2021} we suggested a solid-state reaction selectivity metric based on the difference in elemental chemical potentials between precursors and targets, measured by a distance along the chemical potential diagram (i.e., the ``total chemical potential distance''). This metric was used to rationalize the unique selectivity of the Na-based precursor in synthesizing pyrochlore \ce{Y2Mn2O7} from \ce{YOCl} and \ce{$A$MnO2} ($A$=Li, Na, K). While straightforward to compute using just the chemical potential diagram, the distance metric operates in the space of chemical potentials rather than reaction energies, rendering it less intuitive and more difficult to precisely discern the specific competing reactions. From the results here, we generally recommend that $C_2$ be used instead of total chemical potential distance where possible. Both selectivity metrics capture similar characteristics of the competing phase space: each effectively involves the summation of competing phase stabilities through inverse hull distances or the corresponding chemical potential stability ranges. More precisely, both quantities are correlated (Figure \ref{supp_fig:correlations}) because chemical potentials are mathematical derivatives of the convex hull in energy-composition space. However, the total chemical potential distance is biased, particularly by competing phases with defective elemental-like compositions (e.g., \ce{Mg149Cl}). Due to the increased weight of the entropic ($-TS$) term in the definition of Gibbs free energy, chemical potential diagrams featuring these compositions as competing phases may yield very high (unfavorable) total chemical potential distance values for synthesis reactions.

We acknowledge that while $C_1$ and $C_2$ are meant to capture different, independent mechanisms by which competing phases form, these metrics are at least \textit{partially} correlated due to the geometric constraints of the convex hull. In particular, one situation is geometrically limited from occurring: high $C_2$ and low $C_1$ (Figure \ref{supp_fig:competition_extremes}). Stated explicitly: if a competing phase lies significantly below the tie-line formed by the target and a precursor (i.e., high $C_2$), then both the target and that competing phase necessarily have similar reaction energies to form from the precursors, leading to high $C_1$.  In general, however, this restriction does not make the metrics redundant; while there is some correlation between the two, the correlation is not particularly strong (Figure \ref{supp_fig:correlations}). Therefore, we generally recommend the tandem use of both selectivity metrics.

The major limitation of our current synthesis planning workflow is the assumption that optimal synthesis reactions can be predicted with the thermodynamic energy landscape alone. While this is not the case for all chemistries, we show that, at least for chemical systems that exhibit practical solid-state reaction kinetics, the energy landscape alone can provide much of the rationalization for the observation of impurity phases. To say that impurity and secondary phases are inherently ``kinetic'' products is a misnomer. Rather, these phases may be the thermodynamic minima of smaller ``local'' interface systems, distinct from the thermodynamic products of the entire reaction mixture (the global thermodynamic solution). Furthermore, these impurities are often not easily convertible to final products without long-range mass transport or intervention (e.g., via re-grinding or subsequent heating). This explains why impurities are often pervasive and challenging to remove in chemical systems with lower driving forces and/or slower kinetics (e.g., \ce{BiFeO3}).

Although our current study focuses on the synthesis of oxides, we expect our synthesis planning approach to be suitable to other chemistries where solid-state synthesis can be employed. This includes the chemistries of most ionic compounds: halides, chalcogenides, pnictides, some silicides/carbides, etc. Still, one must ensure that there is enough thermodynamic data available to accurately model phase competition in the chemical system of interest. This is generally true for oxide compounds due to their high prevalence in literature and thermodynamic data; for example, currently, $\sim$53\% of the nearly one-hundred and fifty-thousand compounds in the Materials Project contain oxygen. While the predictive accuracy is currently greatest for oxides, we expect our approach to grow in accuracy and general applicability as computed materials databases grow in size and chemical complexity.

Currently, our workflow focuses exclusively on optimizing product purity; however, there are many issues one must consider when designing a synthesis recipe for a target compound: material cost, safety concerns, stability in air, handling challenges, availability of precursors, etc. Many routes suggested involve the formation of byproducts that are not easily removable from the product mixture (e.g., the formation of \ce{BaTiO3} with byproduct \ce{Ba2TiO4}). To this point, one should be thoughtful in designing criteria by which to filter recommended synthetic routes. For example, one can prioritize the formation of only gaseous byproducts (e.g., \ce{O2} or \ce{CO2}) or those easily removable by a solvent (e.g., \ce{NaCl}). The cost function used to rank reactions can be modified to include other reaction metrics of interest, such as the estimated economic cost of the precursor materials. While not explicitly demonstrated here, the synthesis planning workflow can also be extended for application in multi-step syntheses, allowing one to retrosynthetically sequence reactions to a target material beginning with purchasable, ``off-the-shelf'' precursors.

\section{Conclusions}
Using the interface reaction model for powder reactions, we proposed two thermodynamic selectivity metrics for solid-state reactions: primary ($C_1$) and secondary ($C_2$) competition. To systematically and critically examine the effectiveness of our metrics, we analyzed existing successful synthesis routes available in the literature and, leveraging a massive set of 82,985 synthesis reactions extracted from an 18-element reaction network constructed from Materials Project data, designed and executed nine \ce{BaTiO3} synthesis experiments with a range of selectivity values as compared to conventional precursors (\ce{BaCO3} and \ce{TiO2}). Analysis of reaction pathways in the nine experiments via \textit{ex post facto} synchrotron powder X-ray diffraction reveals that $C_1$ and $C_2$ correlate with the maximum amounts of target and impurity formed, respectively. 

The main advantage of our approach compared to recent, existing approaches\cite{Aykol2021, Chen2023} is the ability to simultaneously consider a wide range of chemistries, including those with unconventional additional elements. These so-called hyperdimensional chemistries \cite{Neilson2023} allow one to bypass commonly encountered intermediates in target systems with many competing phases. This was demonstrated particularly for the \ce{BaTiO3} system studied in this work and is relevant for many other materials in the literature that are conventionally synthesized with theoretically suboptimal precursors (e.g., \ce{Na2Ti3O7}, \ce{NaTaO3}, \ce{LiMn2O4}, etc.).

We anticipate that the selectivity metrics presented here and our computational synthesis planning workflow will significantly reduce the synthesis bottleneck, providing more rapid development of synthesis approaches for new, predicted materials. Our workflow provides a theoretical rationale for using certain precursors and synthesis conditions over other options, which promises to optimize existing synthesis procedures for current technologically important materials. 

We envision our approach to be particularly useful in aiding high-throughput automated laboratory exploration efforts.\cite{Szymanski2021} Predictions can be used to design and downselect the synthesis reactions tested, reducing the cost and current trial-and-error approach to inorganic materials synthesis. The future inclusion of models for the kinetic behavior of reactions, such as estimates of the reactivity of precursors based on solid-state diffusivities, will further enhance predictions.

\section{Methods}
\subsection{Thermodynamic Data}
Gibbs free energies of formation, $\Delta G_\text{f}(T)$, were acquired or approximated in a similar approach to those of previous works.\cite{McDermott2021, Todd2021} We acquired experimental $\Delta G_\text{f}(T)$ values from the NIST-JANAF thermochemical tables \cite{Chase1998} where available. Experimental values were limited to compounds with low melting points (i.e., $T_m \leq 1500$ \textcelsius{}), as these systems demonstrate more complex phase change behavior over the temperature range studied here. For predominantly solid compounds (i.e., those with melting points above this threshold), as well as for all other phases not available in the NIST-JANAF thermochemical tables, we estimated $\Delta G_\text{f}(T)$ using the machine-learned Gibbs free energy descriptor identified by Bartel et al.\cite{Bartel2018} This descriptor was applied using formation enthalpies, $\Delta H_\text{f}(T=298$ K), acquired from the Materials Project (MP) database,\cite{Jain2013} version 2022.10.28.

Due to the well-known and systematic formation energy error of carbonate compounds calculated with GGA exchange-correlation functionals,\cite{Aykol2021, Huo2022} we applied an energy correction of 0.830 eV per \ce{CO3^{2-}} anion to all carbonate compounds acquired from MP. This value was determined by fitting the mean error between computed and experimental $\Delta G_\text{f}(T=300$ K) values for 15 metal carbonate compounds (Figure \ref{supp_fig:carbonates}).

\subsection{Synthesis Planning Workflow}

The synthesis reaction calculation and ranking procedure was implemented as a Python-based workflow in the existing \textit{reaction-network} package.\cite{McDermott2021} The code is available on GitHub at \url{https://github.com/materialsproject/reaction-network}. The workflow was constructed and launched on computing resources using the \textit{jobflow}\cite{Ganose2023} and \textit{fireworks}\cite{Jain2015} workflow packages.

The synthesis planning workflow consists of three sequential steps. First, phases and their formation energies for the chemical system of interest are acquired as previously described. The total number of phases can be optionally reduced by setting a threshold for the maximum energy above hull ($\Delta G_{\text{hull}}$). In this work, we used a moderately large threshold of $\Delta G_{\text{hull}} \leq 50$ meV/atom, evaluated at ambient temperature ($T=300$ K). Second, reaction enumeration is performed for the acquired phases using the combinatorial and free energy minimization approaches described in our previous work on solid-state reaction networks.\cite{McDermott2021} Note that the combinatorial approach allows one to identify reaction product combinations above the hull (i.e., ``metastable'' products), which makes the analysis more robust to numerical error in the thermodynamic data. For systems with an open element (e.g., \ce{O2} gas), this reaction enumeration step is performed again using grand potential energies, where the open element has been assigned a user-defined value for the chemical potential (often the standard state, $\mu=\mu^0$). Finally, $C_1$ and $C_2$ scores are calculated for all target synthesis reactions (i.e., those that form the desired target composition). To do this, the relevant competing reactions are extracted from the full set of enumerated reactions. We define a competing reaction as one whose precursors are a subset of the target reaction's precursors. These competing reactions are then used to compute the interface reaction hull, from which $C_1$ and $C_2$ are calculated via Equations \ref{eq:c1} and \ref{eq:c2}. For open systems, this selectivity calculation procedure is performed again, including any additional enumerated open reactions and ensuring that all reactions are calculated with grand potential energies at the corresponding chemical potential.

\subsection{Secondary Competition Algorithm}
The secondary competition score, $C_2$, is defined as the negative sum of the mean secondary reaction sequence energies to the left and right of the target on the interface reaction hull (Equation \ref{eq:c2}). One approach for acquiring these quantities involves using a recursive algorithm to identify all possible sequences and their energies. However, this strategy is too slow for the high-throughput calculation of $C_2$ in systems with many competing reactions.

Instead, we have identified a non-recursive algorithm that takes advantage of the connection between this problem and the recursive construction of binary trees via the use of the Catalan number sequence. Our algorithm reformulates the sum of all secondary reaction sequence energies as a sum of individual secondary reaction energies weighted by their multiplicities, i.e., the total number of appearances of a particular reaction within the set of all possible secondary reaction sequences. The energy of any reaction indexed $k$ can be calculated geometrically as the altitude, $h_k$, of the triangle formed by its product vertex and two reactant vertices on the interface reaction hull. We find that the altitude multiplicity, $m_{h_k}$, is determined to be the product of three Catalan numbers, $u_n$, such that
\begin{equation}
\label{eq:altitude}
m_{h_k} = u_{n_l} \cdot u_{n_r} \cdot u_{(n-n_l-n_r-1)},
\end{equation}
where $n_l$ and $n_r$ refer to the number of interior vertices (i.e., within the triangle) to the left and right of the vertex of interest, respectively, and $n$ is the total number of interior vertices for the entire hull subsection. For example, for secondary reactions between nearest neighbors, $n_l=0$ and $n_r=0$, resulting in an altitude multiplicity of $m_h=u_{n-1}$. 

The mean secondary reaction sequence energy for the hull subsection can then be calculated as
\begin{equation}
\label{eq:mean_decomp}
\overline{\Delta G_{2}} = \frac{1}{N}\sum_k^V{m_{h_k}\cdot h_k},
\end{equation}
where the sum occurs over all of the $V$ unique reaction energies (altitudes), which is the number of unique triangles that can be constructed for the hull subsection, including the two exterior vertices: $V = \binom{n+2}{3}$. The total number of unique secondary reaction sequences equals the corresponding Catalan number, $N=u_n$. Finally, once this process has been performed for both the left and right hull subsections, the secondary competition ($C_2$) can be calculated via Equation \ref{eq:c2}.

\subsection{Literature Reactions}
Solid-state literature reactions studied in this work were acquired from the text-mined dataset of 31,782 inorganic materials synthesis recipes originally extracted from the literature by Kononova et al.\cite{Kononova2019} and available at \url{https://github.com/CederGroupHub/text-mined-synthesis_public} (version 2020-07-13). The original dataset was filtered down to 8,530 reactions that contain: 1) precursors composed of $\leq2$ solids and $\leq1$ elemental gases (i.e., \ce{O2}, \ce{H2}, and \ce{N2}), 2) no elements with an atomic number greater than 94 or for which the Gibbs free energy descriptor does not apply (e.g., Ne, Ar, Pm, Ra), 3) ten or fewer total elements due to limitations in the convex hull algorithm, and 4) no ions. Finally, these reactions were required to be stoichiometrically balanceable after adjusting compositions for hydrates and fractional formulas. For reactions containing variable compositions with one open variable (e.g., \ce{Nd_{1-$x$}Sr_$x$CoO_3}), we attempted to substitute all extracted values of $x$ and retained the reactions that could be successfully balanced.

Competition metrics and free energies were assessed for each of the remaining reactions. For the enumerated competing reactions, metastable phases were considered up to a maximum threshold of $\Delta G_{\text{hull}}=50$ meV/atom, evaluated at ambient temperature ($T=300$ K). Interface reaction hulls were constructed at the maximum temperature reported during synthesis, $T_\textrm{syn}$. If this was not provided, a temperature of 800 \textcelsius{} was assumed. Formation energies, $\Delta G_\text{f}(T_\text{syn})$, were assigned based on the ground-state energy for a given composition; i.e., we selected the lowest available formation energy of all polymorphs with the composition of interest. For increased accuracy, we did not include a reaction if any of its entries were missing from our thermodynamic data. For reactions with an open gas (i.e., \ce{O2}, \ce{H2}, \ce{N2}), we assigned a chemical potential of $\mu_\text{gas} = 0$ eV (i.e., standard state at $T_\text{syn}$) for that element. For reactions completed in air, we assumed an \ce{O2} partial pressure of 0.21 atm and thus assigned a chemical potential of $\mu_\text{O} = \frac{1}{2}k_bT_\text{syn}\ln(0.21)$ eV. Finally, we removed duplicates with the same reaction equation and temperature/environment, as well as identity reactions (e.g., $A \rightarrow A$). These filtering steps yielded a total of 3,520 unique literature reactions.

\subsection{Precursor Materials}
Precursors for all experiments were purchased from chemical providers or prepared via known solid-state synthesis approaches, as necessary. Precursors acquired from chemical providers include barium carbonate (\ce{BaCO3}, J.T.Baker 99.9\%), titanium(IV) oxide (anatase \ce{TiO2}, Acros Organics 99.9\%), barium sulfate (\ce{BaSO4}, J.T.Baker 99.9\%), barium hydroxide hydrate (\ce{Ba(OH)2*8H2O}, Mathsen Colman \& Bell 98\%), barium chloride hydrate (\ce{BaCl2*2H2O}, Fisher Scientific 99.9\%), and titanium metal (\ce{Ti}, annealed foil, Alfa Aesar 99.7\%).

Precursors prepared via solid-state synthesis include barium orthotitante (\ce{Ba2TiO4}), \ce{BaTi2O5}, barium sulfide (\ce{BaS}), and sodium metatitanate (\ce{Na2TiO3}). Phase purities were assessed via laboratory powder x-ray diffraction (PXRD) analysis performed with a Bruker D8 Discover diffractometer using Cu K$\alpha$ radiation.

\ce{Ba2TiO4} was prepared using stoichiometric amounts of \ce{BaCO3} and anatase \ce{TiO2}.\cite{McSloy2018} The chemicals were mixed, ground using a mortar and pestle, placed in an alumina boat inside of a mullite process tube with self-sealing endcaps, and then heated at 950 \textcelsius{} for 16 hrs under Ar flow with a heating rate of 10 \textcelsius{}/min. The powder was then reground and reheated at 1100 \textcelsius{} for another 16 hrs at a heating rate of 10 \textcelsius{}/min. Handling operations were completed in an Ar glovebox due to the hygroscopic nature of \ce{Ba2TiO4}. The product was phase pure $\beta$-\ce{Ba2TiO4} with no observed impurities.

\ce{BaTi2O5} was prepared using stoichiometric amounts of \ce{BaCO3} and anatase \ce{TiO2}.\cite{Zhu2010} The chemicals were mixed, ground using a mortar and pestle, and heated in an alumina boat at 900 \textcelsius{} for 5 hrs as a pre-treatment step. The powder was then reground and reheated at 1220-1225 \textcelsius{} for 24 hrs with heating and cooling steps of 3 hrs. The product was mostly phase pure with minor impurities, including a small amount of unreacted \ce{BaCO3} precursor ($<$3 mol \%) and \ce{Ba6Ti17O40} ($\sim$3 mol \%). The latter phase was similarly observed in Ref.\ \citenum{Zhu2010}, where its formation was attributed to the thermodynamic instability of \ce{BaTi2O5} at temperatures outside a very narrow range (1220-1230 \textcelsius{}).

\ce{BaS} was prepared using \ce{BaSO4} and activated carbon (\ce{C}, J.T.Baker 99.9\%).\cite{Murthy2012} The chemicals were mixed, ground using a mortar and pestle, pressed into a 0.5'' diameter pellet with two tons of force, and heated in an alumina boat at 1100 \textcelsius{} for 7-10 min in air, with a heating rate of 10 \textcelsius{}/min and natural cooling in the furnace. The product was phase pure with no detectable impurities.

\ce{Na2TiO3} was prepared using stoichiometric amounts of sodium hydroxide (\ce{NaOH}, Fisher Scientific 99.9\%) and anatase \ce{TiO2}, with a slight excess of \ce{NaOH}.\cite{Meng2016} The chemicals were mixed, ground using a mortar and pestle, and heated in an alumina boat at 500 \textcelsius{} for 2 hrs with a heating rate of 10 \textcelsius{}/min. The product was mostly phase pure with minor impurities. The sodium titanate peaks are best fit by a cubic $\alpha$-\ce{Na2TiO3} structure with a small crystallite size. A minor amount of unreacted anatase \ce{TiO2} was present in the product ($\sim$1 mol \%). \ce{Na2CO3} also appears to be present as an impurity ($\sim$11 mol \%); we suspect this is due to contamination of the \ce{NaOH} precursor via reaction with \ce{CO2} in the air.

\subsection{\textit{Ex Post Facto} SPXRD Reactions}

Synchrotron powder X-ray diffraction (SPXRD) data were collected in transmission (i.e., Debye-Scherrer) geometry on beamline 28-ID-2 (XPD) at the National Synchrotron Light Source-II (NSLS-II). Data were collected on a 2D area detector (PerkinElmer XRD 1621, 2048x2048 pixel array, 200x200 $\mu$m pixel size) at a sample-to-detector distance of 1407.1 mm using an incident X-ray energy of 68.12 keV ($\lambda$ = 0.182 \AA) with a 0.60 x 0.20 mm beam size. A total acquisition time of 1 s was used, summing five subframes collected for 0.2 s each.
 
Samples were packed into 1.1 mm OD/0.9 mm ID quartz capillaries. The capillary ends were filled with a 3 mm plug of powder silicon (\ce{Si}, Strem 99.0\%) followed by a cap of recycled silicon dioxide (\ce{SiO2}). To account for possible gas production, the capillaries for Expts.\ 1, 2, 5, 7, and 8 were left unsealed, and a moderate vacuum ($P_\textrm{gage}=-20$ in Hg) was pulled on the samples during heating. All other sample capillaries (Expts.\ 3, 4, 6, and 9) were flame-sealed under argon.

Experiments were carried out in the gradient furnace described in Ref.\ \citenum{ONolan2020}, which heats samples to different temperatures across a range of spatial positions on the capillary (Figure \ref{supp_fig:gradient_furnace}). The furnace was operated with a Eurotherm 2408 temperature controller and a TDK Lambda 900W (30V/30A) power supply. Furnace heating elements were wound from resistive wire (Kanthal A-1, \#24 awg). A K-type thermocouple (stainless steel, 0.01'' OD) placed at an intermediate position along the sample was used as input for PID control of the furnace. We performed experiments in three temperature ranges with setpoints of $T_H=550$ \textcelsius{} (Expts.\ 1, 3, 4, 8), $T_{L1}=450$ \textcelsius{} (Expts.\ 2, 5, 6, 9), and $T_{L2}=400$ \textcelsius{} (Expt.\ 7). This choice was motivated by differences in reactivity among the samples. Position-dependent temperatures were determined using a fit of measured \textit{in situ} lattice expansion from \ce{NaCl}/\ce{Si} and \ce{Al2O3}/\ce{MgO} standards (Figure \ref{supp_fig:gradient_calibration}). Using the root-mean-square error of the curve fit, the estimated uncertainty for each temperature point is 11.4 \textcelsius{} ($T_H$), 7.9 \textcelsius{} ($T_{L1}$), and 10.9 \textcelsius{} ($T_{L2}$). The experiments spanned a total temperature range of 189-1064 \textcelsius{}. 

The median total time of each experiment was $\sim$67 min. Heating, holding, and cooling times varied among experiments due to differences in sample heat capacities, reactivities, and the sizes of investigated temperature windows; specifically, unreactive samples (Expts.\ 8, 9) and samples with smaller studied temperature windows (Expts.\ 3, 4) were held at elevated temperatures for shorter times. These differences are accounted for in our analysis via the use of relative metrics (i.e., mole fraction) and normalization by reaction progress. The exact heating, holding, and cooling times for each sample are shown in Table \ref{tab:exp_times}.

\subsection{Quantitative Phase Analysis of Powder Diffraction Data}

Quantitative analysis of synchrotron powder X-ray powder diffraction data was carried out with the Rietveld method using either the TOPAS v6 (Expts.\ 1-3, 5, 7-9) or GSAS-II (Expts.\ 4, 6) software packages.\cite{Coelho2018, Toby2013} Atomic displacement parameters were fixed to $B=1$ \r{A}$^2$, and peak broadening was primarily modeled via crystal size broadening using a Lorentzian function. Site occupancies were fixed at one.

\section{Supporting Information}
\begin{itemize}
    \item Description of cost function power transformation; interface reaction hull for \ce{BaO}$|$\ce{TiO2}; open-\ce{O2} reaction data for \ce{BaTiO3}; selected Rietveld refinements from all experiments; interface reaction hull plots for selected experimental reactions; selectivity metric correlations; details of carbonate correction; description of gradient furnace setup and temperature calibration; table of experimental reaction times and temperature setpoints (PDF)
    \item Table of 3,520 reactions extracted from the literature with their calculated performance metrics and DOIs (XLSX)
    \item Table of 82,985 predicted/ranked closed \ce{BaTiO3} synthesis reactions (XLSX)
    \item Table of 62,133 predicted/ranked open-\ce{O2} \ce{BaTiO3} synthesis reactions (XLSX)
    \item Table of 133 Pareto optimal open-\ce{O2} \ce{BaTiO3} synthesis reactions (XLSX)
    \item Table of 478 predicted/ranked closed \ce{BaTiO3} synthesis reactions filtered by commonly available (i.e., purchasable) precursors (XLSX)
    \item Table of 622 predicted/ranked open-\ce{O2} \ce{BaTiO3} synthesis reactions filtered by commonly available (i.e., purchasable) precursors (XLSX)
\end{itemize}

\section{Acknowledgements}
This work was supported as part of GENESIS: A Next Generation Synthesis Center, an Energy Frontier Research Center funded by the U.S. Department of Energy, Office of Science, Basic Energy Sciences under Award Number DE-SC0019212. This research used resources of the National Energy Research Scientific Computing Center (NERSC), a U.S. Department of Energy Office of Science User Facility operated under Contract No. DE-AC02-05CH11231. Additional computational resources were provided by the Swift high-performance computing (HPC) system at the National Renewable Energy Laboratory (NREL). This research used Beamline 28-ID-2 (XPD) of the National Synchrotron Light Source II, a U.S. Department of Energy (DOE) Office of Science User Facility operated for the DOE Office of Science by Brookhaven National Laboratory under Contract No. DE-SC0012704. The authors thank the Analytical Resources Core, Center for Materials and Molecular Analysis at Colorado State University for instrument access and training (RRID: SCR\_021758). Partial support for G.E.K. was provided by an NRT QuADS Fellowship (Quantitative Analysis of Dynamic Structures) under NSF DGE-1922639.

The authors would like to thank Ren Borgia, Dr. Sanjit K. Ghose, Dr. Hui Zhong, and John Trunk for their experimental assistance. M.J.M. would also like to thank Dr. Rishi Kumar, Dr. Andrew Rosen, and Evan Spotte-Smith for helpful discussion and feedback.

\bibliography{selectivity_paper}

\pagebreak

\section{TOC Graphic}

\vspace{3mm} 

\begin{center}
\frame{\includegraphics[]{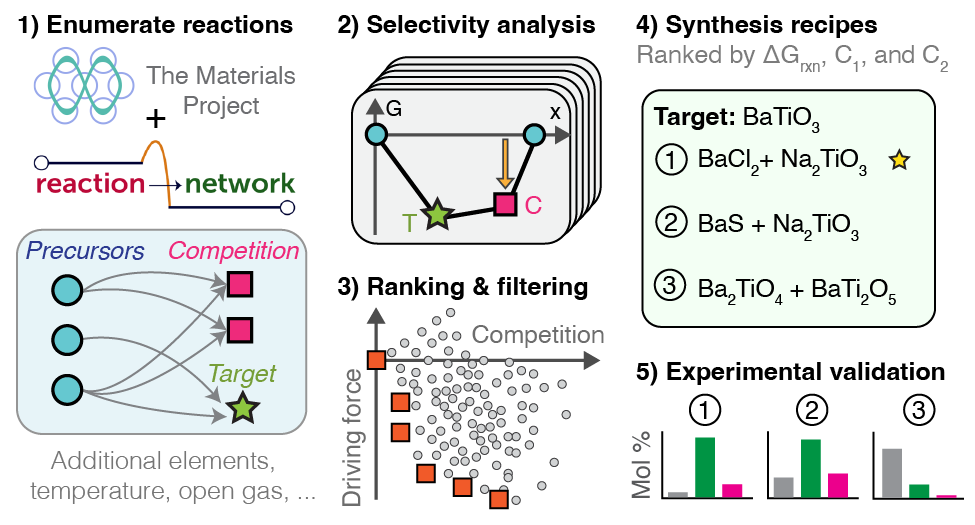}}
\end{center}                              

\section{Synopsis}
Selectivity analysis of massive inorganic chemical reaction networks identifies alternative synthesis recipes for solid-state materials (e.g., \ce{BaTiO3}) that outperform conventional approaches.

\end{document}


\newpage

\section{Power transform of costs in literature reactions}

\begin{figure}[ht!]
\begin{center}
\includegraphics[width=0.90\textwidth]{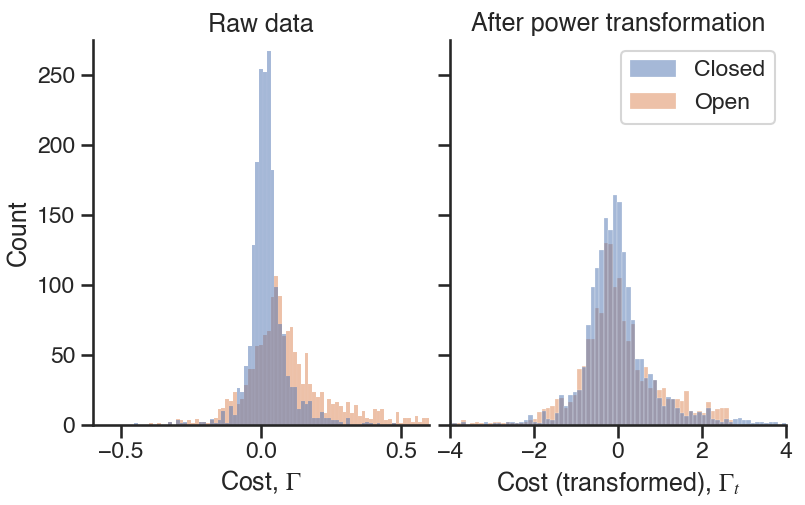}
\caption{\label{supp_fig:power_transform} \textbf{Power transformation applied to literature reaction costs, $\Gamma$}. The power transform monotonically transforms the reaction costs for closed and open reactions such that they resemble standard normal distributions, allowing for better comparison between the two datasets.}
\end{center} 
\end{figure}

\newpage

\section{\ce{BaO}$|$\ce{TiO2} interface reaction hull}

\begin{figure}[ht!]
\begin{center}
\includegraphics[width=0.75\textwidth]{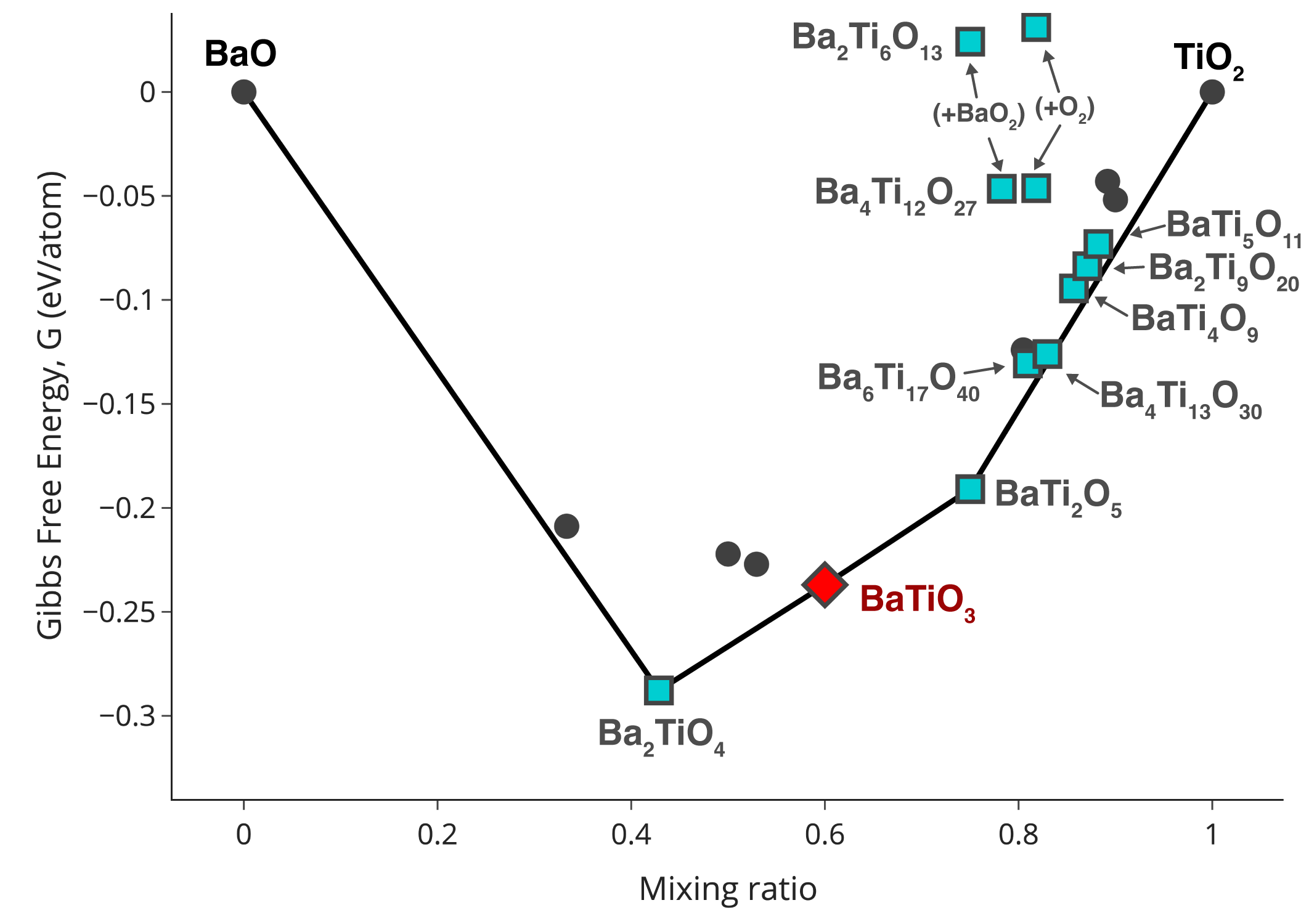}
\caption{\label{supp_fig:batio3_irh} \textbf{Calculated interface reaction hull for \ce{BaO}$|$\ce{TiO2} at \textit{T} = 600 \textcelsius{}}. Blue squares denote experimentally known ternary phases in the system. \ce{Ba2TiO4} is the most commonly observed intermediate/impurity in standard syntheses of \ce{BaTiO3} from \ce{BaO} (or \ce{BaCO3)} and \ce{TiO2}; it is also predicted to be the phase with the greatest driving force to form.}
\end{center} 
\end{figure}

\pagebreak

\section{\ce{BaTiO3} synthesis reactions: open \ce{O2}}

\begin{table}[ht!]
\footnotesize
\caption{Selected experimental reactions to \ce{BaTiO3} and their associated grand potential energies ($\mu_O=-0.1001$ eV), $\Delta \Phi_{\text{rxn}}$ ($T=600$ \textcelsius{}), primary competition scores, $C_1$, secondary competition scores, $C_2$, and costs, $\Gamma$.}
\label{tab:bto_open_selected}
\begin{tabular}{cccccc}
\toprule
 & Reaction & $\Delta \Phi_{\text{rxn}}$ & C$_1$ & C$_2$ & $\Gamma$ \\
Expt. &  & (eV/at) & (eV/at) & (eV/at) & (eV/at) \\
\midrule
\textbf{1} & BaCO$_{3}$ + TiO$_{2}$ $\longrightarrow$ BaTiO$_{3}$ + CO$_{2}$ & 0.112 & 0.114 & 0.001 & 0.063 \\
\textbf{2} & BaO$_{2}$ + TiO$_{2}$ $\longrightarrow$ BaTiO$_{3}$ + 0.5 O$_{2}$ & -0.590 & 0.078 & 0.389 & 0.151 \\
\textbf{3} & Ba$_{2}$TiO$_{4}$ + TiO$_{2}$ $\longrightarrow$ 2 BaTiO$_{3}$ & -0.089 & 0.085 & 0.114 & 0.081 \\
\textbf{4} & Ba$_{2}$TiO$_{4}$ + BaTi$_{2}$O$_{5}$ $\longrightarrow$ 3 BaTiO$_{3}$ & -0.002 & -0.002 & 0.000 & -0.001 \\
\textbf{5} & Ba(OH)$_2\cdot$H$_2$O + 3.666 Ti $\longrightarrow$ BaTiO$_{3}$ + 1.333 Ti$_{2}$H$_{3}$ & -0.713 & 8.188 & 8.973 & 7.651 \\
\textbf{6} & BaCl$_{2}$ + Na$_{2}$TiO$_{3}$ $\longrightarrow$ BaTiO$_{3}$ + 2 NaCl & -0.112 & -0.004 & 0.077 & 0.021 \\
\textbf{7} & BaS + Na$_{2}$TiO$_{3}$ $\longrightarrow$ BaTiO$_{3}$ + Na$_{2}$S & -0.077 & 4.032 & 4.174 & 3.685 \\
\textbf{8} & 2 BaS + 3 TiO$_{2}$ $\longrightarrow$ 2 BaTiO$_{3}$ + TiS$_{2}$ & 0.168 & 4.276 & 4.109 & 3.790 \\
\textbf{9} & BaSO$_{4}$ + 2 TiO$_{2}$ $\longrightarrow$ BaTiO$_{3}$ + TiOSO$_{4}$ & 0.533 & 0.533 & -0.000 & 0.293 \\
\bottomrule
\end{tabular}
\end{table}

\pagebreak

\begin{figure}[ht!]
\begin{center}
\includegraphics[width=1.0\textwidth]{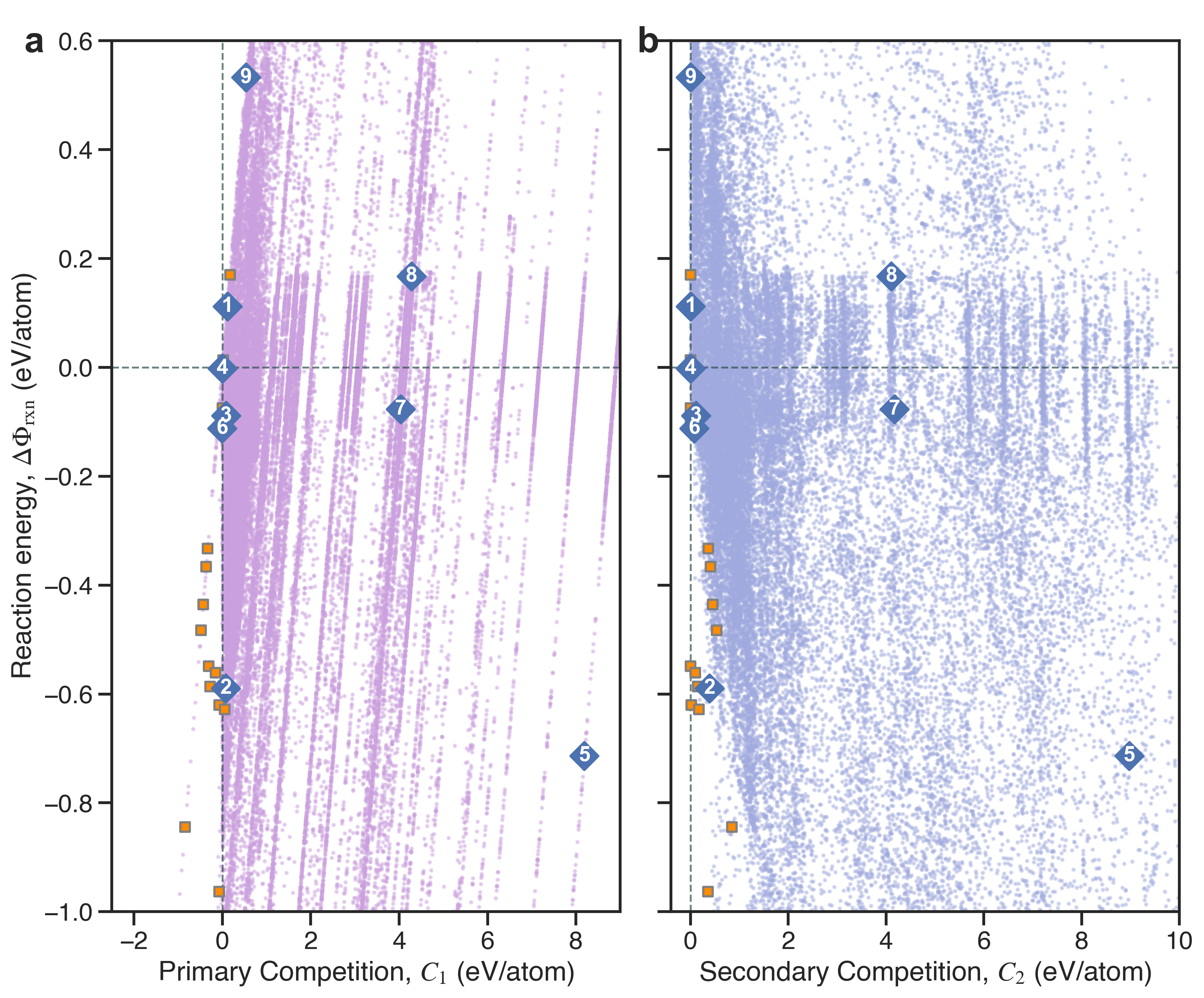}
\caption{\textbf{Synthesis map of 62,133 calculated open-\ce{O2} synthesis reactions producing \ce{BaTiO3}}. Reaction energies and competition scores are calculated assuming a temperature of $T=600$ \textcelsius{} and oxygen chemical potential $\mu_\text{O}=-0.1001$ eV, as approximated by selected experimental conditions (vacuum at $P_\textrm{gage}=-20$ in Hg). As in Figure \ref{fig:batio3_calcs}, reactions are plotted on a shared axis of reaction energy, $\Delta \Phi_\text{rxn}$, and on independent axes of (a) primary competition, $C_1$, and (b) secondary competition, $C_2$. Orange squares represent reactions on the three-dimensional Pareto frontier of $\Delta \Phi_\text{rxn}$, $C_1$, and $C_2$. Blue diamonds indicate selected reactions experimentally tested in this work.}
\label{supp_fig:batio3_open}
\end{center} 
\end{figure}

\pagebreak

\section{Selected Rietveld refinements for \ce{BaTiO3} experiments}

\begin{figure}[ht!]
\begin{center}
\includegraphics[width=1.0\textwidth]{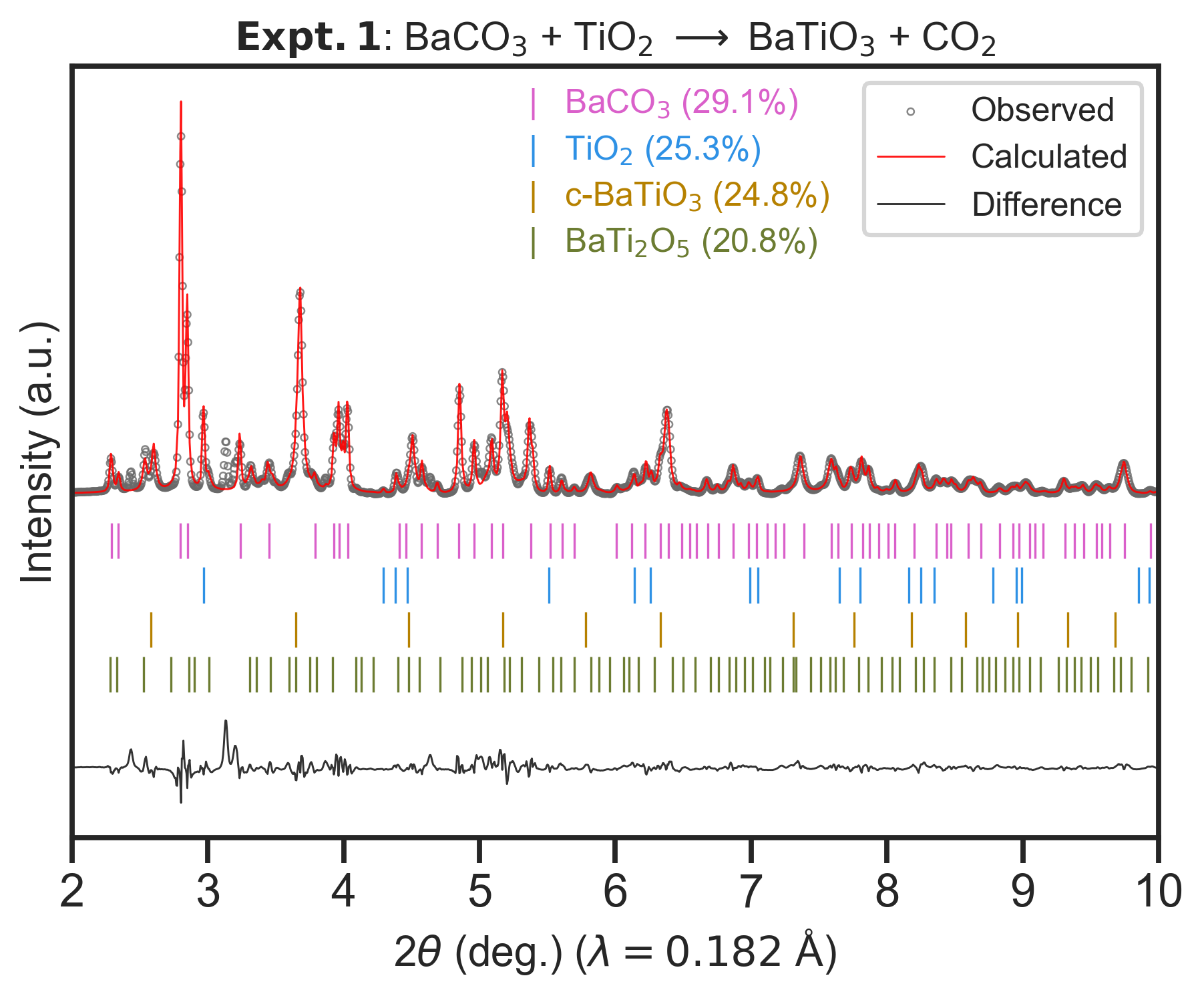}
\caption{\textbf{Selected Rietveld refinement for Experiment 1}. The observed pattern represents \textit{ex post facto} synchrotron powder X-ray diffraction data (SPXRD) captured following reaction at $T=1040$ \textcelsius{}, which corresponds to the temperature with the highest \ce{BaTiO3} yield. Phase fractions are shown in units of mole percent.}
\label{supp_fig:refinement1} 
\end{center} 
\end{figure}

\pagebreak

\begin{figure}[ht!]
\begin{center}
\includegraphics[width=1.0\textwidth]{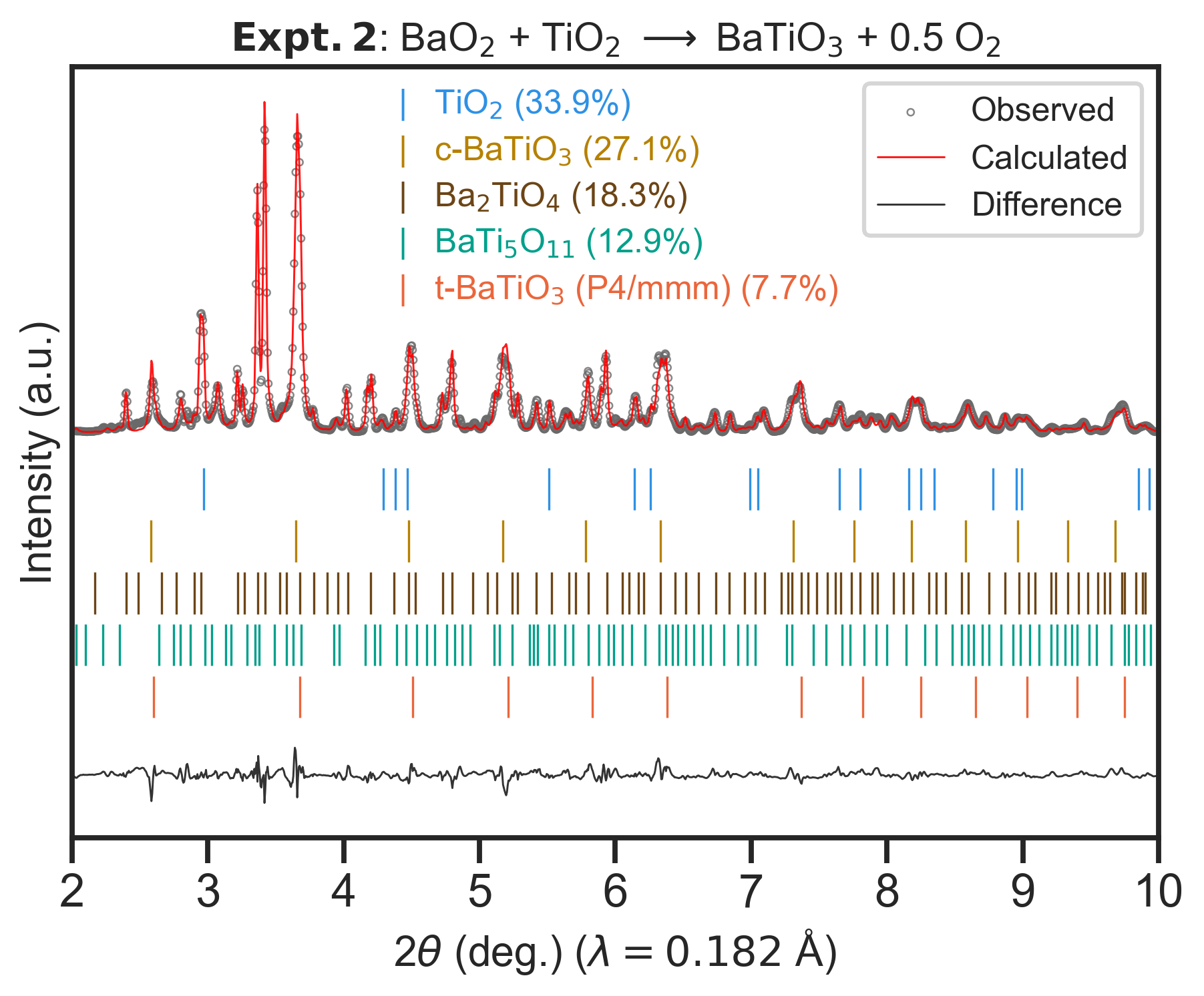}
\caption{\textbf{Selected Rietveld refinement for Experiment 2}. The observed pattern represents \textit{ex post facto} SPXRD data captured following reaction at $T=762$ \textcelsius{}, which corresponds to the temperature with the highest \ce{BaTiO3} yield. Phase fractions are shown in units of mole percent.}
\label{supp_fig:refinement2} 
\end{center} 
\end{figure}

\pagebreak

\begin{figure}[ht!]
\begin{center}
\includegraphics[width=1.0\textwidth]{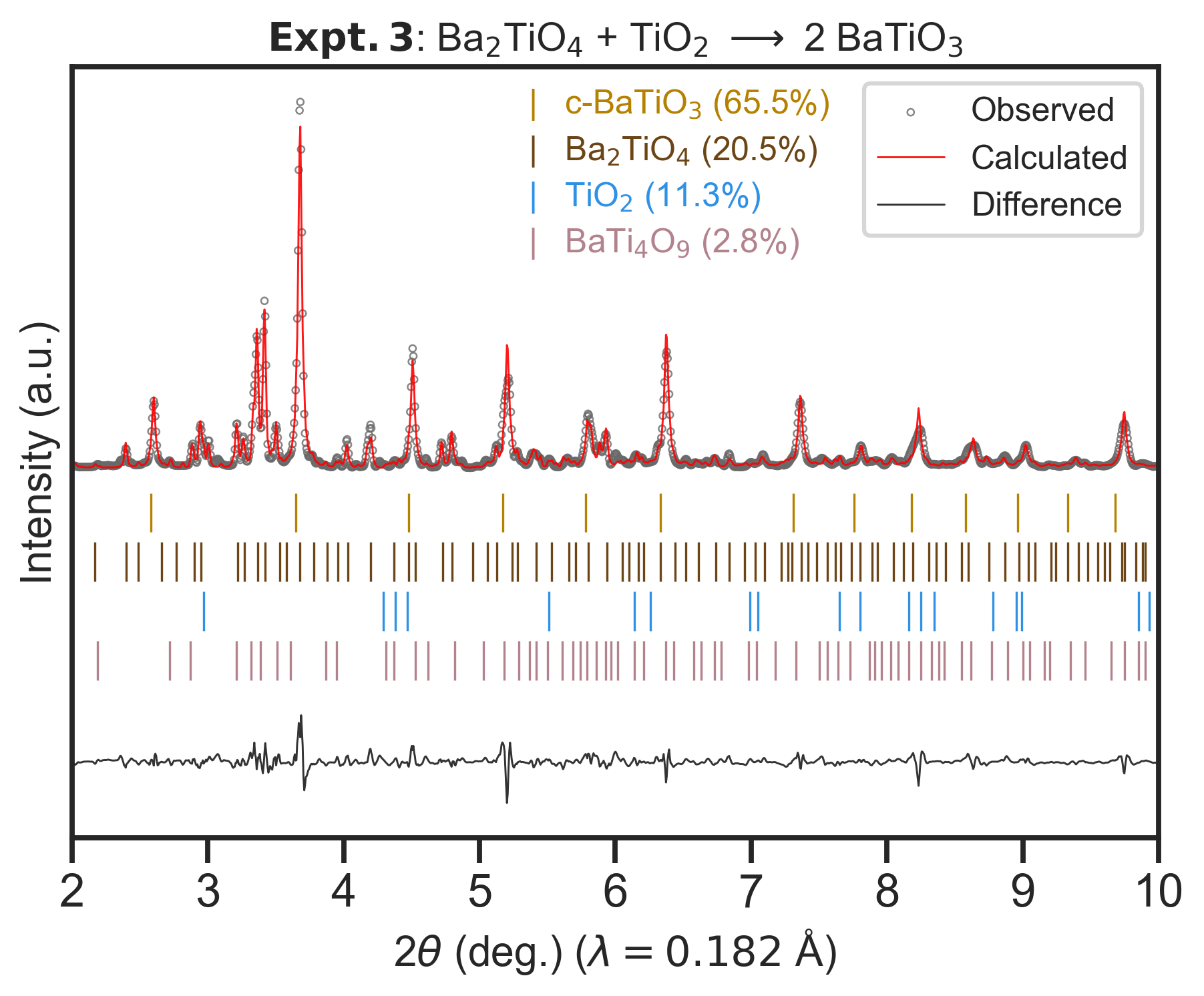}
\caption{\textbf{Selected Rietveld refinement for Experiment 3}. The observed pattern represents \textit{ex post facto} SPXRD data captured following reaction at $T=1045$ \textcelsius{}, which corresponds to the temperature with the highest \ce{BaTiO3} yield. Phase fractions are shown in units of mole percent.}
\label{supp_fig:refinement3} 
\end{center} 
\end{figure}

\pagebreak

\begin{figure}[ht!]
\begin{center}
\includegraphics[width=1.0\textwidth]{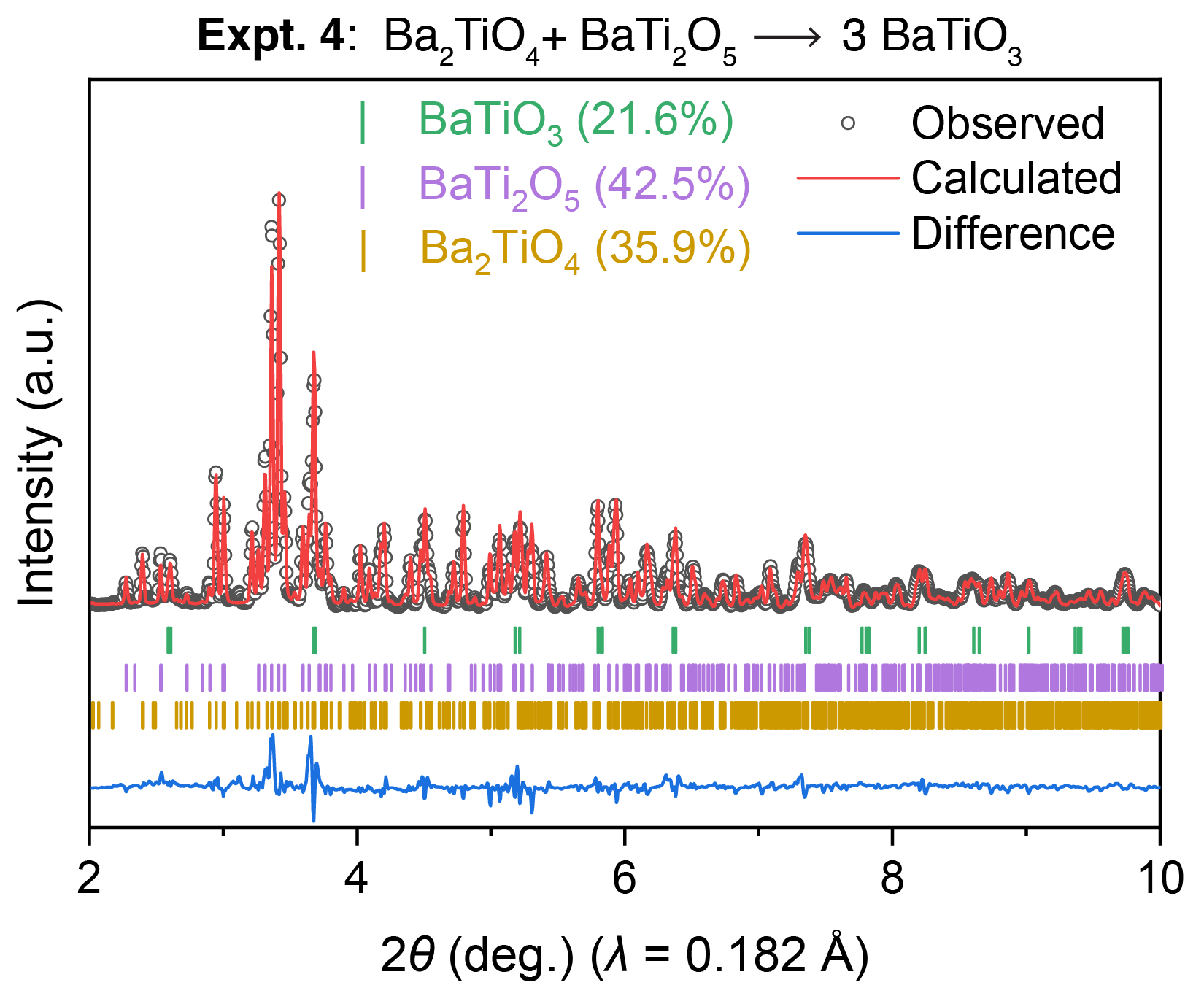}
\caption{\textbf{Selected Rietveld refinement for Experiment 4}. The observed pattern represents \textit{ex post facto} SPXRD data captured following reaction at $T=1025$ \textcelsius{}, which corresponds to the temperature with the highest \ce{BaTiO3} yield. Phase fractions are shown in units of mole percent.}
\label{supp_fig:refinement4} 
\end{center} 
\end{figure}

\pagebreak

\begin{figure}[ht!]
\begin{center}
\includegraphics[width=1.0\textwidth]{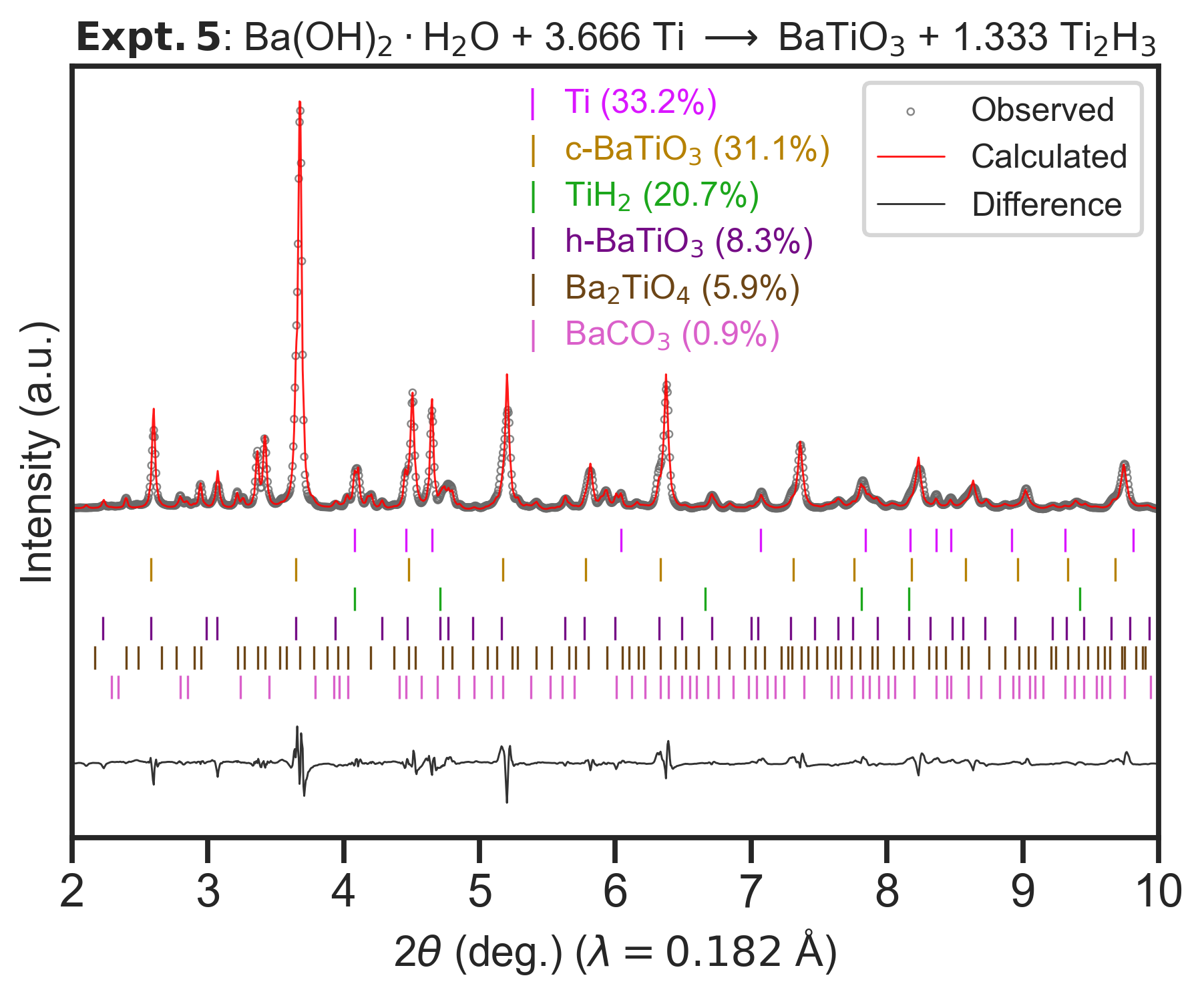}
\caption{\textbf{Selected Rietveld refinement for Experiment 5}. The observed pattern represents \textit{ex post facto} SPXRD data captured following reaction at $T=474$ \textcelsius{}, which corresponds to the temperature with the highest \ce{BaTiO3} yield. Phase fractions are shown in units of mole percent.}
\label{supp_fig:refinement5} 
\end{center} 
\end{figure}

\pagebreak

\begin{figure}[ht!]
\begin{center}
\includegraphics[width=1.0\textwidth]{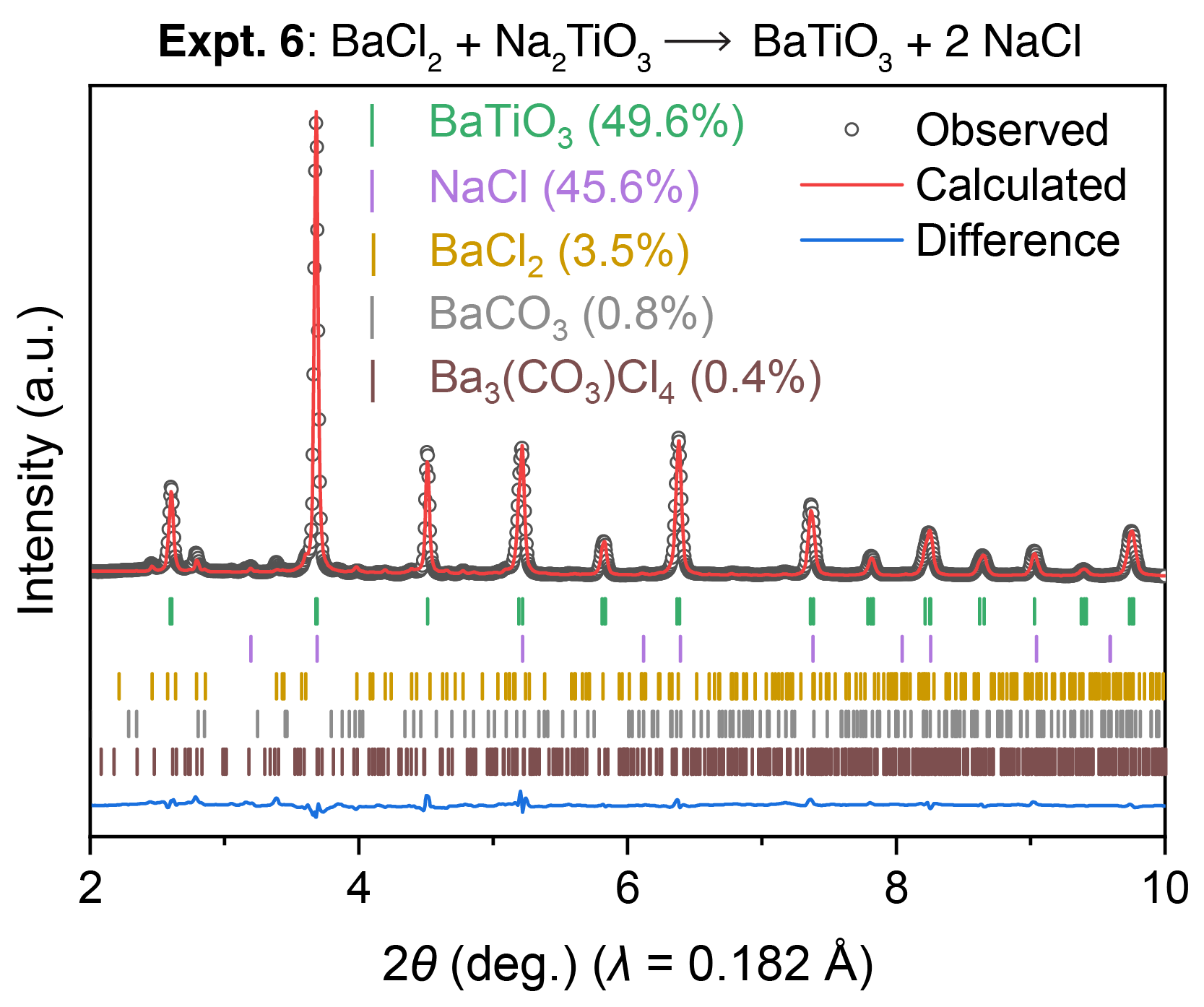}
\caption{\textbf{Selected Rietveld refinement for Experiment 6}. The observed pattern represents \textit{ex post facto} SPXRD data captured following reaction at $T=625$ \textcelsius{}, which corresponds to the temperature with the highest \ce{BaTiO3} yield. Phase fractions are shown in units of mole percent.}
\label{supp_fig:refinement6} 
\end{center} 
\end{figure}

\pagebreak

\begin{figure}[ht!]
\begin{center}
\includegraphics[width=1.0\textwidth]{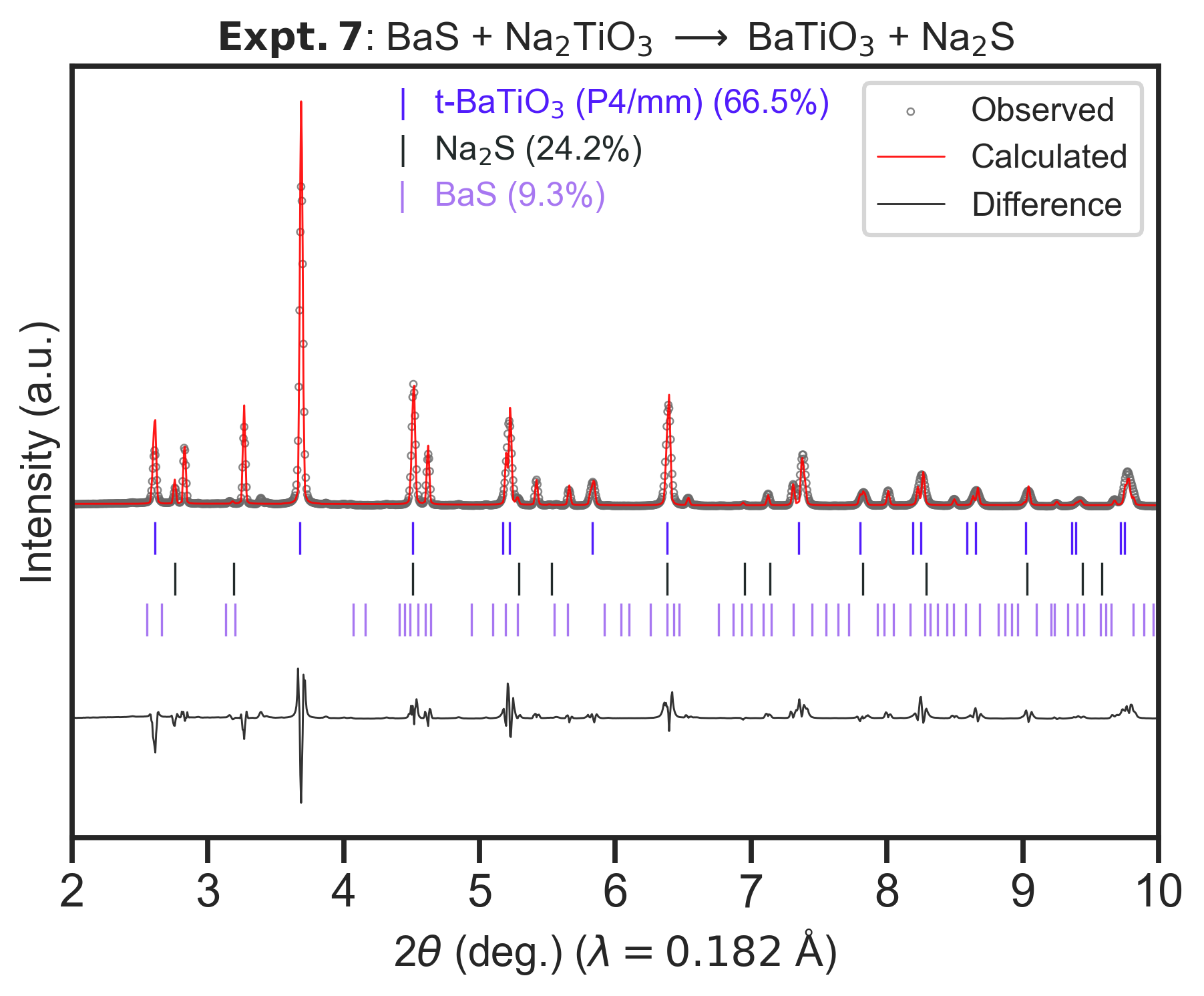}
\caption{\textbf{Selected Rietveld refinement for Experiment 7}. The observed pattern represents \textit{ex post facto} SPXRD data captured following reaction at $T=727$ \textcelsius{}, which corresponds to the temperature with the highest \ce{BaTiO3} yield. Phase fractions are shown in units of mole percent.}
\label{supp_fig:refinement7} 
\end{center} 
\end{figure}

\pagebreak

\begin{figure}[ht!]
\begin{center}
\includegraphics[width=1.0\textwidth]{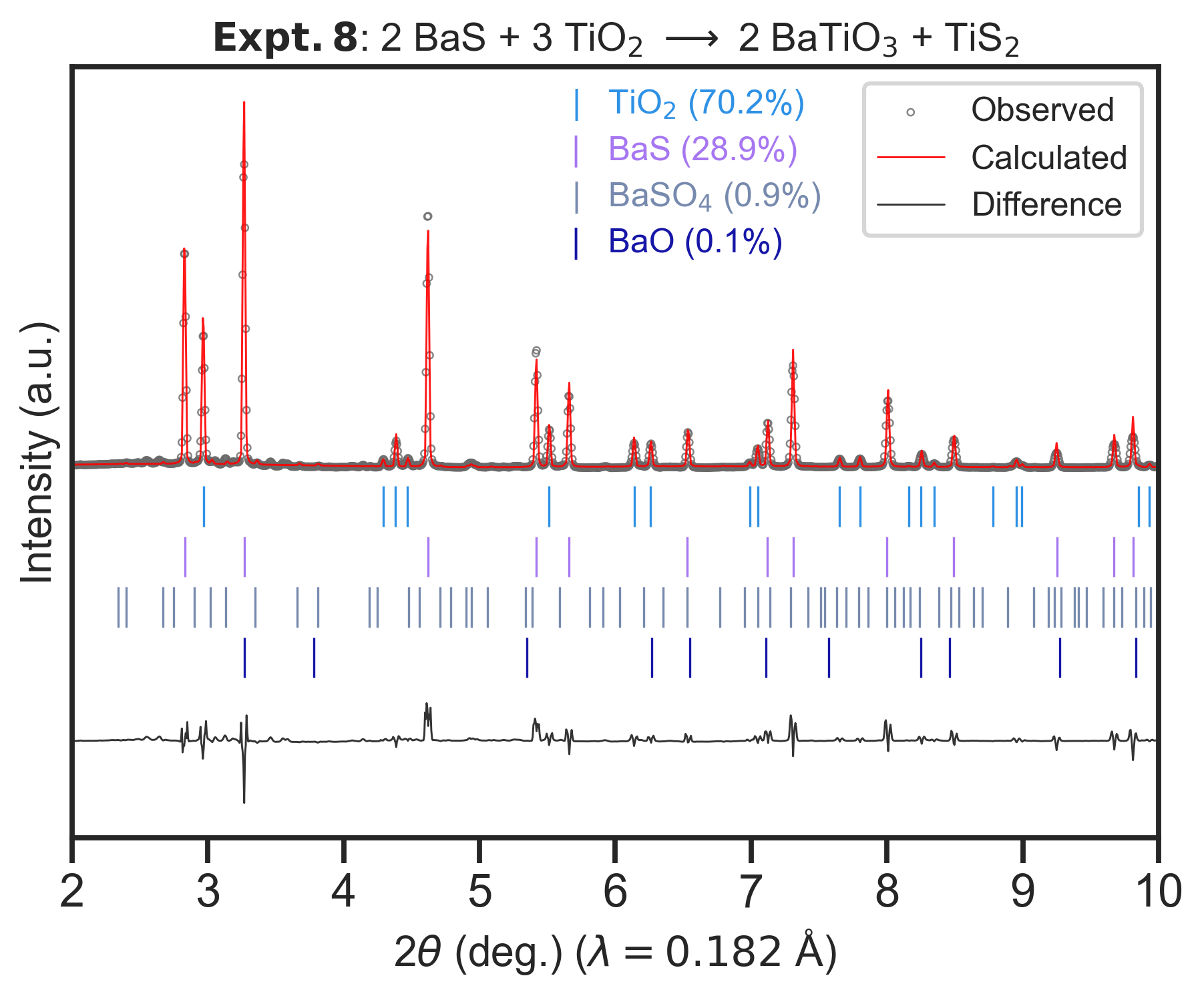}
\caption{\textbf{Selected Rietveld refinement for Experiment 8}. The observed pattern represents \textit{ex post facto} SPXRD data captured following reaction at $T=603$ \textcelsius{}. This value corresponds to the median temperature studied since \ce{BaTiO3} was not observed at any temperature. Phase fractions are shown in units of mole percent.}
\label{supp_fig:refinement8} 
\end{center} 
\end{figure}

\pagebreak

\begin{figure}[ht!]
\begin{center}
\includegraphics[width=1.0\textwidth]{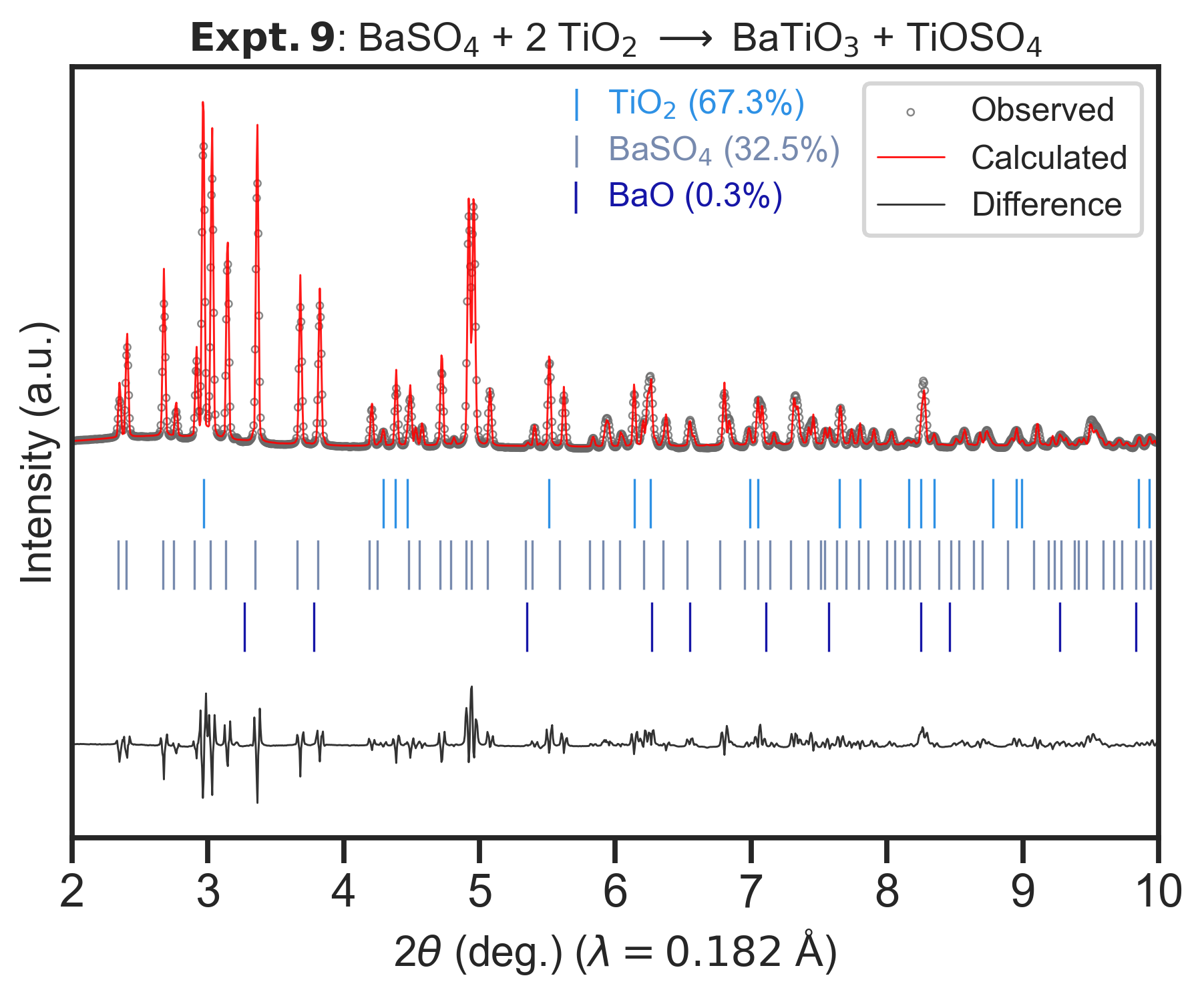}
\caption{\textbf{Selected Rietveld refinement for Experiment 9}. The observed pattern represents \textit{ex post facto} SPXRD data captured following reaction at $T=594$ \textcelsius{}. This value corresponds to the median temperature studied since \ce{BaTiO3} was not observed at any temperature. Phase fractions are shown in units of mole percent.}
\label{supp_fig:refinement9} 
\end{center} 
\end{figure}

\pagebreak
\section{Interface reaction hulls for selected \ce{BaTiO3} experiments}

\begin{figure}[ht!]
\begin{center}
\includegraphics[width=1.0\textwidth]{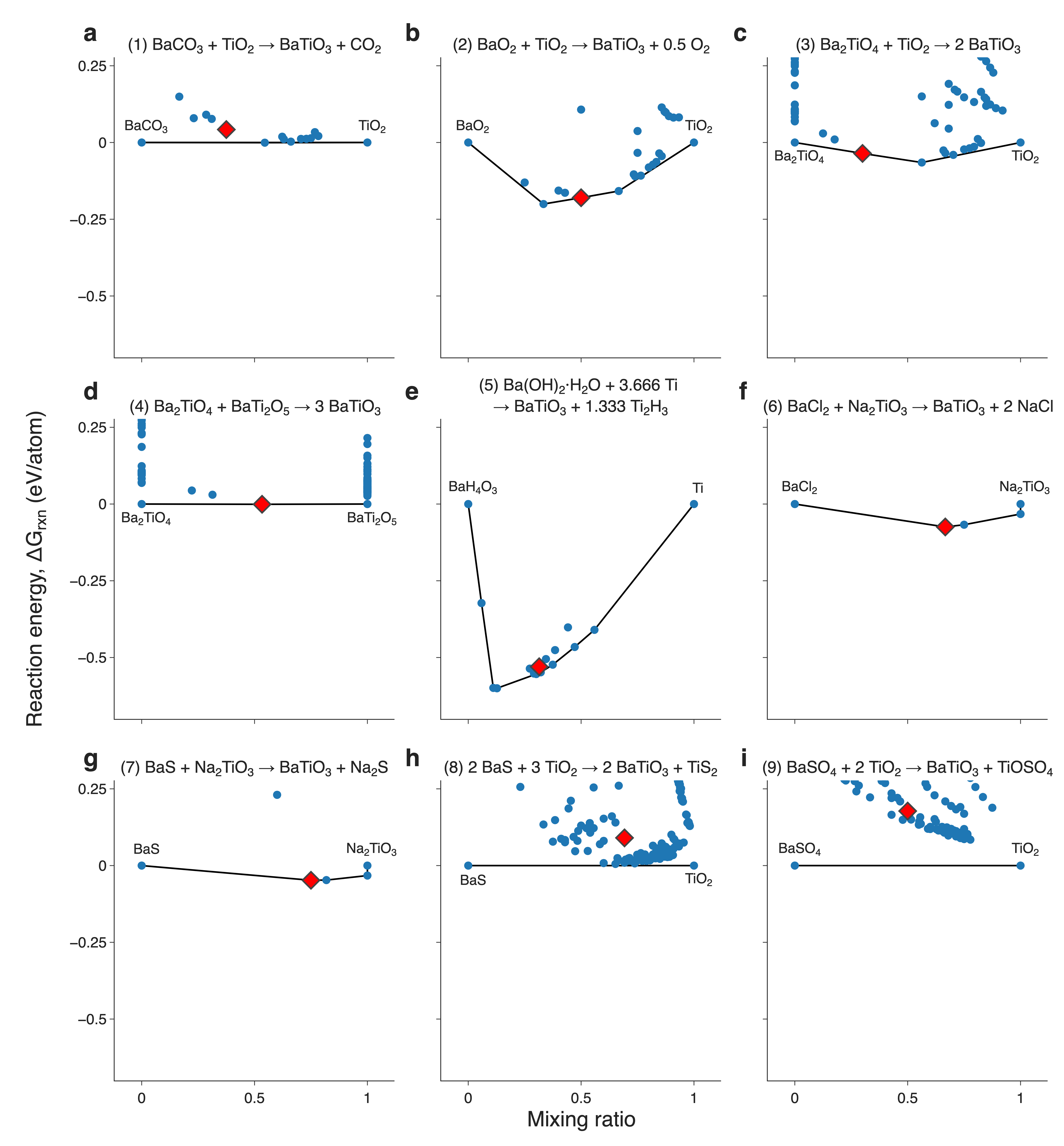}
\caption{\textbf{Interface reaction hulls for selected \ce{BaTiO3} experiments}. (a-i) Hulls for Experiments 1-9, as extracted from the full reaction network calculated during the synthesis planning workflow. Red diamonds mark the selected reactions of interest. All hulls are plotted on a uniform energy scale to facilitate comparison.}
\label{supp_fig:all_irh} 
\end{center} 
\end{figure}

\pagebreak

\section{All pairwise correlation plots between reaction metrics and experimental outcomes}
\begin{figure}[ht!]
\begin{center}
\includegraphics[width=1.0\textwidth]{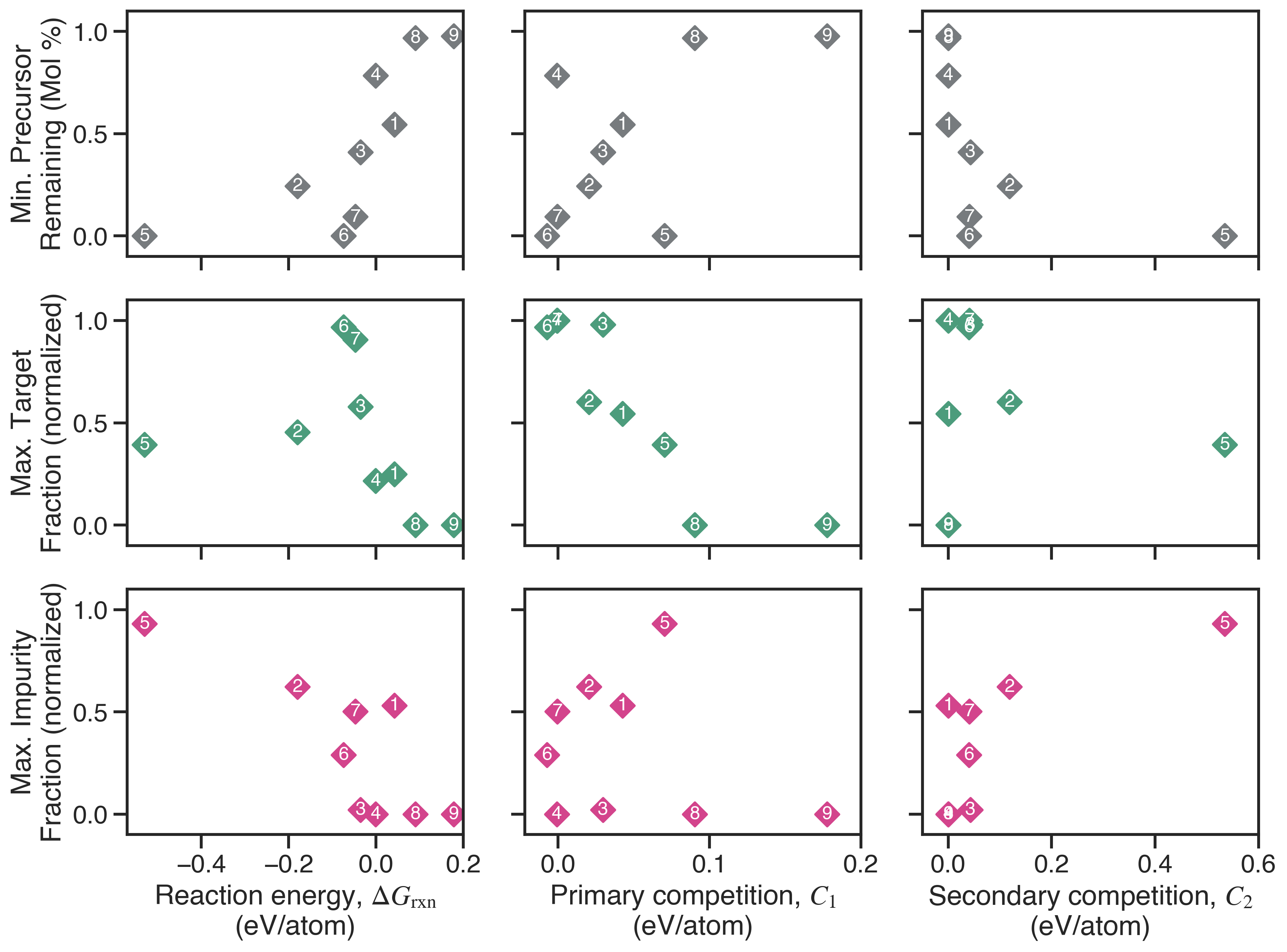}
\caption{\label{supp_fig:metric_performance} \textbf{Pairwise correlation plots between reaction metrics and experimental outcomes}. Plots of minimum remaining precursor ($P$), maximum target ($T$), and maximum impurity ($I$) formed as a function of reaction energy, primary competition, and secondary competition. The target and impurity plots have been normalized by the amount of precursor consumed ($1-P$).}
\end{center} 
\end{figure}

\pagebreak
\section{Correlations between selectivity metrics}

\begin{figure}[ht!]
\begin{center}
\includegraphics[width=1.0\textwidth]{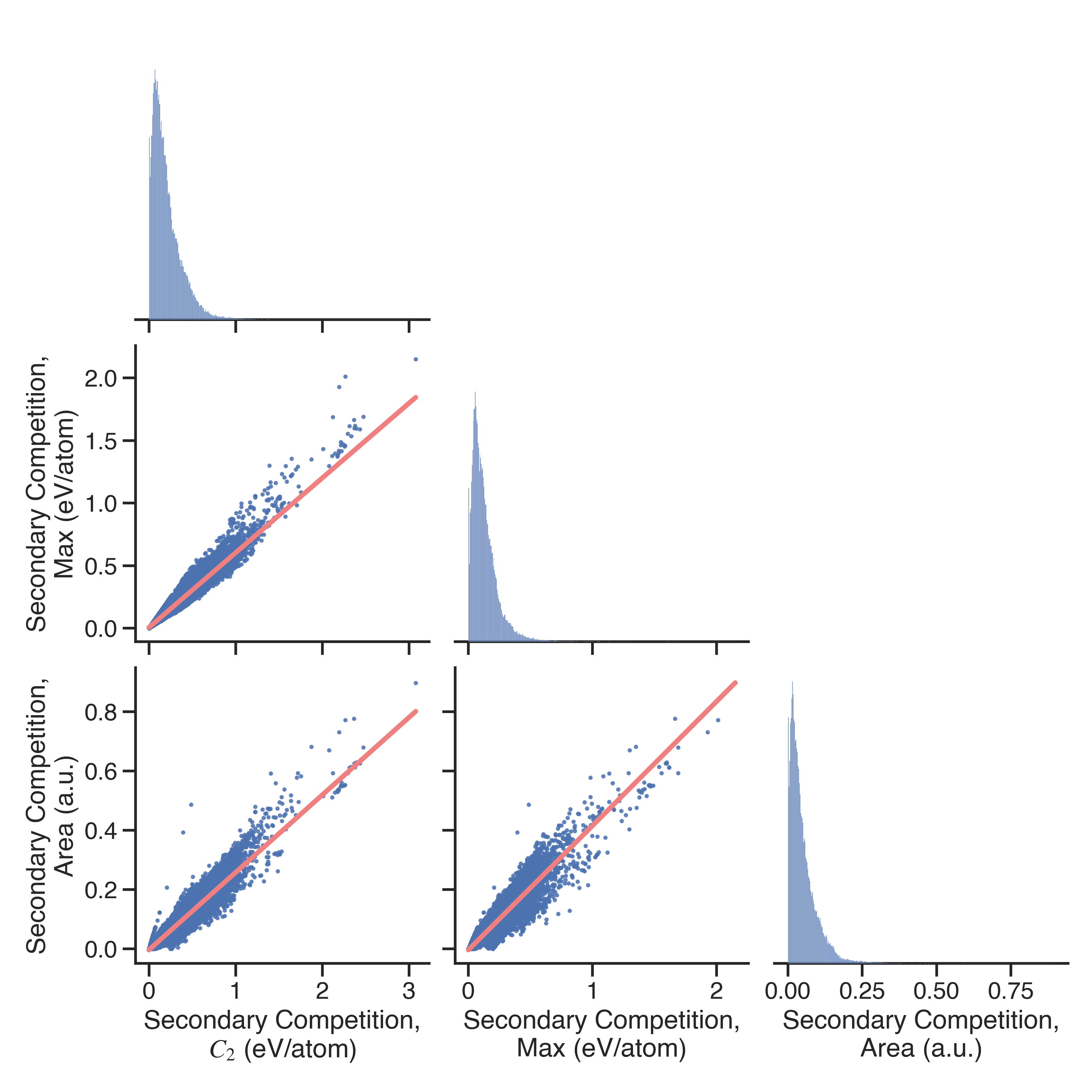}
\caption{\label{supp_fig:correlations_secondary} \textbf{Pairwise correlations of alternative secondary competition metrics.} Secondary competition (max) is defined as the sum of the energies of only the secondary reactions with the highest driving forces on either side of the target. Secondary competition (area) is the enclosed area of the interface reaction hull.}
\end{center} 
\end{figure}

\pagebreak

\begin{figure}[ht!]
\begin{center}
\includegraphics[width=1.0\textwidth]{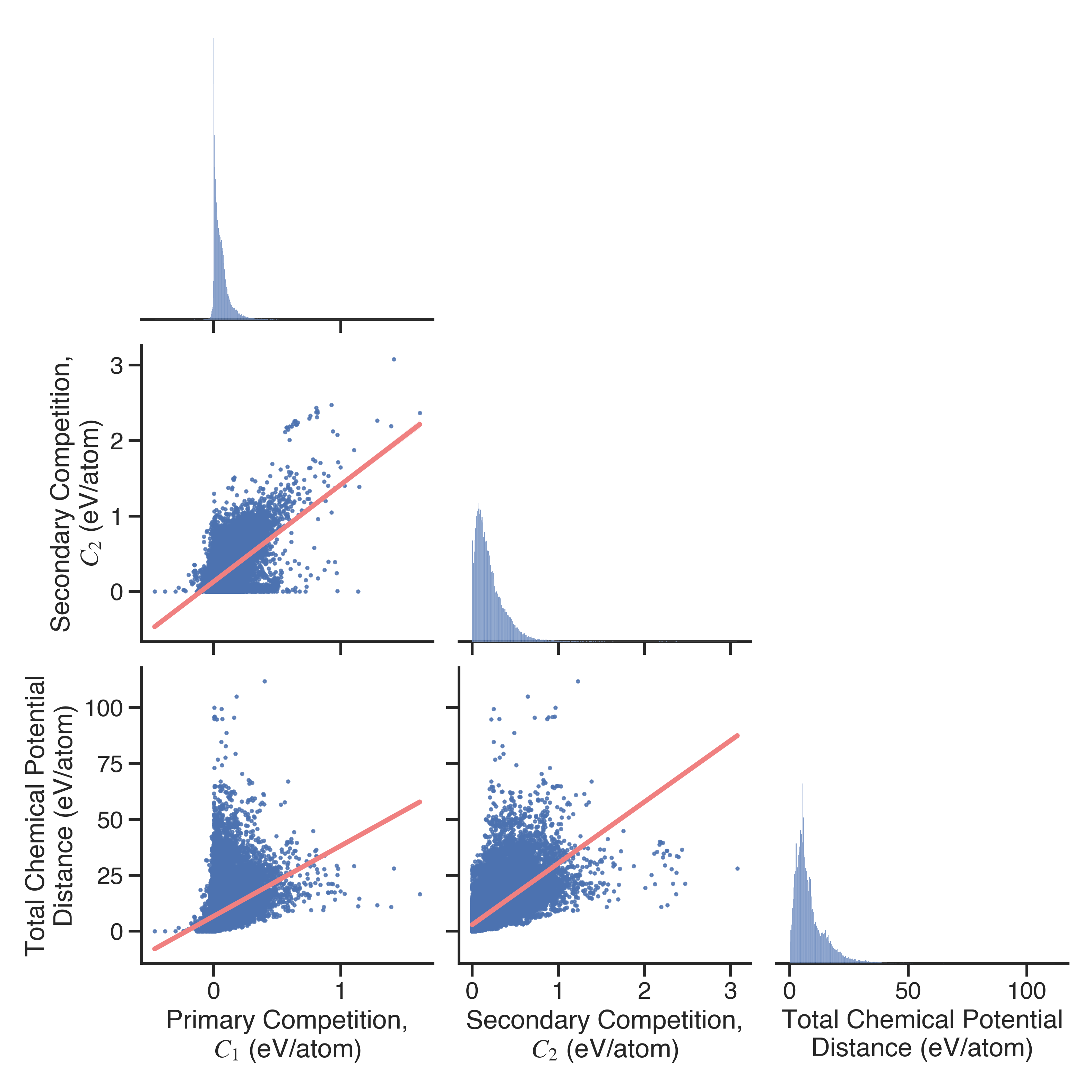}
\caption{\label{supp_fig:correlations} \textbf{Pairwise correlations of selectivity metrics.} The total chemical potential distance is calculated using the methodology outlined in Ref. \citenum{Todd2021_SI}.}
\end{center} 
\end{figure}

\pagebreak

\begin{figure}[ht!]
\begin{center}
\includegraphics[width=0.8\textwidth]{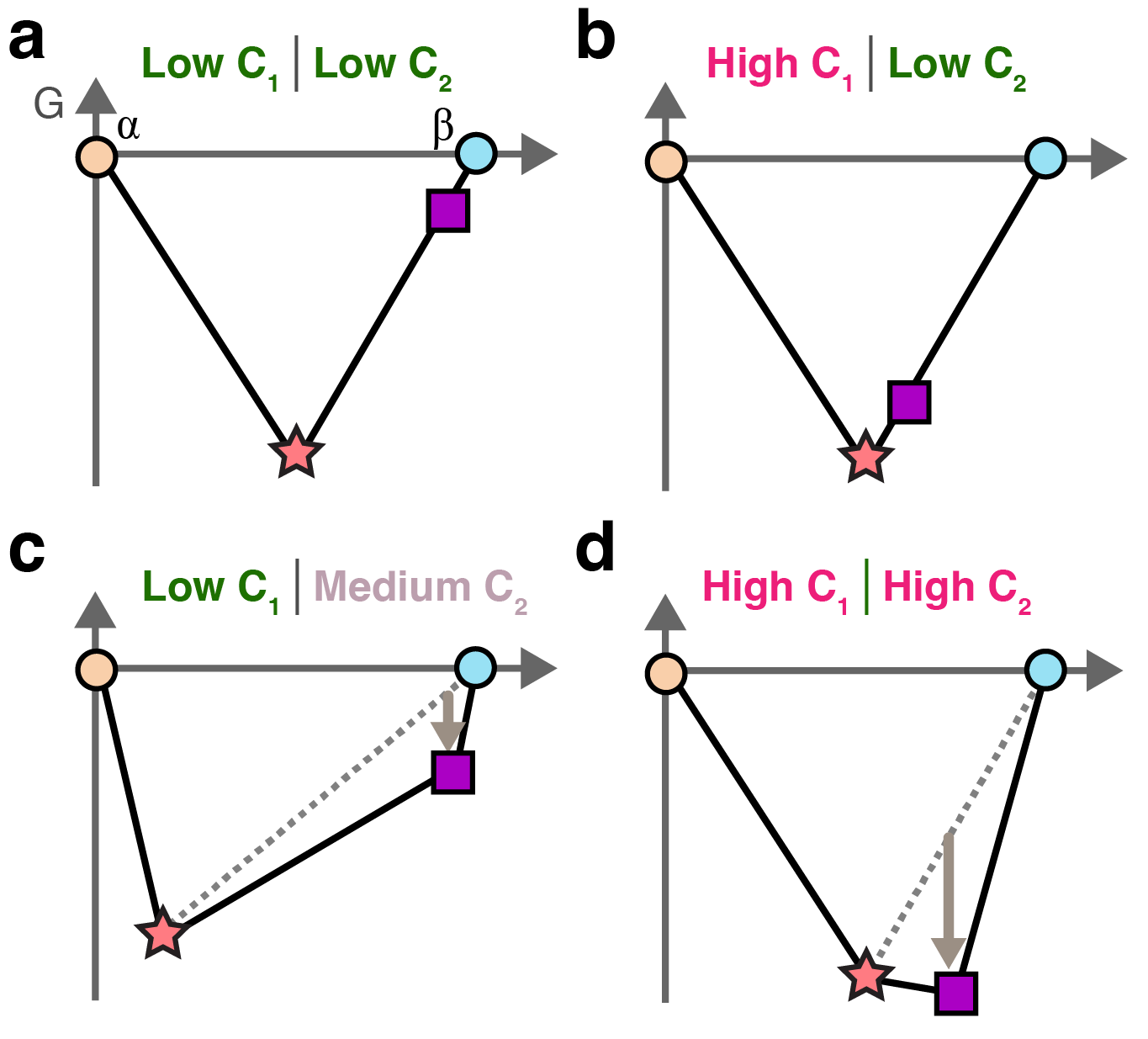}
\caption{\label{supp_fig:competition_extremes} \textbf{Examples of varying $C_1$ and $C_2$ indicating their partial correlation.} (a, b) When secondary competition ($C_2$) is low, primary competition ($C_1$) can either be low or high. c) However, when $C_1$ is low, $C_2$ can not be high due to the geometric constraints of the hull. d) There are no restrictions for both $C_1$ and $C_2$ to be high.}
\end{center} 
\end{figure}

\pagebreak

\section{Formation energy correction for carbonates}
\begin{figure}[ht!]
\begin{center}
\includegraphics[width=1.0\textwidth]{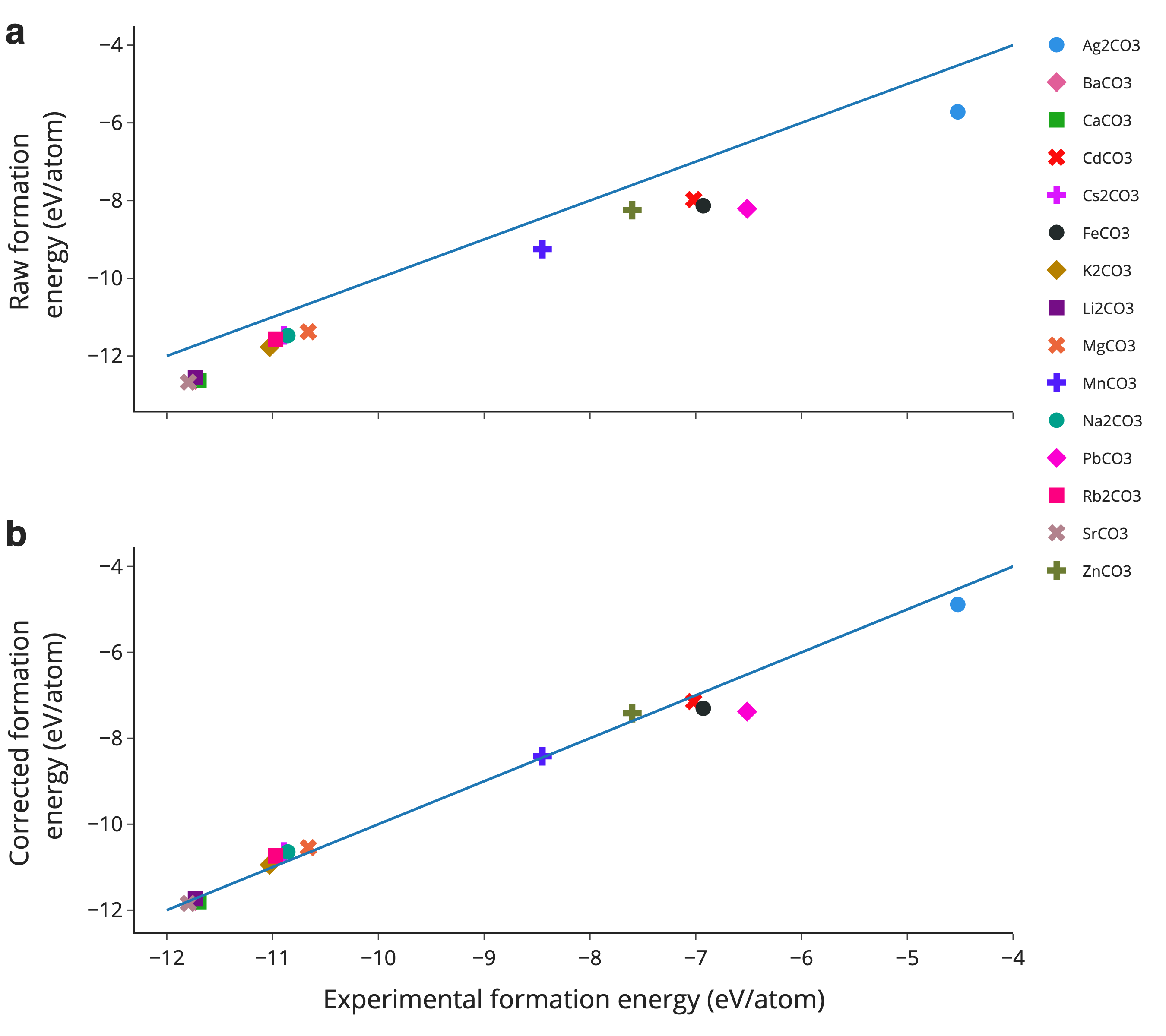}
\caption{\textbf{Fitting procedure for formation energy correction of carbonate compounds.} (a) Carbonates exhibit a systematic negative shift in predicted Gibbs free energy of formation, $\Delta G_f$ ($T=$300 K) compared to the experimental values. (b) Energies of the same carbonate compounds after applying a fit energy correction of 0.830 eV/CO$_3^{2-}$.}
\label{supp_fig:carbonates}
\end{center} 
\end{figure}

\pagebreak

\section{Gradient furnace setup, heating, and calibration}

\begin{figure}[ht!]
\begin{center}
\includegraphics[width=0.8\textwidth]{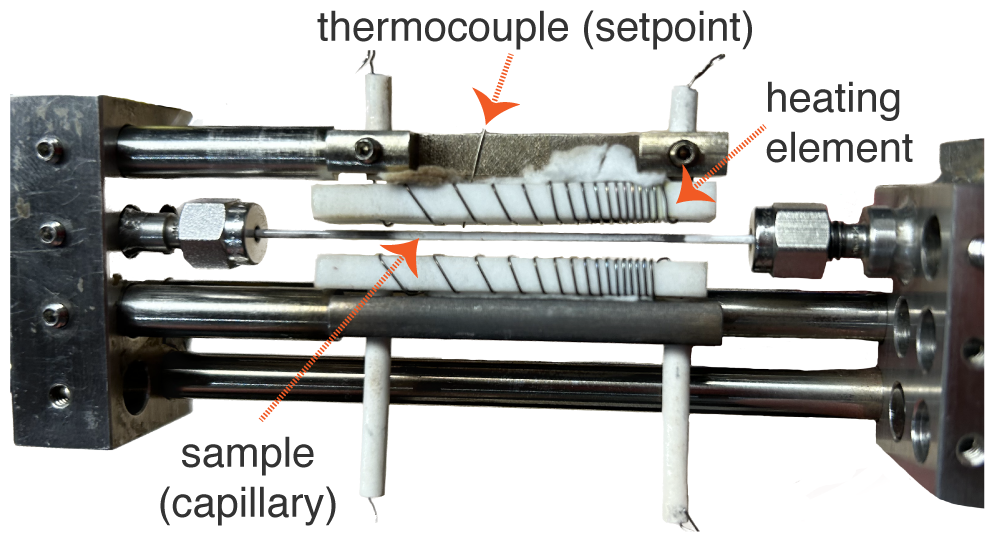}
\caption{\textbf{The gradient furnace used in all \textit{ex post facto} SPXRD experiments}. The pictured device has the same specifications as described in Ref. \citenum{ONolan2020_SI}. The heating wires are wound with variable pitch, resulting in a wide temperature profile over the powder sample (capillary tube). A thermocouple is used as a setpoint for determining the power supplied to the heating elements.}
\label{supp_fig:gradient_furnace}
\end{center} 
\end{figure}

\begin{figure}[ht!]
\begin{center}
\includegraphics[width=0.7\textwidth]{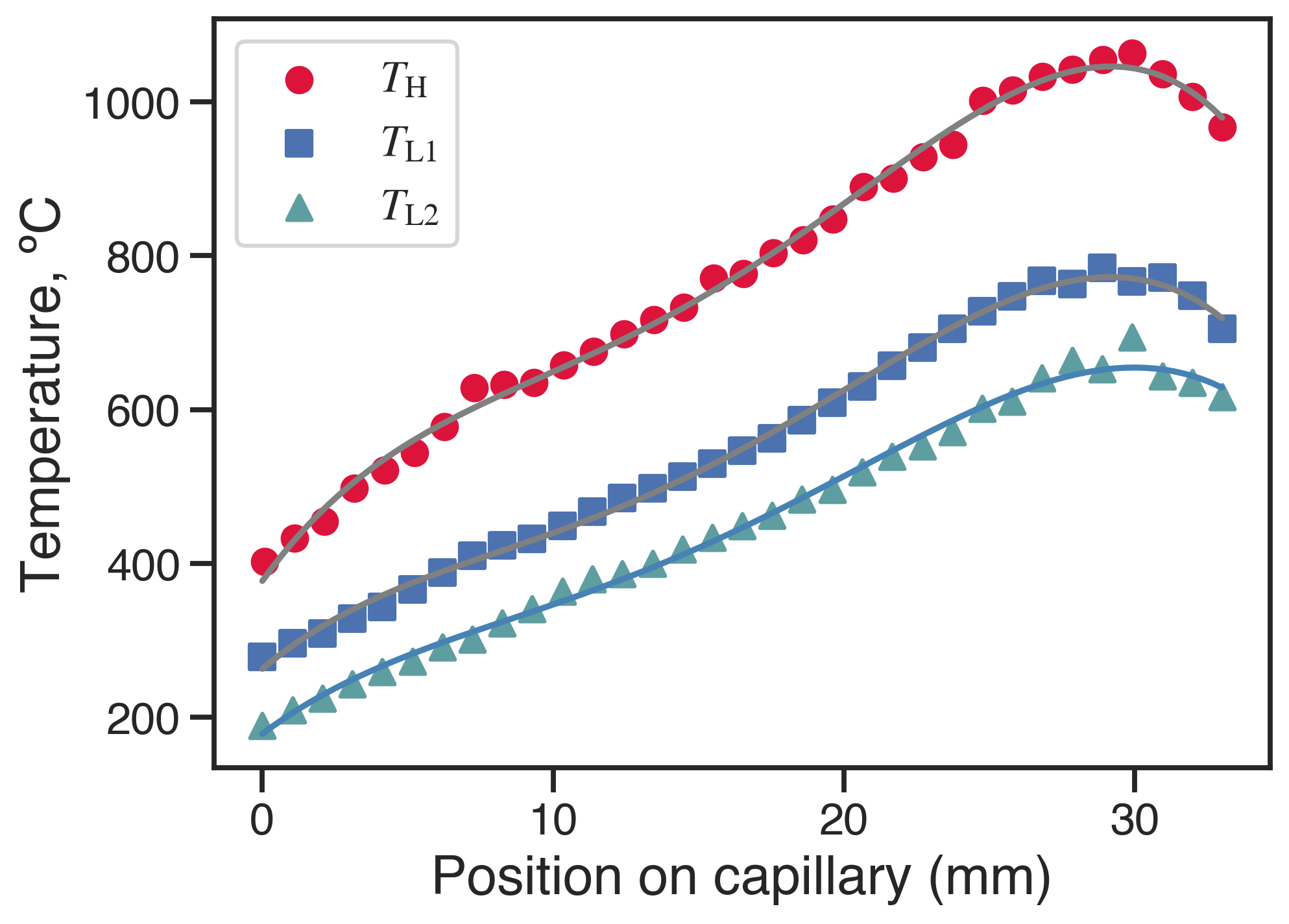}
\caption{\textbf{Measured temperature distributions along the length of sample capillary.} Temperatures were determined by refinement of lattice parameters from NaCl/Si ($T_\text{L1}$, $T_\text{L2}$) or \ce{Al2O3}/\ce{MgO} standards ($T_\text{H}$). The calibration curves correspond to gradient furnace setpoints of $T_H=550$ \textcelsius{}, $T_{L1}=450$ \textcelsius{}, and $T_{L2}=400$ \textcelsius{}.}
\label{supp_fig:gradient_calibration}
\end{center} 
\end{figure}

\begin{table}[ht!]
\caption{Reaction times for selected \ce{BaTiO3} experiments. All times are shown in MM:SS format (minutes/seconds). Temperatures correspond to gradient furnace setpoints of $T_H=550$ \textcelsius{}, $T_{L1}=450$ \textcelsius{}, and $T_{L2}=400$ \textcelsius{}.}
\label{tab:exp_times}
\begin{tabular}{ccccccc}
\toprule
 & Precursors & Heat & Hold & Cool & Total & Temp. \\
Expt. &  &  &  &  &  &  \\
\midrule
\textbf{1} & BaCO$_{3}$-TiO$_{2}$ & 09:08 & 43:28 & 14:05 & 66:41 & $T_H$ \\
\textbf{2} & BaO$_{2}$-TiO$_{2}$ & 01:47 & 64:38 & 13:00 & 79:25 & $T_{L1}$ \\
\textbf{3} & Ba$_{2}$TiO$_{4}$-TiO$_{2}$ & 06:44 & 18:22 & 11:48 & 36:54 & $T_H$ \\
\textbf{4} & Ba$_{2}$TiO$_{4}$-BaTi$_{2}$O$_{5}$ & 12:07 & 14:30 & 12:35 & 39:12 & $T_H$ \\
\textbf{5} & Ba(OH)$_2\cdot$H$_2$O-Ti & 07:15 & 58:30 & 14:33 & 80:18 & $T_{L1}$ \\
\textbf{6} & BaCl$_{2}$-Na$_{2}$TiO$_{3}$ & 01:36 & 65:04 & 22:43 & 89:23 & $T_{L2}$ \\
\textbf{7} & BaS-Na$_{2}$TiO$_{3}$ & 02:14 & 61:26 & 28:16 & 91:56 & $T_{L1}$ \\
\textbf{8} & BaS-TiO$_{2}$ & 09:03 & 14:24 & 11:46 & 35:13 & $T_H$ \\
\textbf{9} & BaSO$_{4}$-TiO$_{2}$ & 07:15 & 25:54 & 13:46 & 46:55 & $T_{L1}$ \\
\bottomrule
\end{tabular}
\end{table}

\bibliography{si_refs}